\documentclass[compsoc,journal]{IEEEtran}

\usepackage{hyperref}
\usepackage{xcolor,colortbl}
\newcommand{\fig}[1]{Figure~\ref{fig:#1}}

\newcommand{\tion}[1]{\S\ref{sec:#1}}
\usepackage{balance}
\usepackage{wrapfig}
\usepackage{multirow}
\usepackage{blindtext, graphicx}
\usepackage{boldline}
\usepackage{subfig}
\usepackage{booktabs}
\usepackage{ragged2e}
\usepackage{dblfloatfix}
\usepackage[para,online,flushleft]{threeparttable}
\usepackage{multirow}
\newcommand{\IT}{\textbf{Jitterbug}}

\usepackage{lipsum}

\usepackage[tikz]{bclogo}
\newenvironment{RQ}[1]%
{\noindent\begin{minipage}[c]{\linewidth}%
\begin{bclogo}[couleur=gray!25,%
                arrondi=0.1,%
                logo=\bctrombone,%
                ombre=true]{{\small ~#1}}}%
{\end{bclogo}\end{minipage}\vspace{2mm}}

\newcommand{\bi}{\begin{itemize}}
\newcommand{\ei}{\end{itemize}}

\usepackage{chngcntr}

\usepackage{enumitem}
\setlist[itemize]{leftmargin=*}
\setlist[enumerate]{leftmargin=*}
\usepackage{makecell}
\usepackage[linesnumbered,ruled,vlined]{algorithm2e}
\setlist{nolistsep}

\setlist[1]{itemsep=0pt}

\usepackage{amsmath}

\usepackage{color}
\usepackage{listings}
\usepackage{times}
\usepackage{colortbl}
\usepackage{tikz}

\makeatletter
\let\th@plain\relax
\makeatother

\definecolor{Gray}{rgb}{0.88,1,1}
\definecolor{Gray}{gray}{0.85}
\definecolor{lightgray}{gray}{0.8}
\usepackage[framed]{ntheorem}
\usepackage{framed}
\usepackage{tikz}
\usetikzlibrary{shadows}
\theoremclass{Lesson}
\theoremstyle{break}
\usepackage{newtxtext,newtxmath,amsmath}
\tikzstyle{thmbox} = [rectangle, rounded corners, draw=black,
fill=gray!25,  drop shadow={fill=black, opacity=1}]

\newcolumntype{P}[1]{>{\centering\arraybackslash}p{#1}}

\definecolor{myGray}{RGB}{220, 220, 220}
\definecolor{myRed}{RGB}{156, 4, 4}
\definecolor{myBlue}{RGB}{108, 156, 236}
\definecolor{myGreen}{RGB}{108, 172, 76}

\usepackage{xcolor}

\bstctlcite{IEEEexample:BSTcontrol}

\begin{document}

\SetKwProg{Fn}{Function}{}{}
%

\author{
    \IEEEauthorblockN{\IEEEauthorrefmark{1}Zhe Yu\IEEEcompsocitemizethanks{\IEEEcompsocthanksitem 
E-mail: zxyvse@rit.edu},~\IEEEauthorrefmark{2}Fahmid Morshed Fahid,~
        \IEEEauthorrefmark{2}Huy Tu,~
        and \IEEEauthorrefmark{2}Tim Menzies,~\IEEEmembership{Fellow,~IEEE}\IEEEcompsocitemizethanks{\IEEEcompsocthanksitem 
E-mail: \{ffahid, hqtu\}@ncsu.edu, timm@ieee.org.}}\\
    \IEEEauthorblockA{\IEEEauthorrefmark{1}Department of Software Engineering, Rochester Institute of Technology, USA}\\
    \IEEEauthorblockA{\IEEEauthorrefmark{2}Department
of Computer Science, North Carolina State University, USA}
}




\title{Identifying Self-Admitted Technical Debts with {\IT}: A Two-Step Approach}

\IEEEtitleabstractindextext{%
{\justify\begin{abstract}
Keeping track of and managing Self-Admitted Technical Debts (SATDs) are important to maintaining a healthy software project. This requires much time and effort from human experts to identify the SATDs manually. The current automated solutions do not have satisfactory precision and recall in identifying SATDs to fully automate the process. To solve the above problems, we propose a two-step framework called {\IT} for identifying SATDs. {\IT} first identifies the ``easy to find'' SATDs automatically with close to 100\% precision using a novel pattern recognition technique. Subsequently, machine learning techniques are applied to assist human experts in manually identifying the remaining ``hard to find'' SATDs with reduced human effort. Our simulation studies on ten software projects show that {\IT} can identify SATDs more efficiently (with less human effort) than the prior state-of-the-art methods.

\end{abstract}}
\begin{IEEEkeywords}
Technical debt, software engineering, machine learning, pattern recognition.
\end{IEEEkeywords}}

\IEEEdisplaynontitleabstractindextext

\maketitle
\section{Introduction}
\label{sec:introduction}

Recently,  much  research  has  been  focused on identifying the  “Self-Admitted  Technical  Debts”  (SATDs) from source code comments. Keeping track of and managing these SATDs are important to maintaining a healthy software project as they (1) are diffused in the codebase; (2) increase over time and accumulate interests when not fixed in time; (3) even when fixed, it survives long time (over 1,000 commits on average) in the system~\cite{bavota2016large}; and (4)  make the code more difficult to change in the future~\cite{wehaibi2016examining}. What we found in this work is that there are two types of SATDs:
\bi
\item
\textbf{The ``easy to find'' SATDs}, which can be automatically identified without human verification in almost 100\% precision. As an example, the comments  containing  keywords  like ``{\em fixme,  todo}'' are  almost always related to SATDs.
\item
\textbf{The ``hard to find'' SATDs}, which only human experts can accurately decide whether they are SATDs or not. As an example, the comment ``{\em Modify the system class loader instead - horrible! But it works!}'' can be easily classified as an SATD by human experts but remains a hard problem for algorithms.
\ei
The most important message we want to convey is
\begin{lesson*}
\textbf{Do not waste effort on finding the ``easy to find'' SATDs, focus more on identifying the ``hard to find'' SATDs.}
\end{lesson*}

Current solutions for identifying SATDs do not separate the two types of SATDs, and belong to either {\em pattern-based approaches} or {\em machine learning approaches}. Researchers exploring {\em pattern-based approaches} first manually inspect code comments and label each one as SATD or non-SATD, then manually analyze the labeled items and summarize patterns for SATDs, e.g. if a comment has keywords like ``{\em hack, fixme, probably a bug}'', then it has a high chance of being related to a SATD. On the other hand, {\em machine learning approaches} first train a classification model on the manually labeled comments, then predict for which comments are related to SATDs (usually on a ``hold-out'' test set so that performance metrics like precision and recall can be calculated). Limitations exist in both approaches:
\bi
\item
{\em Pattern-based approaches} require large amounts of human effort in analyzing and summarizing effective patterns.
\item
Since not all SATDs are ``easy to find,'' many of the patterns identified by the {\em pattern-based approaches} from some source projects could be ineffective in a new, unseen project.
\item
Even the state-of-the-art {\em machine learning approaches} can only reach around 78\% F1 score~\cite{ren2019neural} and 74\% AUC~\cite{zampetti2019automatically} (to which the ``easy to find'' SATDs contribute greatly). That means the process cannot be fully automated without human experts checking the algorithms' decisions and making the final call.
\ei

\begin{figure*}
\begin{center}
\includegraphics[width=\textwidth]{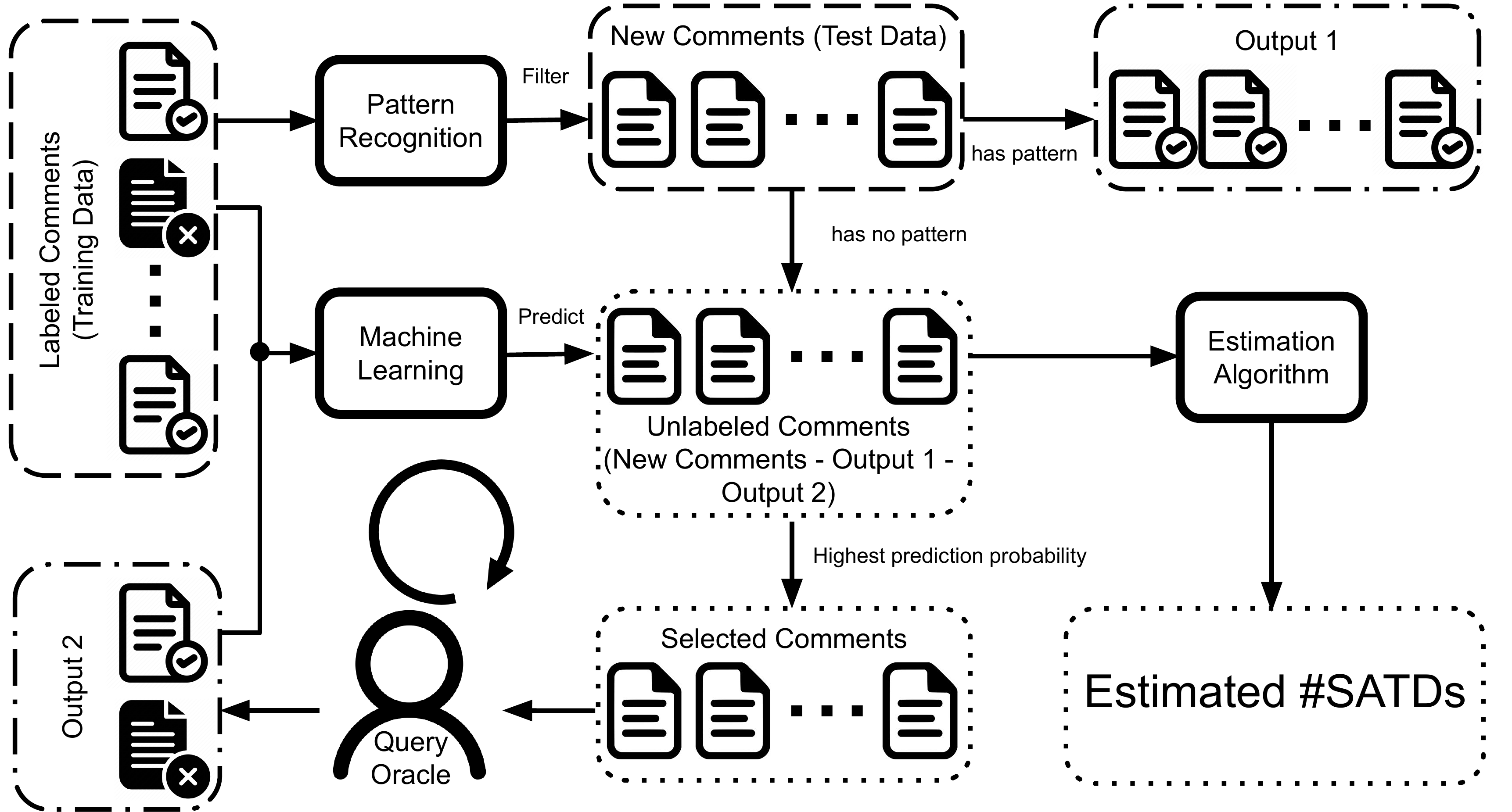}
\end{center}
\caption{Workflow of {\IT}.}\label{fig:workflow}
\end{figure*}

Acknowledging the existence of the two types of SATDs, we address the SATD identification problem in two steps:
\bi
\item
{\em\textbf{Step 1: identify the ``easy to find'' SATDs automatically}}. The comments containing keywords like {\em``fixme, todo''} are almost always related to SATDs. This suggests that there exist strong patterns that could be used to identify such ``easy to find'' SATDs automatically, with very high precision. The key challenge of this step is to automatically identify these strong patterns with close to 100\% precision so that human experts do not need to verify the results.
\item
{\em\textbf{Step 2: guide human experts to manually read the comments without strong patterns looking for the remaining ``hard to find'' SATDs}}. The remaining ``hard to find'' SATDs cannot be accurately identified through machine learning algorithms. Human efforts are essential for identifying such SATDs. Therefore, the key challenge of this step is to (1) guide the human effort to the comments that most likely contain SATDs, and (2) provide information such as an estimation of the number of undiscovered SATDs to help human experts make trade-off choices, such as deciding on when to stop the process.
\ei
As shown in \fig{workflow}, We designed {\IT}, a two-step framework. In \textbf{Step 1}, {\IT} utilizes a novel pattern recognition technique to identify patterns that could yield very high precision (if one comment has the recorded patterns then there is a close to 100\% chance it is related to SATDs). In \textbf{Step 2}, Then machine learning models are trained to guide humans to discover SATDs from comments that do not have high precision patterns, as well as to estimate the number of SATDs left in the comments. This idea of separating the SATD identification problem into two steps provides the following advantages: (1) Mining patterns in \textbf{Step 1} becomes easy since high recall is no longer a requirement. (2) Human efforts are only spent in \textbf{Step 2} on the ``hard to find'' SATDs--- there is zero human effort and consequently negligible cost for finding the ``easy to find'' SATDs in \textbf{Step 1}.

Simulated on the latest SATD dataset from Maldonado and Shihab~\cite{maldonado2015detecting}, we ask and answer the following research questions.

\textbf{RQ1: How to find the strong patterns of the ``easy to find'' SATDs in Step 1?} First, on 9 training projects, a pattern recognizer named \textbf{Easy}, with fitness function specifically designed to achieve high precision, is applied to identify patterns with precision higher than 80\%. Then, the identified patterns are used on the holdout project to test the performance. We also conduct a validation study by manually analyzing the comments containing the strong patterns found with \textbf{Easy} but were labeled as Non-SATDs in the dataset. Interesting findings were discovered during our exploration of this step:
\begin{enumerate}
\item
\textbf{Easy} detects the same set of strong patterns--- {\em ``todo, fixme, hack, workaround''} for every target project.
\item
\textbf{Easy} achieves close to 100\% precision (100\% on eight projects and 99\% on two projects) on identifying the ``easy to find'' SATDs. These results are higher than the human-derived set of patterns--- {\em ``todo, fixme, hack, xxx''} from Guo et al.~\cite{guo2019mat} (MAT).
\item
\textbf{Easy} is even more accurate than human experts in identifying the ``easy to find'' SATDs since 98\% of the conflicting comments, which were labeled as Non-SATDs by humans but contain the patterns from \textbf{Easy}, are identified as SATDs in our validation study. 
\item
Although \textbf{Easy} is an algorithm with close to 100\% precision and barely any cost (training takes seconds), it alone can only identify 20\% to 90\% of the SATDs. Thus it is necessary for \textbf{Step 2} to find the remaining ``hard to find'' SATDs.
\end{enumerate}

\textbf{RQ2: How to find the ``hard to find'' SATDs efficiently with human experts?} After all the ``easy to find'' SATDs are filtered out, only the ``hard to find'' SATDs persist in the dataset. First, we will show in RQ2.1 that the “hard to find” SATDs cannot be automatically detected without human oracles. Then the human-in-the-loop approach \textbf{Hard} is designed in RQ2.2 and RQ2.3. As shown in \fig{workflow}, a machine learning model is trained to rank the remaining comments. Human oracles are queried for the top rank comments and then those oracles are used to update the machine learning model. This loop will iterate until the target level of recall has been reached in estimation. The advantage of this human-in-the-loop strategy is that the information in both source projects and newly labeled data in the target project can be utilized to better direct human effort towards comments that are more likely to contain SATDs. Our results in RQ2.2 show that this strategy finds more SATDs with fewer human oracles than the state-of-the-art supervised learning techniques where only information from the training datasets is utilized. Meanwhile, RQ2.3 shows that the number of undiscovered SATDs can be accurately estimated by \textbf{Hard}, thus helping the human experts make decisions on whether to spend more time looking for the ``hard to find'' SATDs or to stop at that point. 

\textbf{RQ3: Overall how does {\IT} perform?} 
We evaluate the overall performance of {\IT} with three sub RQs. In RQ3.1, it is shown that {\IT} can always find more SATDs with less human effort compared to other state-of-the-art methods as well as \textbf{Easy} or \textbf{Hard} alone. RQ3.2 evaluates {\IT} when it stops at 90\% target recall. Results show that {\IT} always achieves higher recall than another more complex baseline algorithm CNN~\cite{ren2019neural}, and {\IT} binary dominates CNN~\cite{ren2019neural} in terms of recall and cost on 2 of the 10 projects. This suggests that {\IT} is a better framework than CNN~\cite{ren2019neural} while the model of CNN~\cite{ren2019neural} has the potential to further improve the performance of {\IT}. In RQ3.3, the computational overhead of {\IT} is similar to that of a traditional supervised learning model (34 seconds), and is much lower than training a deep learning model (3,548 seconds~\cite{ren2019neural}). With a human reviewing 3 comments per minute (estimated with our own experience in classifying SATD comments), on a medium-sized project with 5,000 comments, {\IT} finds in median 97\% of the SATDs in 4.5 hours while reviewing all the comments would have cost 28 hours. Therefore, more than 23 hours of human work can be saved for each project when using {\IT}.

\subsection{Contributions of this Paper}
\begin{enumerate}
\item
In this paper, we show that there are two types of SATDs: the ``easy to find'' ones that can be identified without human verification, and the ``hard to find'' ones that only humans can make the final decisions on. 
\item
A novel two-step framework {\IT} is proposed to identify the two types of SATDs. This framework first identifies the ``easy to find'' SATDs automatically with a novel pattern recognition technique, then applies machine learning techniques to assist human experts in manually identifying the remaining ``hard to find'' SATDs with reduced human effort.
\item
A novel pattern recognition technique \textbf{Easy} is presented to find strong patterns with close to 100\% precision for the ``easy to find'' SATDs. Results show that its precision is even higher than humans', thus making it reliable to be applied automatically.
\item
A continuous learning framework \textbf{Hard} is shown to outperform other supervised learning models in retrieving the ``hard to find'' SATDs with less human effort, and also to provide information on how many more SATDs there are to be found.
\item
All code and data in this work
are available\footnote{\url{https://github.com/ai-se/Jitterbug}}, allowing other researchers to replicate, improve, or even refute our findings.
\end{enumerate}

The rest of this paper is structured as follows. Background and related work are discussed in \tion{related-work}. Our methodology is described in  \tion{methodology}. This is followed by the details of the SATD datasets in \tion{case-study}. Experiment (simulation) design and answers to the research questions are then presented in \tion{experiments}. In \tion{apply}, {\IT} is applied to identify SATDs from a real-world, unlabeled software project with a human reading the comments to test its generalizability. Threats to the validity of this work are analyzed in \tion{Threats to Validity}. Lastly, conclusion and future work are provided in \tion{conclusion}.

\section{Background and Related Work}
\label{sec:related-work}
\subsection{About Technical Debt}

When developers cut corners and make haste to rush out code, that code
often contains {\em technical debt} (TD), i.e. decisions that must be repaid, later on, with further work.
Technical debt is like dirt in the gears of software production.
As TD accumulates, development becomes harder and slower.  Ever since the term technical debt (TD) was first introduced by Cunningham in 1993~\cite{cunningham1993wycash}, it has been found to be a widespread problem in the software industry damaging many aspects of a system including {\em evolvability} (how fast we can add new functionality) and {\em maintainability} (how well we can keep bugs out of the code)~\cite{cunningham1993wycash,guo2011tracking,nugroho2011empirical}:
\bi
\item
In 2012, after interviewing 35 software developers from diverse projects in different companies, varying both in size and type, Lim et al.~\cite{lim2012balancing}
found  developers generate TD  due to factors like  increased workload,  unrealistic deadline in projects, lack of knowledge, boredom,
 peer-pressure among developers, unawareness or short-term business goals of stakeholders, and reuse of legacy or third party or open-source code. 
\item
After observing five large-scale projects, Wehaibi et al.~\cite{wehaibi2016examining} found that the number of technical debts in a project may be very low (only 3\% on average), yet they create a significant amount of defects in the future (and fixing such technical debts are more difficult than regular defects). 
\item
Another study on five software large-scale companies revealed that TDs contaminate other parts of a software system and most of the future interests are non-linear in nature with respect to time~\cite{martini2015danger}. 
\item
According to the SIG (Software Improvement Group) study of Nugroho et al.~\cite{nugroho2011empirical}, a regular mid-level project owes $\$857,500$ in TD and resolving TD has a Return On Investment (ROI) of 15\% in seven years. 
\item
Guo et al.~\cite{guo2011tracking} also found similar results and concluded that the cost of resolving TD in the future is twice as much as resolving immediately. 
\item
As Ozkaya et al.~\cite{Ozkaya12} revealed
, technical debt  affects
multiple aspects of the software development process and is mostly invisible.
\ei
Therefore, identifying TD has a large impact on software development. However, limited success has been achieved while much research tried to identify TD as part of Code Smells using static code analysis~\cite{marinescu2010incode,marinescu2004detection,zazworka2013case,marinescu2012assessing,fontana2012investigating}. Static code analysis has a high rate of false alarms while imposing complex and heavy structures for identifying TD~\cite{tsantalis2011identification,tsantalis2015assessing,graf2010speeding,ali2012application}.

\begin{table*}[!tbh]
\caption{Differences between our approach and Guo et al.~\cite{guo2019mat}}
\label{tab:diffs}
\begin{center}
\begin{tabular}{l|p{7cm}|p{8cm}|}
& \textbf{How to find patterns for the ``easy to find'' SATDs}   & \textbf{How to find the ``hard to find'' SATDs} \\ \hline
\textbf{Guo et al.~\cite{guo2019mat}}   & Manually find patterns from the test set. Require large amounts of human effort. Performances are tested on the same data used for finding those patterns. & Train a supervised learning model on the training set and test its classification performance on a holdout test set. Users have little control of the recall and precision achieved.     \\ \hline
\textbf{Our approach} & Automatically mine patterns from the training set. No human effort cost. Utilize a holdout set to validate the performance of the mined patterns.  & Continuously train/update a model on both training set and labeled data from the test set, then use the model to select comments for human experts to read, these human decisions are then used as new labeled data for updating the model. Also apply another model to estimate the total number of SATDs in the comments, thus providing information for the user about what level of recall has been achieved. \\\hline
\end{tabular}
\end{center}
\end{table*}

\subsection{Identifying Self-Admitted Technical Debt}

Recently, much more success has been seen in work on the ``self-admitted technical debt'' (SATD).
Technical debt is often ``self-admitted'' by the developer in code comments~\cite{potdar2014exploratory}, thus making it much easier to find. Identifying and tracking these SATDs have large benefits:
\bi
\item
Removing the SATDs early reduces the maintenance cost of a software project. As reported by Wehaibi et al.~\cite{wehaibi2016examining} in 2016, these SATDs have  negative implications on the software
development process in particular by making it more difficult to change in the future.
\item
SATDs can provide cheap training data for learning to identify technical debts (TDs). SATDs are not a specific type of TDs, rather they are the TDs that have been ``admitted'' by the developers. SATDs also cover the different types of TDs such as code, defect, and requirement debts~\cite{bavota2016large}.
\ei
In 2014, after studying four large-scale open-source software projects, Potdar and Shihab~\cite{potdar2014exploratory} concluded that developers intentionally leave traces of TD in their comments (saying things like ``{\em hack, fixme, is problematic, this isn't very solid, probably a bug, hope everything will work, fix this crap}''). 

\subsubsection{Pattern-Based Approaches}
{\em Pattern-based approaches}~\cite{potdar2014exploratory,de2015contextualized,de2016investigating,de2020identifying,maldonado2015detecting} consist of three steps: (1) manually inspect code comments and label each one as SATD or non-SATD; (2) manually analyze the labeled items and summarize patterns for SATDs, e.g. if a comment has keywords like ``{\em hack, fixme, probably a bug}'', then it has a high chance of being related to a SATD; (3) apply the summarized patterns to unlabeled comments to identify SATDs. 

Potdar and Shihar's work is now considered the first {\em pattern-based} approach for identifying SATDs. They found 62 distinct keywords for identifying such TDs~\cite{potdar2014exploratory}
(similar conclusions were made by Faris et al.~\cite{de2015contextualized,de2016investigating,de2020identifying}). 
In 2015, Maldonado et al. used five open-source projects to manually classify different types of SATDs ~\cite{maldonado2015detecting} and found
\bi
\item
SATDs mostly contain Requirement Debt and Design Debt in source code comments; 
\item 
75\% of SATDs get removed, but the median lifetime of SATDs ranges from 18 to 173 days~\cite{maldonado2017empirical}. 
\ei
Another study tried to find the SATD-introducing commits in Github using different features on change level~\cite{yan2018automating}. Instead of using the bag of word approach, a recent study also proposed word embedding as a vectorization technique for identifying SATD~\cite{flisar2018enhanced}. These {\em pattern-based} studies focused on identifying keywords in code comments indicating SATDs and then used those keywords to label comments as SATDs~\cite{sierra2019survey}. 

There are risks and problems to this approach. First, it requires much manual effort from human experts to find those keywords by reading thousands of comments. Second, it is natural to believe that such keywords can vary from projects to projects and will not produce 100\% precision and recall but none of the studies used a holdout set to evaluate the precision and recall of using such keywords to identify SATDs.

\subsubsection{Machine Learning Approaches}

To solve the above mentioned problems, {\em machine learning}~\cite{maldonado2017using,liu2018satd,huang2018identifying,zampetti2019automatically} approaches are proposed for identifying SATDs. In these approaches, supervised learning models are trained on labeled SATD datasets to learn the underlying rules of comments admitting TDs.
For example, Tan et al.~\cite{tan2007icomment,tan2012tcomment} analyzed source code comments using natural language processing to understand programming rules and documentations and indicates comment quality and inconsistency. 
A similar study was done by Khamis et al~\cite{khamis2010automatic}. 
After analyzing and categorizing comments in source code, Steidl et al.~\cite{steidl2013quality} proposed a machine learning technique that can measure the comment quality according to category. 
Malik et al.~\cite{malik2008understanding} used a random forest classifier to understand the lifetime of code comments. 
A similar study on three open-source projects was also done by Fluri et al.~\cite{fluri2007code}. 
In 2017, Maldonado et al.~\cite{maldonado2017using} successfully identified two types of SATD in 10 open-source projects (average 63\% F1 Score) using Natural Language Processing (Max Entropy Stanford Classifier) using only 23\% training data. 
A different approach was introduced by Huang et al.~\cite{liu2018satd} in 2018. 
Using eight datasets, Huang et al. build a Multinomial Naive Bayes sub-classifier for each training dataset using information gain as feature selection.
By implementing a boosting technique using all those sub-classifiers, they have found an average of 73\% F1 scores for all datasets~\cite{huang2018identifying}. 
A recent IDE for Eclipse was also released using this technique for identifying SATD in java projects~\cite{liu2018satd}. More recently, Zampetti et al.~\cite{zampetti2019automatically} reported an average precision of 55\%, recall of 57\%, and AUC of 0.73 with a deep learning-based approach.
Recently, some studies explore different feature engineering for identifying SATDs, e.g. Wattanakriengkrai et al.~\cite{wattanakriengkrai2019automatic} applied N-gram IDF as features, and Flisar and Podgorelec~\cite{flisar2019identification} explored how feature selection with word embedding can help the prediction. The latest progress from Ren et al.~\cite{ren2019neural} utilized a deep convolutional neural network with hyperparameter tuning to achieve a higher F1 score than all the previous solutions.

These machine learning models can be a good indicator for which comments are more likely to be related to SATDs. However, with precision ranging from 60\% to 85\%, it is not reliable to fully automate the process. Human experts are then required to verify every decision the machine learning model made and thus costs a large amount of time and labor but still finding only, say 57\% of the SATDs.

\subsubsection{Two-Step Approaches}

As described in \tion{introduction}, we take a two-step approach to identify SATDs: (1) identify patterns for the ``easy to find'' SATDs with close to 100\% precision and automatically classify comments with the patterns as SATDs (without human verification); (2) then apply machine learning techniques to guide human experts to find the remaining ``hard to find'' SATDs with least number of comments read. Interestingly, during the drafting of this paper, we found a preprint~\cite{guo2019mat} that utilized a similar idea to our two-step approach. Guo et al.~\cite{guo2019mat} used four keywords ({\em ``fixme, todo, hack, xxx''}) to identify the ``easy to find'' SATDs and applied supervised learning models to find the remaining ``hard to find'' SATDs. Although Guo et al. consider their approach as just a strong baseline, it still demonstrates the effectiveness of such two-step approaches. The differences between our approach and Guo et al.'s are listed in Table~\ref{tab:diffs}. More detailed comparisons, along with other state-of-the-art machine learning algorithms will be presented in \tion{experiments}.

\section{Methodology}
\label{sec:methodology}

As shown in \fig{workflow}, {\IT} consists of two operators--- a pattern recognizer \textbf{Easy} and a continuous learning model \textbf{Hard}. To find all possible strong patterns in Step 1, we featurize the data as a term frequency matrix without stemming or stop word removal. This section breakdowns the workflow as shown in Algorithm~\ref{alg:workflow} and introduces the two operators in detail.

\begin{algorithm}[!htp]
\scriptsize
\SetKwInOut{Input}{Input}
\SetKwInOut{Output}{Output}
\SetKwInOut{Parameter}{Parameter}
\SetKwRepeat{Do}{do}{while}
\Input{$X$, set of training data.\\$Y$, set of test data.\\$T_{rec}$, target recall of the "hard to find" SATDs.\\$CL$, the machine learning model applied.
}
\Output{$TD$, set of SATDs identified from test data.}
\BlankLine

\Fn{{\IT} ($X,Y,T_{rec},CL$)}{
    \tcp{Extract patterns from training data.}
    $patterns\leftarrow $Easy($X$)\;
    \tcp{Identify the "easy to find" SATDs.}
    $TD_{easy} \leftarrow $Has\_Pattern($Y,patterns$)\;
    \tcp{Remove "easy to find" SATDs from training and test data.}
    $Y_{hard} \leftarrow Y\setminus TD_{easy} $\;
    $X_{hard} \leftarrow X\setminus$ Has\_Pattern($X,patterns$)\;
    \tcp{Identify the "hard to find" SATDs.}
    $TD_{hard} \leftarrow$Hard($X_{hard},Y_{hard},T_{rec}$)\;
    $TD \leftarrow TD_{easy} \cup TD_{hard}$\;
    \Return{$TD$}\;
}

\caption{Psuedo Code for {\IT}.}\label{alg:workflow}
\end{algorithm}

\subsection{Easy}

\begin{algorithm}[!htp]
\scriptsize
\SetKwInOut{Input}{Input}
\SetKwInOut{Output}{Output}
\SetKwInOut{Parameter}{Parameter}
\SetKwRepeat{Do}{do}{while}
\Input{$X$, set of training data.
}
\Output{$patterns$, list of identified patterns.}
\BlankLine

\Fn{Easy ($X$)}{
    \tcp{Set precision threshold as a stopping rule.}
    $thres\leftarrow 0.8$\;
    $patterns\leftarrow []$\;
    \BlankLine
    \While{$True$}{
       \tcp{Find the pattern of highest fitness score.}
       $scores \leftarrow $ \{ p : FitnessFunction ($X,p$)    \textbf{foreach} $p \in $All\_Patterns($X$) \}\;
       \BlankLine
       $p \leftarrow $ argmax($scores$)\;
       \BlankLine
       \tcp{Check if highest precision is below the threshold.}
       \If{Precision($X, p$)$< thres$}{
            \textbf{break}\;
        }
       \tcp{Add p as one of the strong patterns.}
       $patterns$.append($p$)\;
       \tcp{Remove comments that contain p.}
       $X$.remove(Has\_Pattern($X,p$))\;
    }
    \Return{$patterns$}\;
}
\BlankLine
\Fn{FitnessFunction ($X, p$)}{
    \tcp{Calculate the fitness score of input pattern.}
    $P,TP \leftarrow$ Metrics($X,p$)\;
    $score \leftarrow TP^4 / P^3$\;
    \Return{$score$}\;
}
\BlankLine
\Fn{Precision ($X, p$)}{
    \tcp{Calculate the precision of input pattern.}
    $P,TP \leftarrow$ Metrics($X,p$)\;
    $prec \leftarrow TP / P$\;
    \Return{$prec$}\;
}
\BlankLine
\Fn{Metrics ($X, p$)}{
    \tcp{Calculate \# Positives and \# True Positives.}
    $Ps \leftarrow $Has\_Pattern($X,p$)\;
    $TPs \leftarrow$ Is\_SATD($P$)\;
    \Return{$|Ps|$, $|TPs|$}\;
}

\caption{Psuedo Code for \textbf{Easy}.}\label{alg:PR}
\end{algorithm}

Pattern Recognition is an engineering application of Machine Learning. Machine Learning deals with the construction and study of systems that can learn from data, rather than follow only explicitly programmed instructions whereas Pattern recognition is the recognition of patterns and regularities in data~\cite{Siva}. Here in {\IT}, the task of the pattern recognizer \textbf{Easy} is to find the strong patterns of the ``easy to find'' SATDs (\textbf{RQ1}). For each potential pattern (a keyword in the comments in our case), we measure two metrics:
\bi
\item
$P(p)$: the number of comments containing the pattern $p$ (positives).
\item
$TP(p)$: the number of SATD comments containing the pattern $p$ (true positives).
\ei
Derived from the above two metrics, we also have
\bi
\item
$Prec(p) = TP(p) / P(p)$: precision of the pattern $p$.
\ei
To achieve high reliability and thus fully automate the process, we want to find those patterns with very high precision. On the other hand, we also want to avoid rare patterns, e.g. if a pattern only appears once, it is not useful even with 100\% precision. As a result, we define our fitness function as
\begin{equation}
\label{eq:fitness}
Fitness(p) = Prec(p)^N\cdot P(p) = TP(p)^N / P(p)^{N-1}
\end{equation}
We set $N=4$ to find patterns with close to 100\% precision. Using our labeled training data, the pattern recognizer looks for the pattern with the highest fitness score, then removes comments containing that pattern from the training data and finds the next pattern with the highest fitness score (re-calculated). The detailed algorithm is shown in Algorithm~\ref{alg:PR}. 

\begin{algorithm}[!tbhp]
\scriptsize
\SetKwInOut{Input}{Input}
\SetKwInOut{Output}{Output}
\SetKwInOut{Parameter}{Parameter}
\SetKwRepeat{Do}{do}{while}
\Input{$X_{hard}$, labeled training data containing ``hard to find'' SATDs.\\$Y_{hard}$, unlabeled test data containing ``hard to find'' SATDs.\\$T_{rec}$, target recall (as stopping rule).}
\Output{$TD_{hard}$, ``hard to find'' SATDs identified.}
\BlankLine
\Fn{Hard ($X_{hard},Y_{hard},T_{rec}$)}{
    $Y_{labeled} \leftarrow \emptyset$\;
    $TD_{hard} \leftarrow 0$\;
    \tcp{Each time query the oracle for 10 comments.}
    $K \leftarrow 10$\;
    \While{$True$}{
        \tcp{Train the machine learning model.}
        $CL$.fit($X_{hard} \cup Y_{labeled}$)\;
        \tcp{Estimate \# "hard to find" SATDs.}
        $|R_E| \leftarrow$Estimate($CL,Y_{hard},Y_{labeled}$)\;
        \tcp{Check if target recall has been reached.}
        \If{$|TD_{hard}| / (|TD_{hard}|+|R_E|) \ge T_{rec}$}{
            break\;
        }
        \tcp{Select comments with top K prediction probability.}
        $Q\leftarrow$argsort($CL$.decision\_function($Y_{hard}\setminus Y_{labeled}$))[:$K$]\;
        $Y_{labeled} \leftarrow Y_{labeled}\cup Q$\;
        \tcp{Query oracles for the selected comments.}
        $TD_{hard}\leftarrow TD_{hard} \cup$Is\_SATD($Q$)\;
    }
    \Return{$TD_{hard}$}\;
}
\BlankLine
\Fn{Estimate ($CL,Y_{hard},Y_{labeled}$)}{
    \If{$|Y_{labeled}|==0$}{
        \Return{$NaN$}\;
    }
    $|R_E|_{last}\leftarrow 0$\;
    $Y_{unlabeled} \leftarrow Y_{hard} \setminus Y_{labeled}$\;
    \BlankLine
    \ForEach{$x \in Y_{hard}$}{
        $D(x) \leftarrow CL.decision\_function(x)$\;
        \If{$x \in Y_{labeled}$ \textbf{and} Is\_SATD($x$)}{
            $L(x)\leftarrow 1$\;
        }
        \Else{
            $L(x)\leftarrow 0 $\;
        }
    }
    \BlankLine
    $|R_E| \leftarrow \sum\limits_{x\in Y_{hard}} L(x)$\;
    \BlankLine
    \While{$|R_E|\neq |R_E|_{last}$}{
    \BlankLine
       \tcp{Fit and transform Logistic Regression}
       LogisticRegression.fit($D(Y_{hard}),L(Y_{hard})$)\;
       $LReg(Y_{unlabeled}) \leftarrow$ LogisticRegression.predict\_proba($D(Y_{unlabeled}$)\;
       \BlankLine
       $L \leftarrow TemporaryLabel(LReg,L)$\;
       \BlankLine
       $|R_E|_{last}\leftarrow |R_E|$\;
       \BlankLine
       \tcp{Estimation based on temporary labels}
       $|R_E| \leftarrow \sum\limits_{x\in Y_{hard}} L(x)$\;
    }
    \Return{$|R_E|$}\;
}
\BlankLine
\Fn{TemporaryLabel ($LReg,L$)}{
    $count \leftarrow 0$\;
    $target \leftarrow 1$\;
    $can \leftarrow []$\;
    \BlankLine
    \tcp{Sort $Y_{unlabeled}$ by descending order of $LReg$}
    $Y_{unlabeled} \leftarrow$ argsort(LReg)[::-1]\;
    \BlankLine
    \ForEach{$x \in Y_{unlabeled}$}{
        $count \leftarrow count+LReg(x)$\;
        $can$.append($x$)\;
        \If{$count \geq target$}{
            $L(can[0]) \leftarrow 1$\;
            $target \leftarrow target+1$\;
            $can \leftarrow []$\;
        }
    }
    \Return{$L$}\;
}
\caption{Psuedo Code for \textbf{Hard}}\label{alg:Hard}
\end{algorithm}

\subsection{Hard}
\label{sec:learners}

As shown in Algorithm~\ref{alg:Hard}, \textbf{Hard} utilizes a machine learner to continuously learn from both labeled data in the source projects and human decisions of comments in the target project. This machine learner in \textbf{Hard} can be any supervised learner in theory. However, since it will be updated/re-trained frequently, we only consider models that can be trained within seconds (users will not wait for more than a few seconds every time they finish a batch of comments). For this reason, we only test the following fast and simple learners listed below.

\textbf{Logistic Regression:} Logistic regression is a statistical model that in its basic form uses a logistic function to model a binary dependent variable~\cite{wright1995logistic}. A standard logistic function is a common ``S'' shape with Equation~\eqref{eq:LR}:
\begin{equation}
\label{eq:LR}
p(x) = \frac{1}{1+e^{-(\beta_0+\beta_1x)}}
\end{equation}
where $p(x)\in (0,1)$ for all $t$. Through fitting on the training data, logistic regression looks for the best parameter $\beta$ to classify input data $x$ into two target classes $\{0,1\}$. 

\textbf{Decision Tree:} Decision tree learning is a method commonly used in data mining which uses a decision tree (as a predictive model) to go from observations about an item (represented in the branches) to conclusions about the item's target value (represented in the leaves). Algorithms for constructing decision trees usually work top-down, by choosing a variable at each step that best splits the set of items~\cite{rokach2008data}. Two metrics are commonly applied to determine the best split:
\bi
\item
\textbf{Gini impurity: }$I_G(p) = \sum\limits_{i=1}^{J}p_i(1-p_i)$.
\item
\textbf{Entropy: }$I_E(p) = \sum\limits_{i=1}^{J}p_i\log_2p_i$.
\ei
Where $J$ is the number of classes and $p_i$ is the fraction of items labeled with class $i$ in the training dataset.
The algorithm will find the best split after which the value of $I_G(p)$ or $I_E(p)$ decreases the most. In this paper, we use Gini impurity.

\textbf{Random Forest:} Random forest classifier is an ensemble learning method that operates by constructing a multitude of decision trees at training time and outputting the class that is the mode of the classes of the individual trees~\cite{ho1995random}. Each decision tree from the random forest model is independently trained on all the training data but with only a subset of the features. In this way, these decision trees are 100\% accurate on training data and yet have different generalization errors. When used together for inference, these decision trees correct for each other's generalization errors and thus avoid overfitting on the training data.

\textbf{Naive Bayes:} Naive Bayes classifiers are a family of simple ``probabilistic classifiers'' based on Bayes' theorem with strong (naïve) independence assumptions between the features~\cite{rish2001empirical}. With the strong assumption that all features are mutually independent, a Naive Bayes classifier predicts the conditional probability of data $x$ belonging to class $C_i$ to be
\begin{equation}
\label{eq:NB}
p(C_{k}\mid x_{1},\dots ,x_{n}) \propto p(C_{k})\prod_{i=1}^{n}p(x_{i}\mid C_{k})
\end{equation}
where $p(C_{k})$ and $p(x_{i}\mid C_{k})$ are counted from the training data. Multinomial Naive Bayes model assumes that each $p(x_{i}\mid C_{k})$ is a multinomial distribution, which works well for text data.

\textbf{Support Vector Machine:} A Support Vector Machine (SVM) is a discriminative classifier formally defined by a separating hyperplane~\cite{suykens1999least}. Soft-margin linear SVMs are commonly used in text classification given the high dimensionality of the feature space. A soft-margin linear SVM looks for the decision hyperplane that maximizes the margin between training data of two classes while minimizing the training error (hinge loss):
\begin{equation}
\label{eq:SVM}
\min {\lambda \lVert w\rVert ^{2}+\left[{\frac {1}{n}}\sum _{i=1}^{n}\max \left(0,1-y_{i}(w\cdot x_{i}-b)\right)\right]}
\end{equation}
where the class of $x$ is predicted as $sgn(w\cdot x-b)$.

\textbf{Hard} also utilizes an estimator to estimate the number of ``hard to find'' SATDs and thus determine when to stop. This estimator, also described in Algorithm~\ref{alg:Hard}, is adopted from our previous work~\cite{Yu2019} where it was shown to outperform any other state-of-the-art estimators. The idea behind this estimator is that it (1) assigns temporary labels to unlabeled data points following the probability prediction from a logistic regression model, (2) then updates that logistic regression model on the temporary labeled data, (3) iterates the above two steps until convergence (when the number of temporarily assigned labels stays unchanged).

\section{Datasets}
\label{sec:case-study}

While \tion{methodology} shows how {\IT} should be applied in practice with human reading source code comments looking for the ``hard to find'' SATDs, it is too expensive for humans to test different treatments and answer all the research questions. As a result, the performance of {\IT} is tested through simulations on a publicly available SATD dataset originally collected by Maldonado and Shihab~\cite{maldonado2015detecting}. This dataset contains ten open-source java projects on different application domains (five of these projects were added by the same authors later after its first release), varying in size and the number of developers and most importantly, in the number of comments in source code. 
All of these ten projects, namely Apache-Ant-1.7.0, Apache-Jmeter-2.10, ArgoUML, Columba-1.4-src, EMF-2.4.1, Hibernate-Distribution-3.3.2.GA, jEdit-4.2, jFreeChart-1.0.19, jRuby-1.4.0, SQL12 were collected from GitHub. 
The provided dataset contains project names, classification type (if any) with actual comments. 
Note that, our problem does not concern with the type of SATD, rather we care about a binary problem of being a SATD or not. 
So, we have changed the final label into a binary problem by defining $WITHOUT\_CLASSIFICATION$ as $no$ and the rest (for example $DESIGN$) as $yes$. 
A few examples from the dataset are given in Table~\ref{table examples from dataset} for readers' ease.

\begin{table}[!htb]
\caption{Examples from Dataset}
\centering
\setlength\tabcolsep{4pt}
\begin{tabular}{|p{.13\linewidth}|p{.27\linewidth}|p{.40\linewidth}|p{.06\linewidth}|}
\hline
\textbf{project} & 
\textbf{classification} & 
\textbf{commenttext} & 
\textbf{label} \\
\hline
Apache Ant & 
DEFECT & 
// FIXME formatters are not thread-safe & 
yes \\
\hline
EMF & 
IMPLEMENTATION & 
// TODO Binary incompatibility; an old override must override putAll. & 
yes \\
\hline
JFreeChart & 
DESIGN & 
// calculate the adjusted data area taking into account the 3D effect... this assumes that there is a 3D renderer, all this 3D effect is a bit of an ugly hack... & 
yes \\
\hline
JRuby & 
WITHOUT CLASSIFICATION & 
// build first node (and ignore its result) and then second node & 
no \\
\hline
Columba & 
WITHOUT CLASSIFICATION & 
// get message header & 
no \\
\hline
JMeter & 
WITHOUT CLASSIFICATION & 
// parameters to pass to script file (or script) & 
no \\
\hline

\end{tabular}
\label{table examples from dataset}
\end{table}

\subsection{Independent Variables}
\label{sec:independent}

When Maldonado and Shihab~\cite{maldonado2015detecting} created this dataset, jDeodrant~\cite{fokaefs2011jdeodorant} was applied, which is an Eclipse plugin for extracting comments from the source code of java files. 
After that, Maldonado and Shihab~\cite{maldonado2015detecting} used four filtering heuristics to the comments. 
A short description of the filtering heuristics is given below. 
\begin{itemize}
    \item Removed licensed comments, auto-generated comments, etc. because according to the dataset authors, they do not contain SATD by developers.
    \item Removed commented source codes as commented source codes do not contain any SATD.
    \item Removed Javadoc comments that do not contain the words such as ``todo'', ``fixme'', ``xxx'' etc. because according to the dataset authors, the rest of the comments rarely contain any SATDs.
    \item Multiple single-line comments are grouped into a single comment because they all convey a single message and it is easy to consider them as a group.
\end{itemize}

After Maldonado and Shihab~\cite{maldonado2015detecting} applied these heuristics, the number of comments in each project reduced significantly (for example, the number of comments in Apache Ant reduced from $21,587$ to $4140$, almost 19\% of the original size).

\begin{table}[!htb]
\caption{Dataset Details}
\centering
\begin{tabular}{|p{.12\linewidth}|p{.09\linewidth}|p{.15\linewidth}|p{.14\linewidth}|p{.1\linewidth}|p{.09\linewidth}|}
\hline
\textbf{Project} & 
\textbf{Release / Year} & 
\textbf{Domain} & 
\textbf{Comments} &
\textbf{SATDs} &
\textbf{Ratio}\\
\hline
Apache Ant & 
1.7.0 / 2006& 
Automating Build&
4098 &
131 &
3.2\%\\
\hline
JMeter & 
2.10 / 2013 & 
Testing &
8057 &
374 &
4.64\%\\
\hline
ArgoUML & 
- & 
UML Diagram &
9452 &
1413 &
14.95\% \\
\hline
Columba & 
1.4 / 2007 & 
Email Client &
6468 &
204 &
3.15\% \\
\hline
EMF & 
2.4.1 / 2008& 
Model Framework &
4390 &
104 &
2.37\% \\
\hline
Hibernate & 
3.3.2 / 2009 & 
Object Mapping Tool &
2968 &
472 &
15.90\% \\
\hline
JEdit & 
4.2 / 2004 & 
Java Text Editor &
10322 &
256 &
2.48\% \\
\hline
JFreeChart & 
1.0.19 / 2014 & 
Java Framework &
4408 &
209 &
4.74\% \\
\hline
JRuby & 
1.4.0 / 2009 & 
Ruby for Java &
4897 &
622 &
12.70\% \\
\hline
SQuirrel & 
- & 
Database &
7215 &
286 &
3.96\% \\
\hline
\textbf{SUM} & 
 & 
 &
\textbf{62275} &
\textbf{4071} &
\textbf{6.54\%} \\
\hline
\textbf{MEDIAN} & 
 & 
 &
\textbf{5682.5} &
\textbf{271} &
\textbf{4.77\%} \\
\hline
\end{tabular}
\label{tab:details}
\end{table}

\subsection{Dependent Variables}
\label{sec:dependent}

In the work of Maldonado and Shihab~\cite{maldonado2015detecting}, two humans then manually classified each comment according to the six different types of TD mentioned by Alves et al.~\cite{alves2014towards} if they contained any SATD at all, else marked them $WITHOUT\_CLASSIFICATION$. 
Stratified sampling of the dataset was applied to check personal bias and found a $99\%$ confidence level with a confidence interval of $5\%$. 
A third human verified the agreement between the two using stratified sampling and reported a high level of agreement, using Cohen's Kapp~\cite{cohen1968weighted} coefficient of $+0.81$. Such a high confidence level, as well as a higher level of agreement indicates that the dataset is unbiased and reliable. A detailed description of the dataset is given in in Table ~\ref{tab:details}.

\section{Experiments and Results}
\label{sec:experiments}

Experiments are conducted on the SATD dataset with 10 projects described in \tion{case-study}. Each time, one project is selected as a target project (with labels unknown) and the rest 9 datasets are treated as source projects (with labels known). In Step 2, when oracles are queried for the target project, the ground truth labels are applied to label the queried comments, thus simulating the human-in-the-loop process without a real human in the loop. The rest of this section will provide details on the experiments and results for answering the research questions listed in \tion{introduction}.

\begin{table*}[!htp]
\caption{Experimental results for Step 1 on every targeting project. \textbf{Easy} represents the pattern recognizer in {\IT} while \textbf{MAT} is a baseline approach from Guo et al.~\cite{guo2019mat}} that uses human-derived patterns--- {\em ``todo, fixme, hack, xxx''} to find SATDs. The column \textbf{Better} summarizes how many times one treatment is better than the other on the given metric. This table presents results on two sets of ground truth labels: (1) \textbf{Original}: the ground truth labels provided by Maldonado  and  Shihab~\cite{maldonado2015detecting}, and (2) \textbf{Corrected}: labels after validating the conflicts between \textbf{Easy} and original ground truth, as shown in \fig{validate}.
\centering
\setlength\tabcolsep{4.8pt}
\begin{tabular}{|P{.045\linewidth}|p{.05\linewidth}|p{.06\linewidth}|P{.06\linewidth}|P{.04\linewidth}|P{.04\linewidth}|P{.04\linewidth}|P{.07\linewidth}|P{.06\linewidth}|P{.05\linewidth}|P{.07\linewidth}|P{.06\linewidth}|P{.04\linewidth}|P{.035\linewidth}|}\hline
\textbf{Ground Truth} & \textbf{Metrics}   & \textbf{Treatment} & \textbf{SQuirrel} & \textbf{JMeter} & \textbf{EMF} & \textbf{Apache Ant} & \textbf{ArgoUML} & \textbf{Hibernate} & \textbf{JEdit} & \textbf{JFreeChart} & \textbf{Columba} & \textbf{JRuby} &  \textbf{Better}\\\hline
 \multirow{6}{*}{\rotatebox[origin=c]{90}{\bf Original}} & \multirow{2}{*}{Precision} & Easy  & 0.85 & 0.87 & \cellcolor{gray!40}0.69 & 0.89 & 0.85 & 0.94 & \cellcolor{gray!40}0.95 & 0.72 & 0.91 & \cellcolor{gray!40}0.93 & \textbf{3}\\
         &  & MAT & 0.85 & 0.87 & 0.67 & \cellcolor{gray!40}0.90 & 0.85 & 0.94 & 0.81 & 0.72 & 0.91 & 0.92 & \textbf{1}\\\cline{2-14}
         & \multirow{2}{*}{Recall}    & Easy  & 0.54 & 0.75 & \cellcolor{gray!40}0.33 & 0.24 & 0.88 & \cellcolor{gray!40}0.74 & \cellcolor{gray!40}0.21 & \cellcolor{gray!40}0.47 & \cellcolor{gray!40}0.87 & 0.52 & \textbf{5}\\
         &     & MAT & 0.54 & 0.75 & 0.29 & \cellcolor{gray!40}0.47 & 0.88 & 0.73 & 0.19 & 0.46 & 0.86 & \cellcolor{gray!40}0.90 & \textbf{2} \\\cline{2-14}
         & \multirow{2}{*}{F1}        & Easy  & 0.66 & 0.80 & \cellcolor{gray!40}0.44 & 0.38 & \cellcolor{gray!40}0.87 & \cellcolor{gray!40}0.83 & \cellcolor{gray!40}0.35 & \cellcolor{gray!40}0.57 & \cellcolor{gray!40}0.89 & 0.67 & \textbf{6}\\
         &         & MAT & 0.66 & 0.80 & 0.40 & \cellcolor{gray!40}0.62 & 0.86 & 0.82 & 0.30 & 0.56 & 0.88 & \cellcolor{gray!40}0.91 & \textbf{2}\\\hline
 \multirow{6}{*}{\rotatebox[origin=c]{90}{\bf Corrected}} & \multirow{2}{*}{Precision} & Easy        & 1.00     & \cellcolor{gray!40}1.00                  & 1.00         & \cellcolor{gray!40}0.99                & \cellcolor{gray!40}1.00       & \cellcolor{gray!40}1.00                               & \cellcolor{gray!40}1.00         & 1.00                 & 0.99            & \cellcolor{gray!40}1.00           & \textbf{6}\\
& & MAT       & 1.00     & 0.99                  & 1.00         & 0.96             & 0.99    & 0.99                            & 0.85      & 1.00                 & 0.99            & 0.99           & \textbf{0}\\\cline{2-14}
&\multirow{2}{*}{Recall}    & Easy        & 0.58  & 0.77               & \cellcolor{gray!40}0.41      & 0.27             & 0.90     & \cellcolor{gray!40}0.75                            & \cellcolor{gray!40}0.22      & 0.55              & \cellcolor{gray!40}0.88            & 0.90         & \textbf{4}\\
 & & MAT       & 0.58  & 0.77               & 0.38      & \cellcolor{gray!40}0.49             & 0.90     & 0.74                            & 0.19      & 0.55              & 0.87            & \cellcolor{gray!40}0.91        & \textbf{2}\\\cline{2-14}
&\multirow{2}{*}{F1}   & Easy        & \cellcolor{gray!40}0.74  & 0.87               & \cellcolor{gray!40}0.58      & 0.42             & 0.94    & \cellcolor{gray!40}0.86                            & \cellcolor{gray!40}0.37      & 0.71              & 0.93            & 0.95        & \textbf{4}\\
 &  & MAT       & 0.73  & 0.87               & 0.55      & \cellcolor{gray!40}0.65             & 0.94    & 0.85                            & 0.31      & 0.71              & 0.93            & 0.95       & \textbf{1}\\\hline
\end{tabular}
\label{tab:step1}
\end{table*}


\subsection{RQ1: How to find the strong patterns of the ``easy to find'' SATDs in Step 1?} 

In this experiment, we compare the performance of the following two treatments:
\bi
\item
\textbf{Easy:} The pattern recognizer of {\IT} described in Algorithm~\ref{alg:PR}, which iteratively selects the pattern with the highest fitness score in \eqref{eq:fitness} until the selected pattern has lower than 80\% precision on the training data.
\item
\textbf{MAT:} a baseline approach from Guo et al.~\cite{guo2019mat} where a set of human-derived patterns--- {\em ``todo, fixme, hack, xxx''} is applied to find the ``easy to find'' SATDs.
\ei
on three different performance metrics
\bi
\item
\textbf{Precision:} $Precision = TP / (TP+FP)$.
\item
\textbf{Recall:} $Recall = TP / (TP+FN)$.
\item
\textbf{F1 score:} $F_1 = 2\cdot Precision \cdot Recall / (Precision+Recall)$.
\ei
where $TP$ is the number of true positives (SATD comments predicted as SATDs), $FP$ is the number of false positives (non-SATD comments predicted as SATDs), and $FN$ is the number of false negatives (SATD comments predicted as non-SATDs).

Table~\ref{tab:step1} (\textbf{Original}) shows the results of the experiment on the original dataset. Our observation from the results are
\begin{enumerate}
\item
Choosing any project as the holdout set, the strong patterns discovered by the pattern recognizer are always the same--- {\em ``todo, fixme, hack, workaround''}, except for {\em JRuby} where the strong patterns are {\em ``fixme, hack''}.
\item
Compared with manually discovered patterns--- {\em ``todo, fixme, hack, xxx''} from Guo et al.~\cite{guo2019mat} (MAT), the patterns automatically learned by \textbf{Easy} showed higher or similar precision and recall on 8 out of 10 target projects.
\end{enumerate}

\begin{figure}
\begin{center}
\includegraphics[width=\linewidth]{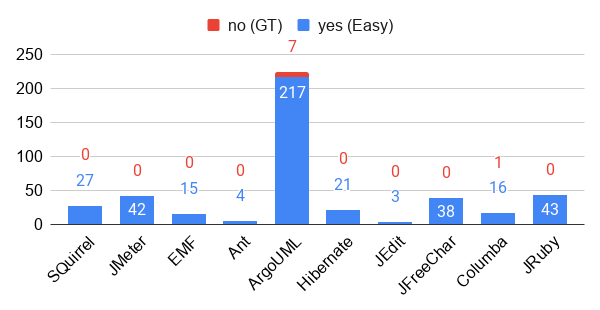}
\end{center}
\caption{Validation results for double-checking false positives of \textbf{Easy}. GT means the original ground truth label and DC means the double-checking result. The values of ``yes (Easy)'' show the number of comments that the double-checking result agrees with the \textbf{Easy} results (Easy=yes AND GT=no AND DC=yes) while the values of ``no (GT)'' show the number of comments that the double-checking result agrees with the original ground truth labels (Easy=yes AND GT=no AND DC=no). This graph shows that most (426 out of 434) of the false positives are actually true positives that were previously wrongly labeled in the original dataset.}\label{fig:validate}
\end{figure}

The results above suggest that our automated pattern recognizer \textbf{Easy} performed better than the human-derived patterns from Guo et al.~\cite{guo2019mat} (MAT). However, it did not reach close to 100\% precision on many target projects as we expected. One possible reason for this is human errors--- labels in the original dataset may not always be correct. 
Therefore, we manually analyzed the false positives (comments containing the strong patterns but were labeled as Non-SATDs) of \textbf{Easy} to double-check their labels. Two graduate students were employed to classify the 434 (out of 62,275, 7\textperthousand) comments where the original ground truth labels (GT) are \textbf{no} but the \textbf{Easy} predictions are \textbf{yes}. Surprisingly, the two graduate students found the comments very easy to classify and both made the same classification. Table~\ref{tab:example} shows some example comments whose labels were flipped. As shown in \fig{validate}, most of the false positives (98\%) were wrongly labeled in the original dataset. That means these strong patterns identified by {\IT} are even more accurate than human experts in finding the ``easy to find''.

After the ground truth labels were corrected, we reran the experiments and collected results in Table~\ref{tab:step1} (\textbf{Corrected}). This time, we observe:
\begin{enumerate}
\item
\textbf{Easy} detects the same set of strong patterns--- {\em ``todo, fixme, hack, workaround''} for every target project including {\em JRuby} after correcting the human errors. This also greatly increases the recall on {\em JRuby}.
\item
\textbf{Easy} achieves close to 100\% precision (100\% on eight projects and 99\% on two projects) on identifying the ``easy to find'' SATDs. These results are higher than the human-derived set of patterns--- {\em ``todo, fixme, hack, xxx''} from Guo et al.~\cite{guo2019mat} (MAT).
\item
\textbf{Easy} achieves much lower recall and F1 score than \textbf{MAT} on {\em Apache Ant}. This is because only on the {\em Apache Ant} project, ``xxx'' is a strong pattern of fitness score 25. On the other projects, fitness scores for ``xxx'' range from 0 to 2. Therefore, the pattern of ``xxx'' can help only on {\em Apache Ant} and will damage the precision when used on other projects. This is exactly the advantage of \textbf{Easy} over \textbf{MAT}--- to avoid such ``trap'' patterns like ``xxx'' by training on a collection of projects.
\item
\textbf{Easy} is even more accurate than human experts in identifying the ``easy to find'' SATDs since 98\% of the conflicting comments, which were labeled as Non-SATDs by humans but contain the patterns from \textbf{Easy}, are identified as SATDs in our validation study. 
\item
Although \textbf{Easy} is an algorithm with close to 100\% precision and barely any cost (training takes seconds), it alone can only identify 20\% to 90\% of the SATDs. Thus it is necessary for \textbf{Step 2}.
\end{enumerate}

\begin{table}[!tbh]
\caption{Examples of Corrected Labels}
\centering
\setlength\tabcolsep{4pt}
\begin{tabular}{|p{.13\linewidth}|p{.6\linewidth}|p{.06\linewidth}|p{.06\linewidth}|}
\hline
\textbf{Project} &  
\textbf{Comment Text} & 
\textbf{GT}  &
\textbf{Easy}\\
\hline
Apache Ant & 
//TODO Test on other versions of weblogic //TODO add more attributes to the task, to take care of all jspc options //TODO Test on Unix & 
no & 
yes \\
\hline
ArgoUML & 
// skip backup files. This is actually a workaround for the cpp generator, which always creates backup files (it's a bug).
 & no & yes \\
\hline
JFreeChart & 
// FIXME: we've cloned the chart, but the dataset(s) aren't cloned and we should do that  & 
no & 
yes \\
\hline
JRuby & 
// All errors to sysread should be SystemCallErrors, but on a closed stream Ruby returns an IOError.  Java throws same exception for all errors so we resort to this hack... & 
no & 
yes \\
\hline
Columba & 
// FIXME r.setPos(); & 
no & 
yes \\
\hline
\end{tabular}
\label{tab:example}
\end{table}

\subsection{RQ2: How to find the ``hard to find'' SATDs efficiently with human experts?}

After all the ``easy to find'' SATDs are filtered out from the datasets, it is now a problem to find the remaining ``hard to find'' SATDs (which are 10-80\% of all the SATDs). To solve this problem, we first ask the following question.

\subsubsection{RQ2.1: Can the ``hard to find'' SATDs be automatically detected without human oracles?}
\begin{figure}
\begin{center}
\includegraphics[width=\linewidth]{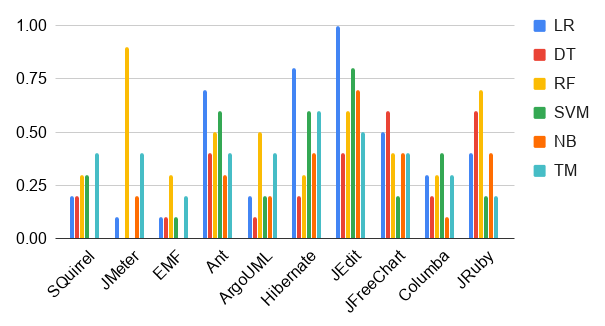}
\end{center}
\caption{P@10 results for supervised learning models on the ``hard to find'' SATDs. At most three out of ten projects have P@10 higher than 0.5. This suggests that all these supervised learners are not precise enough to fully automate the process of identifying the ``hard to find'' SATDs. Human oracles have to be queried to make the final decisions.}\label{fig:precs}
\end{figure}

To answer this question, we trained supervised learning models on source projects and tested them on the target project (both source projects and target projects do not contain comments that have the patterns identified by Step 1). The following models are evaluated in this experiment:
\bi
\item
\textbf{LR}: logistic regression model described in \tion{learners}. Implemented with scikit-learn\footnote{\url{https://scikit-learn.org/}} package {\em LogisticRegression} in Python with balanced class weight. 
\item
\textbf{DT}: decision tree model described in \tion{learners}. Implemented with scikit-learn package {\em DecisionTreeClassifier} in Python with balanced class weight. 
\item
\textbf{RF}: random forest model described in \tion{learners}. Implemented with scikit-learn package {\em RandomForestClassifier} in Python with class\_weight = balanced\_subsample. 
\item
\textbf{SVM}: linear soft-margin support vector machine model described in \tion{learners}. Implemented with scikit-learn package {\em SGDClassifier} in Python with balanced class weight. 
\item
\textbf{NB}: Multinomial Naive Bayes model described in \tion{learners}. Implemented with scikit-learn package {\em MultinomialNB} in Python. 
\item
\textbf{TM}: ensemble model from Huang et al.~\cite{huang2018identifying} where a Multinomial Naive Bayes model is trained on selected features with information gain on each source project and the majority vote of the predictions from each model on the target project is utilized to make the final prediction. \textbf{TM} is considered as one of the state-of-the-art solutions for identifying SATDs so here we apply it as a baseline algorithm.
\ei
To assess whether the above supervised learning models can identify the ``hard to find'' SATDs precisely without human oracles, we apply the following \textbf{P@10} metric:
\bi
\item
\textbf{P@10}: precision for the top 10 predictions. For example, if there are 6 SATDs amongst the 10 comments with the highest prediction probability, $P@10=0.6$. We choose this metric since a model is definitely imprecise if the precision for its top 10 confident predictions is already low.
\ei
\fig{precs} shows that at most three out of ten projects have \textbf{P@10} higher than 0.5. That means all of the machine learning models predict wrongly with higher than 50\% probability even for the 10 most confident predictions on most of the projects.
Therefore, we conclude that directly using the model predictions as classification results will result in a large number of false positives. As a result, human experts have to verify each prediction result to decide whether it is SATD or not. This leads to our next research question.

\begin{table*}[!htp]
\caption{APFD (higher the better) results for different models with or without the \textbf{Hard} strategy on the ``hard to find'' SATDs. Medians and IQRs (delta between 75th percentile and 25th percentile, lower the better) are calculated for easy comparisons. If \textbf{Hard} = no, human oracles on the target project are not utilized, the model is just a one-time trained supervised learning model. On the other hand, if \textbf{Hard} = yes, human oracles on the queried comments are utilized to update the model before it is applied to find its next highest predictions for humans to verify. {\IT} utilizes \textbf{Hard} = yes. A threshold of Cohen'd small effect size (0.02) is applied to determine which treatment performs best in each target project and color them in \colorbox{gray!40}{gray}. The column \textbf{\#Best} shows the number of projects each treatment performs the best in.}

\centering
\setlength\tabcolsep{4pt}
\begin{tabular}{|P{.04\linewidth}|p{.035\linewidth}|P{.06\linewidth}|P{.045\linewidth}|P{.04\linewidth}|P{.05\linewidth}|P{.065\linewidth}|P{.065\linewidth}|P{.04\linewidth}|P{.07\linewidth}|P{.06\linewidth}|P{.05\linewidth}|P{.05\linewidth}|P{.03\linewidth}|P{.03\linewidth}|}\hline
\textbf{Model} & \textbf{Hard} & \textbf{SQuirrel} & \textbf{JMeter} & \textbf{EMF} & \textbf{Apache Ant} & \textbf{ArgoUML} & \textbf{Hibernate} & \textbf{JEdit} & \textbf{JFreeChart} & \textbf{Columba} & \textbf{JRuby} & \textbf{Median} & \textbf{IQR} & \textbf{\#Best}\\\hline
\multirow{2}{*}{\textbf{LR}}  &no   & 0.69 & \cellcolor{gray!40}0.85 & 0.80  & \cellcolor{gray!40}0.88 & 0.93 & 0.80  & 0.80  & 0.80  & 0.90  & 0.83 & \textbf{0.82} & \textbf{0.07} & \textbf{2} \\
 &yes & 0.74 & \cellcolor{gray!40}0.84 & 0.83 & \cellcolor{gray!40}0.89 & 0.93 & \cellcolor{gray!40}0.81 & 0.85 & 0.81 & 0.92 & 0.83 & \textbf{0.84} & \textbf{0.07}  & \textbf{3} \\\hline
\multirow{2}{*}{\textbf{DT}}  &no   & 0.78 & 0.78 & 0.84 & 0.78 & 0.88 & 0.76 & 0.86 & 0.71 & 0.87 & 0.86 & \textbf{0.81} & \textbf{0.08}  & \textbf{0} \\
 &yes & 0.71 & 0.64 & 0.71 & 0.77 & 0.8  & 0.77 & 0.75 & 0.53 & 0.78 & 0.81 & \textbf{0.76} & \textbf{0.07}  & \textbf{0} \\\hline
\multirow{2}{*}{\textbf{RF}}  &no   & 0.82 & 0.79 & 0.81 & 0.83 & 0.88 & 0.78 & 0.82 & 0.73 & 0.91 & 0.83 & \textbf{0.82} & \textbf{0.03}  & \textbf{0} \\
 &yes & \cellcolor{gray!40}0.91 & \cellcolor{gray!40}0.85 & \cellcolor{gray!40}0.93 & \cellcolor{gray!40}0.90  & \cellcolor{gray!40}0.96 & \cellcolor{gray!40}0.83 & 0.92 & \cellcolor{gray!40}0.89 & \cellcolor{gray!40}0.98 & \cellcolor{gray!40}0.91 & \textbf{0.91} & \textbf{0.04}  & \textbf{9} \\\hline
\multirow{2}{*}{\textbf{SVM}} &no   & 0.58 & 0.83 & 0.75 & 0.83 & 0.90  & 0.75 & 0.71 & 0.74 & 0.84 & 0.74 & \textbf{0.75} & \textbf{0.09}  & \textbf{0} \\
&yes & \cellcolor{gray!40}0.91 & \cellcolor{gray!40}0.86 & \cellcolor{gray!40}0.91 & \cellcolor{gray!40}0.90  & 0.92 & \cellcolor{gray!40}0.83 & \cellcolor{gray!40}0.95 & 0.79 & \cellcolor{gray!40}0.97 & \cellcolor{gray!40}0.92 & \textbf{0.91} & \textbf{0.05}  & \textbf{8} \\\hline
\multirow{2}{*}{\textbf{NB}}  &no   & 0.41 & 0.74 & 0.67 & 0.62 & 0.75 & 0.57 & 0.37 & 0.63 & 0.65 & 0.67 & \textbf{0.64} & \textbf{0.09}  & \textbf{0} \\
 &yes & 0.47 & 0.75 & 0.67 & 0.66 & 0.79 & 0.59 & 0.42 & 0.64 & 0.66 & 0.68 & \textbf{0.66} & \textbf{0.08}  & \textbf{0} \\\hline
\textbf{TM}   &no  & 0.73 & 0.7  & 0.72 & 0.80  & 0.69 & 0.75 & 0.77 & 0.69 & 0.77 & 0.89 & \textbf{0.74} & \textbf{0.07} & \textbf{0} \\\hline
\end{tabular}
\label{tab:step2}
\end{table*}

\subsubsection{RQ2.2: How to more efficiently utilize human oracles to find the ``hard to find'' SATDs?}

\begin{figure}
\begin{center}
\includegraphics[width=\linewidth]{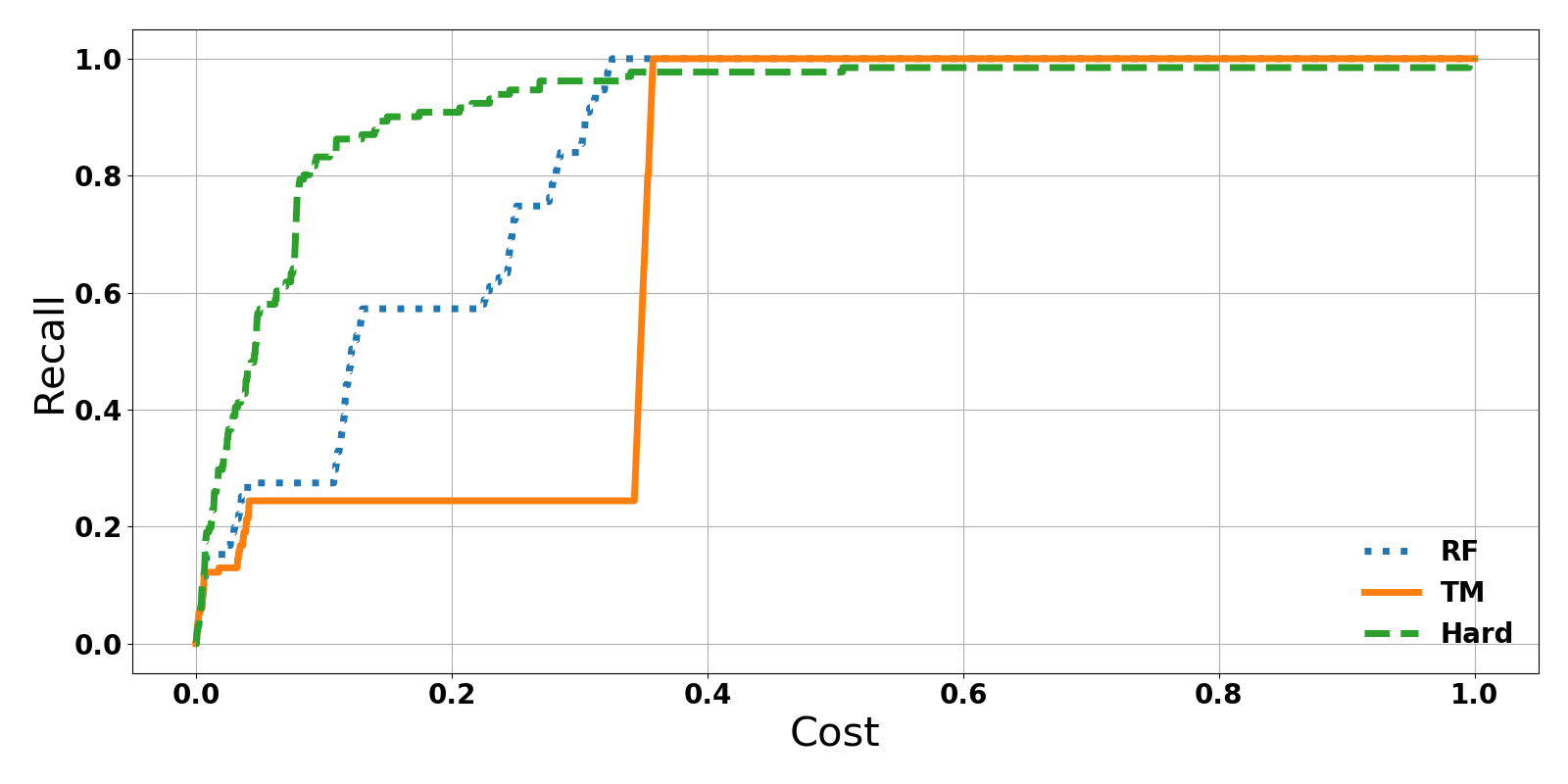}
\end{center}
\caption{Recall-cost curves for three different treatments on finding the ``hard to find'' SATDs on target project SQuirrel. \textbf{Hard} represents \textbf{RF} with \textbf{Hard}=yes while \textbf{RF} represents \textbf{RF} with \textbf{Hard}=no. APFD results in Table~\ref{tab:step2} were calculated as the area under these curves. Figures on other target projects are shown in the Appendix as \fig{rest_all}.}\label{fig:rest_sql}
\end{figure}

Since it is inevitable to spend human effort on verifying the prediction results in Step 2, the \textbf{Hard} strategy is applied to learn from this incrementally acquired information and update its model for better predictions, as described in  Algorithm~\ref{alg:Hard}. In this experiment, we record the recall and its corresponding cost (of human effort) for each algorithm as the cost increases.
\begin{equation}
\label{eq:recall}
Recall = \frac{|\{\text{SATDs}\}\cap\{\text{human verified comments}\}|}{|\{\text{SATDs}\}|}
\end{equation}
\begin{equation}
\label{eq:cost}
Cost = \frac{|\{\text{human verified comments}\}|}{|\{\text{comments}\}|}
\end{equation}
To simplify the comparison between different algorithms, we calculate the area under the recall-cost curve as a performance metrics:
\bi
\item
\textbf{APFD}: first proposed in test case prioritization~\cite{167}, APFD calculates the area under the recall-cost curve. Ranging from 0.0 to 1.0, a larger APFD means higher recall can be achieved at a lower cost, thus the better. An APFD of 0.5 can be achieved by randomly select the next item each time. We choose to evaluate the overall performance with \textbf{APFD} since it is a single metric that evaluates the entire recall-cost curve, while other metrics (e.g. precision, recall, F1 score) only evaluate a single point of the curve. In this way, we do not need to decide the stopping point before comparing different treatments.
\ei
Table~\ref{tab:step2} shows the APFD results for different  models  with  or  without  the \textbf{Hard} strategy. Given that most of the results are deterministic and are close to each other, we applied Cohen'd effect size test to determine which results are similar. To that end, we calculated: 
\begin{equation}
\label{eq:cohen_step2}
Small_{step2} = 0.2 \cdot StdDev(\text{All APFD results}) = 0.02.
\end{equation}
We then consider all the results that are higher than the best result minus the $Small_{step2}$ as the best results on each target project (colored in \colorbox{gray!40}{gray} in Table~\ref{tab:step2}). From these results we can see
\bi
\item
Continuously updating the model (\textbf{Hard}) helps improve the performance on 4 out of 5 models (except for the decision tree model).
\item
Random forest and support vector machine models with the \textbf{Hard} strategy achieved the highest median APFD of 0.91. Given that the \textbf{\#Best} of random forest is higher, we choose random forest as the internal model of \textbf{Hard} for the rest of the experiments.
\item
Random forest model with the \textbf{Hard} strategy outperformed the baseline algorithm \textbf{TM} on finding the ``hard to find'' SATDs.
\ei
For a more intuitive comparison, \fig{rest_sql} shows the recall-cost curves of three different treatments on target project SQuirrel. The APFD results in Table~\ref{tab:step2} were calculated as the area under these curves. As we can see, \textbf{Hard} (\textbf{RF} with \textbf{Hard}=yes) has the highest APFD score of 0.91. Also, in \fig{rest_sql} it almost always reaches the same recall at a lower cost than RF (\textbf{RF} with \textbf{Hard}=no) with APFD score of 0.82 and TM with APFD score of 0.73. Recall-cost curves on other target projects can be found in \fig{rest_all} from the Appendix.

\begin{figure}
\begin{center}
\includegraphics[width=\linewidth]{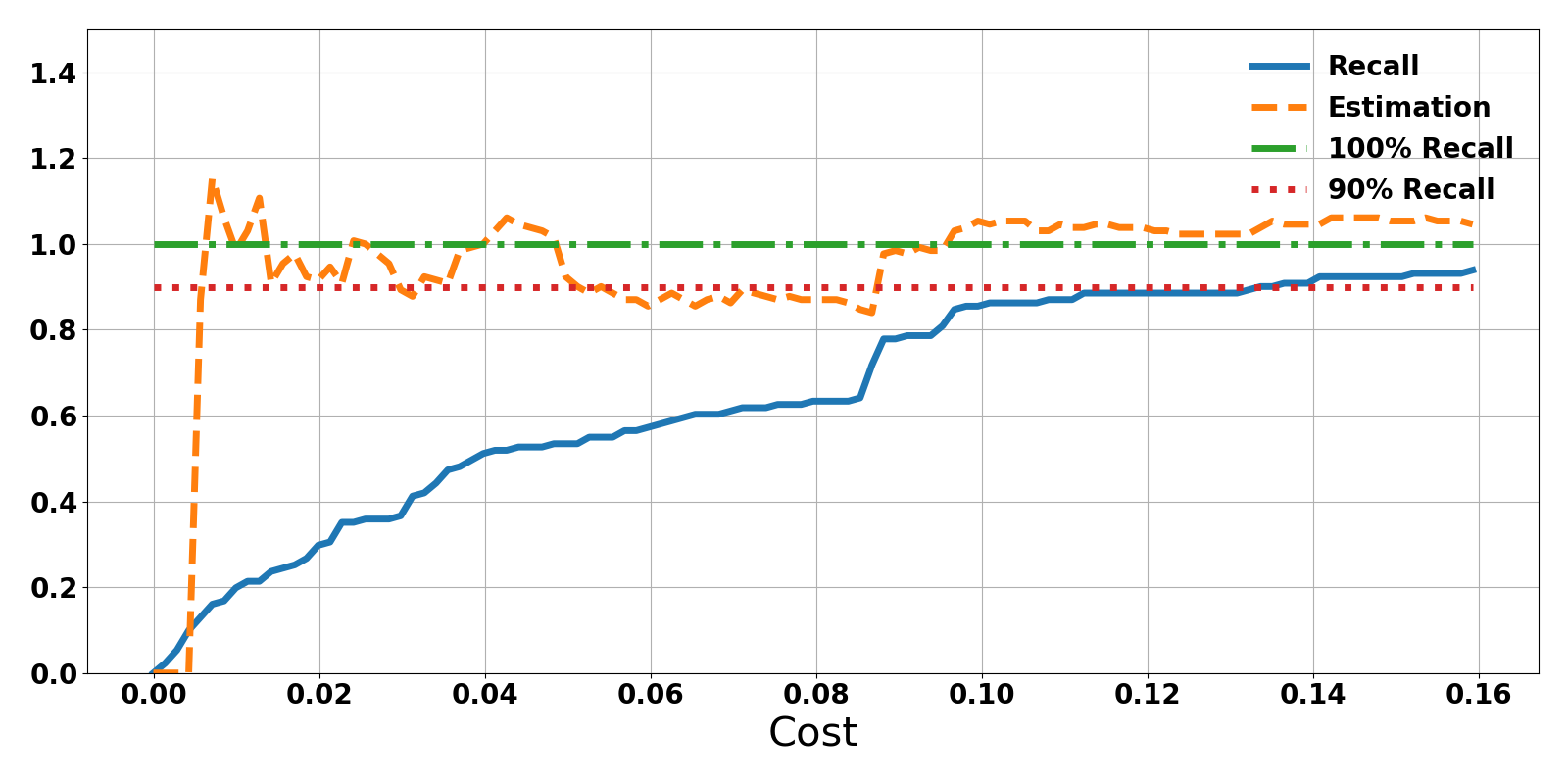}
\end{center}
\caption{Recall-cost and estimation-cost curves for finding 90\% of the ``hard to find'' SATDs with \textbf{Hard} on target project SQuirrel. Results shown in Table~\ref{tab:est_stop} were derived from these curves. Figures on other target projects are shown in the Appendix as \fig{est_all}.}\label{fig:est_sql}
\end{figure}

\begin{table*}[!htp]
\caption{Performance of \textbf{Hard} aiming to find 90\% of the ``hard to find'' SATDs with the estimator.}
\centering
\setlength\tabcolsep{4.8pt}
\begin{tabular}{|p{.08\linewidth}|P{.06\linewidth}|P{.045\linewidth}|P{.045\linewidth}|P{.05\linewidth}|P{.07\linewidth}|P{.06\linewidth}|P{.04\linewidth}|P{.08\linewidth}|P{.06\linewidth}|P{.05\linewidth}|P{.05\linewidth}|P{.04\linewidth}|}\hline
\textbf{Targeting 90\% Recall} & \textbf{SQuirrel} & \textbf{JMeter} & \textbf{EMF} & \textbf{Apache Ant} & \textbf{ArgoUML} & \textbf{Hibernate} & \textbf{JEdit} & \textbf{JFreeChart} & \textbf{Columba} & \textbf{JRuby} & \textbf{Median} & \textbf{IQR}\\ \hline
\textbf{Recall}                & 0.92  & 0.93               & 0.99      & 0.87             & 0.88    & 0.81                            & 0.94      & 0.99              & 1.00            & 0.92        & \textbf{0.92}   & \textbf{0.08} \\\hline
\textbf{Cost}                  & 0.18  & 0.35               & 0.28      & 0.24             & 0.12    & 0.25                            & 0.17      & 0.19              & 0.23            & 0.27        & \textbf{0.24}   & \textbf{0.08} \\\hline
\end{tabular}
\label{tab:est_stop}
\end{table*}

\subsubsection{RQ2.3: When to stop Hard in Step 2?} 

RQ2.2 shows that overall, \textbf{Hard} achieves higher recall at a lower cost than other methods. However, it is still necessary to decide a stopping point of \textbf{Hard} in real-world applications. Our solution to this problem is that, with an accurate estimation of the number of undiscovered SATDs, human experts are easier to make decisions on whether to spend more time looking for the ``hard to find'' SATDs or to stop at that point. To assess the accuracy of the estimation, we plot out the recall-cost curves with the following estimation:
\begin{equation}
\label{eq:est}
Estimation = \frac{\text{estimated number of SATDs}}{|\{\text{SATDs}\}|}
\end{equation}
As we can see from \fig{est_sql}, \textbf{Hard} tends to overestimate in the beginning but converges to the actual value after around 10\% cost and it does help to determine when has 90\% recall been reached. These recall-cost and estimation-cost curves on other target projects can be found in \fig{est_all} from the Appendix. Table~\ref{tab:est_stop} shows the final recall and cost when \textbf{Hard} stops at 90\% recall through estimations. Overall, \textbf{Hard} could stop close to the target 90\% recall with the estimations.

\begin{table*}[!htp]
\caption{APFD (higher the better) results for different treatments on finding all the SATDs. Medians and IQRs (delta between 75th and 25th percentile, lower the better) are calculated for easy comparisons. The proposed treatment \textbf{{\IT}=Easy+Hard}. A threshold of Cohen'd small effect size (0.01) is applied to determine which treatment performs best in each target project and color them in \colorbox{gray!40}{gray}. The column \textbf{\#Best} shows the number of projects each treatment performs best in.}
\centering
\setlength\tabcolsep{4pt}
\begin{tabular}{|P{.07\linewidth}|P{.06\linewidth}|P{.04\linewidth}|P{.05\linewidth}|P{.05\linewidth}|P{.07\linewidth}|P{.07\linewidth}|P{.05\linewidth}|P{.075\linewidth}|P{.06\linewidth}|P{.05\linewidth}|P{.05\linewidth}|P{.04\linewidth}|P{.03\linewidth}|}\hline
\textbf{Treatment} & \textbf{SQuirrel} & \textbf{JMeter} & \textbf{EMF} & \textbf{Apache Ant} & \textbf{ArgoUML} & \textbf{Hibernate} & \textbf{JEdit} & \textbf{JFreeChart} & \textbf{Columba} & \textbf{JRuby} & \textbf{Median} & \textbf{IQR} & \textbf{\#Best}\\\hline
\textbf{{\IT}}   & \cellcolor{gray!40}0.97  & \cellcolor{gray!40}0.97               & \cellcolor{gray!40}0.95      & \cellcolor{gray!40}0.93             & \cellcolor{gray!40}1.00       & \cellcolor{gray!40}0.96                            & \cellcolor{gray!40}0.94      & \cellcolor{gray!40}0.97              & \cellcolor{gray!40}1.00               & \cellcolor{gray!40}0.99        & \textbf{0.97}   & \textbf{0.03} & \textbf{10} \\\hline
\textbf{Easy+RF}   & 0.93  & 0.95               & 0.91      & 0.87             &\cellcolor{gray!40} 0.99    & \cellcolor{gray!40}0.95                            & 0.87      & 0.88              & \cellcolor{gray!40}0.99            & \cellcolor{gray!40}0.99        & \textbf{0.94}   & \textbf{0.09} & \textbf{4}\\\hline
\textbf{Hard}      & 0.95  & 0.95               & \cellcolor{gray!40}0.95      & 0.91             & 0.91    & 0.89                            & \cellcolor{gray!40}0.95      & 0.90               & 0.98            & 0.92        & \textbf{0.93}   & \textbf{0.04} & \textbf{2}\\\hline
\textbf{MAT+RF}    & 0.91  & \cellcolor{gray!40}0.97               & 0.88      & \cellcolor{gray!40}0.92             & \cellcolor{gray!40}0.99    & \cellcolor{gray!40}0.96                            & 0.86      & 0.87              & 0.98            & 0.98        & \textbf{0.94}   & \textbf{0.09} & \textbf{4}\\\hline
\textbf{TM}        & 0.83  & 0.94               & 0.90       & 0.88             & 0.91    & 0.88                            & 0.92      & 0.84              & 0.98            & 0.91        & \textbf{0.90 }   & \textbf{0.04} & \textbf{0}\\\hline
\textbf{RF}        & 0.91  & 0.93               & 0.84      & 0.88             & 0.90     & 0.88                            & 0.88      & 0.84              & 0.97            & 0.92        & \textbf{0.89}   & \textbf{0.04}
& \textbf{0} \\\hline
\end{tabular}
\label{tab:overall}
\end{table*}

\begin{figure}
\begin{center}
\includegraphics[width=\linewidth]{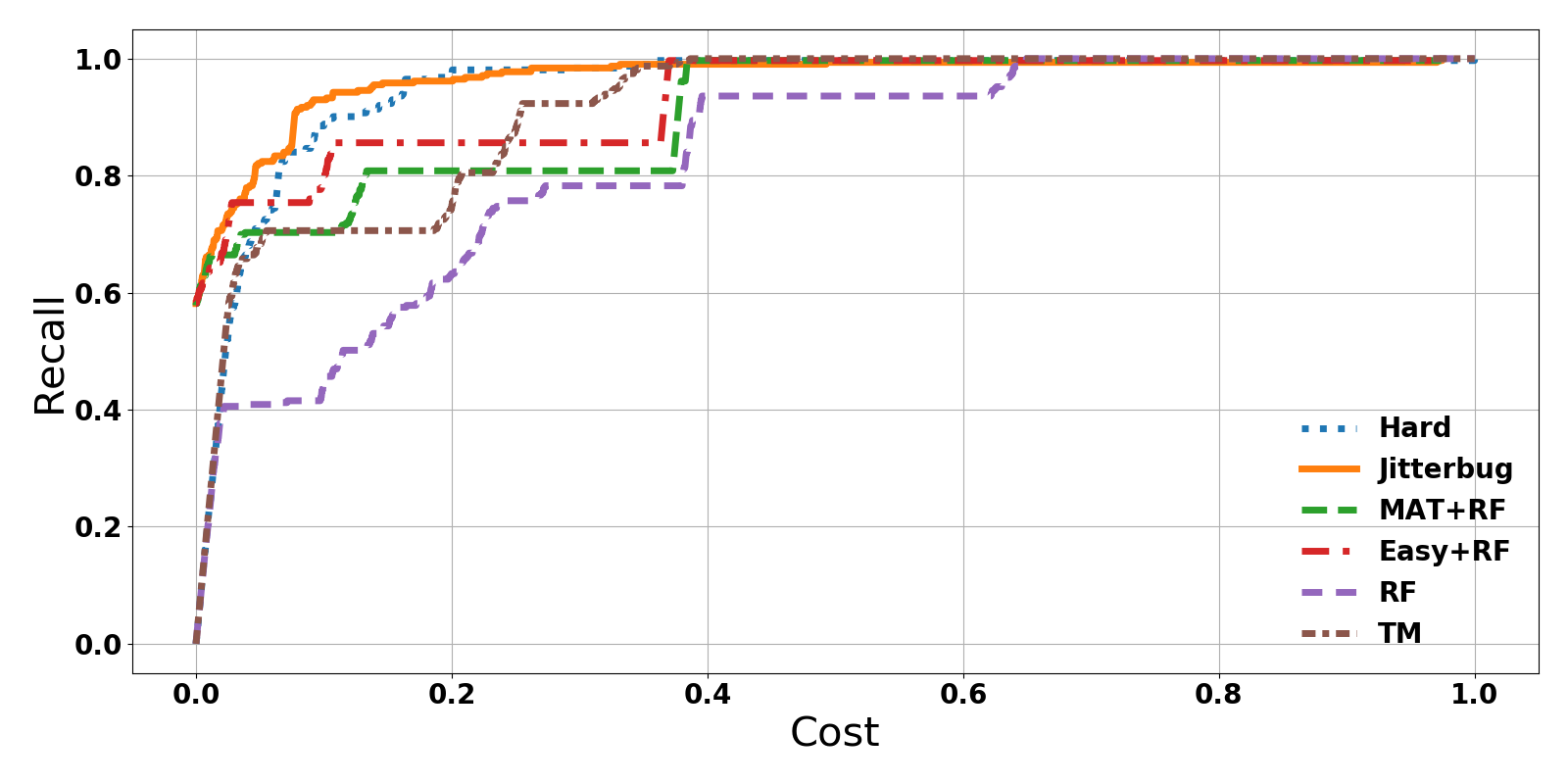}
\end{center}
\caption{Recall-cost curves for finding all SATDs with different treatments on target project SQuirrel. APFD results in Table~\ref{tab:overall} were calculated as the area under these curves. Figures on other target projects are shown in the Appendix as \fig{overall_all}.}\label{fig:overall_sql}
\end{figure}

\subsection{RQ3: Overall how does {\IT} perform?} 

In this research question, we evaluate the overall performance of {\IT} in three aspects: (1) APFD measures its overall efficiency without stopping rules; (2) Precision, Recall, F1 score, and Cost measures its performance when finding 90\% ``hard to find'' SATDs based on estimation; (3) runtime measures the additional computation cost of {\IT} besides the human effort cost.

\subsubsection{RQ3.1: How does {\IT} perform in terms of APFD?}

Table~\ref{tab:overall} shows the overall APFD scores for  finding all the SATDs in the target project. The following treatments are tested in this experiment:
\bi
\item
{\IT}: First apply \textbf{Easy} to automatically identify the ``easy to find'' SATDs with zero human effort, then apply \textbf{Hard} to guide humans in identifying the ``hard to find'' SATDs.
\item
\textbf{Easy+RF}: First apply \textbf{Easy} to automatically identify the ``easy to find'' SATDs with zero human effort, then apply a supervised learner random forest to rank the comments for humans to identify the ``hard to find'' SATDs.
\item
\textbf{Hard}: Directly apply \textbf{Hard} to guide humans in identifying both ``easy to find'' and ``hard to find'' SATDs.
\item
\textbf{MAT+RF}: First apply \textbf{MAT} to automatically identify the ``easy to find'' SATDs with zero human effort, then apply a supervised learner random forest to rank the comments for human to identify the ``hard to find'' SATDs.
\item
\textbf{TM}: Apply \textbf{TM} to rank all comments for humans to identify SATDs.
\item
\textbf{RF}: Apply random forest classifier to rank all comments for humans to identify SATDs.
\ei

Similarly to \textbf{RQ2.2}, we applied the following threshold from Cohen'd small effect size to determine the best treatments in each target project:
\begin{equation}
\label{eq:cohen_all}
Small_{overall} = 0.2 \cdot StdDev(\text{All APFD results}) = 0.01.
\end{equation}
From these results, we observed
\bi
\item
{\IT} outperforms the state-of-the-art solutions \textbf{TM}, \textbf{RF}, and \textbf{MAT+RF}.
\item
{\IT} (\textbf{Easy+Hard}) outperforms \textbf{Easy+RF}, which means the \textbf{Hard} strategy outperforms \textbf{RF} in Step 2 and this contributes to the overall performance.
\item
{\IT} (\textbf{Easy+Hard}) outperforms \textbf{Hard}, which means applying \textbf{Easy} in Step 1 does contribute to the overall performance.
\ei
For a more intuitive comparison, \fig{overall_sql} shows the recall-cost curves of the compared treatments on target project SQuirrel. The APFD results in Table~\ref{tab:overall} were calculated as the area under these curves. 
As we can see, {\IT} has the highest APFD score of 0.97. Also, in \fig{overall_sql} it almost always reaches the same recall at a lower cost than other treatments. Recall-cost curves on other target projects can be found in \fig{overall_all} from the Appendix.

In conclusion, {\IT} outperforms the state-of-the-art solutions (\textbf{TM, RF, MAT+RF}) in identifying SATDs in terms of APFD, and both of the two components \textbf{Easy} and \textbf{Hard} contribute to its good performance.

\begin{table*}[!htp]
\caption{Comparison between \textbf{Easy}, \textbf{Jitterbug}, and the best performing state-of-the-art supervised learning approach--- \textbf{CNN}~\cite{ren2019neural} on the original datasets in terms of precision, recall, F1 score, and cost. Here, \textbf{Jitterbug=Easy+Hard} targets at finding 90\% of the ``hard to find'' SATDs with its estimator and its \textbf{Easy} part costs zero human effort.}
\centering
\setlength\tabcolsep{4pt}
\begin{tabular}{|p{.055\linewidth}|p{.07\linewidth}|P{.06\linewidth}|P{.045\linewidth}|P{.035\linewidth}|P{.045\linewidth}|P{.07\linewidth}|P{.07\linewidth}|P{.045\linewidth}|P{.075\linewidth}|P{.06\linewidth}|P{.05\linewidth}|P{.05\linewidth}|P{.04\linewidth}|}\hline
\textbf{Metrics}   & \textbf{Treatment} & \textbf{SQuirrel} & \textbf{JMeter} & \textbf{EMF} & \textbf{Apache Ant} & \textbf{ArgoUML} & \textbf{Hibernate} & \textbf{JEdit} & \textbf{JFreeChart} & \textbf{Columba} & \textbf{JRuby} & \textbf{Median} & \textbf{IQR} \\\hline
\multirow{3}{*}{Precision} &\cellcolor{gray!40} Easy  & 0.85 & 0.87 & 0.69 & 0.89 & 0.85 & 0.94 & 0.95 & 0.72 & 0.91 & 0.93 & 0.88 & 0.11 \\
& Jitterbug  & 0.21 & 0.13 & 0.10 & 0.12 & 0.62 & 0.48 & 0.16 & 0.23 & 0.33 & 0.65 & 0.22 & 0.39\\
& CNN~\cite{ren2019neural}  & 0.79 & 0.87 & \cellcolor{gray!40}0.79 & 0.58 & 0.82 & 0.93 & 0.77 & 0.69 & 0.83 & 0.81 & 0.8	& 0.09\\
\hline
\multirow{3}{*}{Recall} & Easy  & 0.54 & 0.75 & 0.33 & 0.24 & 0.88 & 0.74 & 0.21 & 0.47 & 0.87 & 0.52 & 0.53	& 0.47\\
& \cellcolor{gray!40}Jitterbug  & 0.97 & 0.98 & 0.96 & 0.93 & 0.99 & 0.95 & 0.96 & 0.98 & 0.99 & 0.96 & 0.97 & 0.02\\
& CNN~\cite{ren2019neural}  & 0.69 & 0.79 & 0.59 & 0.76 & 0.95 & 0.74 & 0.49 & 0.80 & 0.88 & 0.93 & 0.77 & 0.22\\
\hline
\multirow{3}{*}{F1} & Easy  & 0.66 & 0.80 & 0.44 & 0.38 & 0.87 & 0.83 & 0.35 & 0.57 & \cellcolor{gray!40}0.89 & 0.67 & 0.66 & 0.41\\
& Jitterbug  & 0.35 & 0.23 & 0.19 & 0.21 & 0.76 & 0.64 & 0.27 & 0.37 & 0.49 & 0.77 & 0.36 & 0.45\\
& \cellcolor{gray!40}CNN~\cite{ren2019neural}  & 0.74 & 0.83 & 0.68 & 0.66 & 0.88 & 0.83 & 0.60 & 0.74 & 0.85 & 0.86 & 0.78 & 0.18\\
\hline
\multirow{3}{*}{Cost} & \cellcolor{gray!40}Easy  & 0.00 & 0.00 & 0.00 & 0.00 & 0.00 & 0.00 & 0.00 & 0.00 & 0.00 & 0.00 & 0.00 & 0.00\\
& Jitterbug  & 0.16 & 0.31 & 0.21 & 0.24 & \cellcolor{gray!40}0.08 & 0.19 & 0.15 & 0.17 & 0.06 & \cellcolor{gray!40}0.12 & 0.16 & 0.11\\
& CNN~\cite{ren2019neural}  & 0.03 & 0.04 & 0.02 & 0.04 & 0.17 & 0.13 & 0.02 & 0.06 & 0.03 & 0.15 & 0.04 & 0.1\\
\hline
\end{tabular}
\label{tab:baseline}
\end{table*}

\subsubsection{RQ3.2: How does {\IT} perform overall when targeting at finding 90\% of the ``hard to find'' SATDs?}

Table~\ref{tab:baseline} shows the overall performance scores for finding all the SATDs in the target project. To compare against the latest state-of-the-art deep convolutional neural network-based approach~\cite{ren2019neural}, precision, recall, F1 score, and cost are applied to evaluate each treatment on the original (uncorrected) dataset. The following treatments are tested in this experiment:
\bi
\item
\textbf{Easy}: Apply \textbf{Easy} to automatically identify the ``easy to find'' SATDs with zero human effort, then stop.
\item
{\IT}: First apply \textbf{Easy} to automatically identify the ``easy to find'' SATDs with zero human effort, then apply \textbf{Hard} to guide humans until 90\% of the ``hard to find'' SATDs are identified (according to estimation).
\item
\textbf{CNN}: Apply a convolutional neural network (on word2vec features and with hyperparameter tuning) to classify each comment into SATD or non-SATD~\cite{ren2019neural}. Due to the difficulty of reproducing a deep learning solution, we used the same precision, recall, F1 scores reported in Ren et al.~\cite{ren2019neural}, and the cost metric for CNN is calculated as 
$$Cost = \frac{|{SATDs}|\times recall}{|{comments}|\times precision}$$.
\ei
From the results in Table~\ref{tab:baseline}, we observed
\bi
\item
\textbf{Precision}: \textbf{Easy} always achieves the highest precision except for the {\em EMF} project. We know from Table~\ref{tab:step1} that once the labels are corrected, the precision of \textbf{Easy} on the {\em EMF} project will be close to 100\%. Therefore, \textbf{Easy} can reach higher precision than \textbf{CNN} on every project.
\item
\textbf{Recall}: {\IT} achieves the highest recall on every project. Also, since the target of {\IT} is to stop at finding 90\% of the ``hard to find'' SATDs, its final recalls should all be higher than 90\%, which is consistent with the results shown in Table~\ref{tab:step1}.
\item
\textbf{F1}: \textbf{CNN} always achieves the highest F1 score except for the {\em Columba} project. This suggests that the cutoff point of \textbf{CNN} is more balanced (between precision and recall) than \textbf{Easy} and {\IT}.
\item
\textbf{Cost}: While \textbf{Easy} always cost zero human effort, \textbf{CNN} costs less human effort than {\IT} in 8 out of 10 projects due to its higher precision than {\IT}. 
\item
\textbf{Overall}: there is no clear winner except for three projects: (1) on \textbf{ArgoUML} and \textbf{Jruby}, {\IT} achieves both higher recall and a lower cost than \textbf{CNN}; (2) on \textbf{Hibernate}, \textbf{Easy} achieves the same recall at a lower cost than  \textbf{CNN}.
\ei
In conclusion, there is no clear win of {\IT} over \textbf{CNN} except on the \textbf{ArgoUML} and \textbf{Jruby} projects where {\IT} achieves higher recall at a lower cost than \textbf{CNN}. On the rest eight projects, {\IT} always achieves higher recall but also at a higher cost than \textbf{CNN}.

Besides the fact that {\IT} outperforms \textbf{CNN} on two out of ten projects. The advantage of {\IT} is that (1) it separates the ``easy to find'' SATDs from the ``hard to find'' ones so that the ``easy to find'' SATDs can be automatically identified with zero human cost; and (2) it can guarantee a high level of recall on the ``hard to find'' SATDs with accurate estimation. 

On the other hand, \textbf{CNN} also shows its strong prediction capability with its complex model and hyperparameter tuning. It is promising that replacing the random forest classifier in \textbf{Hard} with \textbf{CNN} can further improves the performance of {\IT} in the future.

Therefore, we believe that \textbf{CNN} has the potential to further improve {\IT} in the future but {\IT} is a better solution in terms of framework. Imagine using \textbf{CNN} for SATD identification on a new project, the user will be told that on average 80\% of the predicted comments will contain SATDs. If the user check all the predicted comments, they will be checking on average 4\% of the comments and finding 78\% of the SATDs and this is the end of it. On the other hand, when using {\IT}, the user will first be told that the selected comments by \textbf{Easy} are 100\% related to SATDs. On average, those comments cover 53\% of the total SATDs. If the user wants to find more SATDs, \textbf{Hard} will guide the user to check the comments most likely to be SATDs among the remaining ``hard to find'' ones. During this process, a recall will be estimated to inform the user what percentage of the ``hard to find'' SATDs have been identified. Thus {\IT} can guarantee to reach the user-specified recall.

\subsubsection{RQ3.3: How much computation cost does {\IT} add to the process besides the human effort cost?}

The runtime for {\IT} is split into two parts:
\bi
\item
\textbf{Easy} takes 12 seconds to train on 62,275 comments and to detect the ``easy to find'' SATDs on one target project.
\item
\textbf{Hard} takes on average 20.5 seconds to train the random forest model and 1.5 seconds to estimate recall in each iteration.
\ei
As a result, the total runtime for {\IT} would be $12+22\times N$ seconds where $N$ is the number of iterations (10 comments per iteration). For projects that require many iterations of \textbf{Hard}, the total runtime for {\IT} can be hours. However, given the fact that human needs more than 22 seconds to classify 10 comments in each iteration, in practice, the training time of \textbf{Hard} can hide behind the human classification time. That is to say, we can train the model once, provide the first 20 comments to human, then update the model for every 10 labeled comments and utilize the updated model to provide the next 10 comments for human classification\footnote{Note that the current design of the {\IT} does not implement this. We plan to develop a more interactive and user-friendly tool of {\IT} utilizing this query strategy in our future work.}. In this way, the additional computation cost is 34 seconds for {\IT} which is similar to training a traditional supervised learning model (\textbf{TM}, \textbf{RF}) and is much lower than training a deep learning model (3,548 seconds for training a \textbf{CNN} model in Ren et al.~\cite{ren2019neural}).

\begin{RQ}{In summary:}
{\IT} outperforms the state-of-the-art SATD identification solutions by reaching higher recall at a lower cost and negligible additional computation time (34 seconds). This is attributed to two factors of {\IT}: (1) it first identifies the ``easy to find'' SATDs with close to 100\% precision, thus 20 to 90\% of the SATDs can be found with zero human effort; (2) for the remaining ``hard to find'' SATDs, it utilizes the human classification results to update its prediction model incrementally and thus can make better guidance to which comments are more likely to be SATDs.
\end{RQ}

\section{Apply {\IT} to A New Project}
\label{sec:apply}

To test the generalizability of {\IT} while also demonstrating how researchers and developers can use {\IT}, we applied {\IT} to identify SATDs from a new, unlabeled project Apache httpd-2.4.6\footnote{\url{https://archive.apache.org/dist/httpd/httpd-2.4.6.tar.gz}}.

\textbf{Data collection: }we applied srcML\footnote{\url{https://www.srcml.org/}} and extracted 17,208 code comments from the Apache httpd-2.4.6 project, similar to what Potdar and Shiha~\cite{potdar2014exploratory} did.

\textbf{Easy: }the \textbf{Easy} algorithm trained on the 10 labeled datasets (the four patterns of ``{\em fixme,  todo, hack, workaround}'') was applied to extract the ``easy to find'' SATDs from the 17,208 code comments. As a result, 148 comments were extracted as the ``easy to find'' SATDs. 

\textbf{Validation of Easy: }given the close to 100\% precision of \textbf{Easy}, the 148 comments should be automatically treated as SATDs without human verification. However, to test the generalizability of \textbf{Easy}, we validated these 148 comments by manually inspecting and classifying each one. We found 4 false-positives from these 148 comments, as listed in Table~\ref{tab:fp}. Therefore, the precision of \textbf{Easy} on the Apache httpd-2.4.6 project is 144/148=97\%. The full validation results of all 148 comments are available on the GitHub repository\footnote{\url{https://github.com/ai-se/Jitterbug/blob/master/httpd/httpd\_easy_validated.csv}}.

\begin{table}[!tbh]
\caption{False-Positives of Easy on Apache Httpd}
\centering
\setlength\tabcolsep{2pt}
\begin{tabular}{|p{.75\linewidth}|p{.06\linewidth}|p{.1\linewidth}|}
\hline
\textbf{Comment Text} & 
\textbf{Easy}  &
\textbf{Human}\\
\hline
/\* See TODO in ap\_queue\_info\_set\_idle() \*/ & 
yes & 
no \\
\hline
/\* See TODO in ap\_queue\_info\_set\_idle() \*/ & 
yes & 
no \\
\hline
/\* basedir is either "", or "/\%2f" for the "squid \%2f hack" \*/ &
yes & 
no \\
\hline
/\* Add a link to the root directory (if \%2f hack was used) \*/ & 
yes & 
no \\
\hline
\end{tabular}
\label{tab:fp}
\end{table}

\textbf{Hard: }the remaining 17,060 comments contain only the ``hard to find'' SATDs. To identify such ``hard to find'' SATDs, the \textbf{Hard} model trained on the 10 labeled datasets was applied to sample 10 most informative comments from the 17,060 comments. Then the first author (acting as a user) manually inspected and labeled the sampled 10 comments. Next, the \textbf{Hard} model was re-trained with the newly added 10 labeled data. These processes should iterate until the number of ``hard to find'' SATDs identified meets the desired recall (based on the estimation of the total number of ``hard to find'' SATDs in the 17,060 comments). However, with limited human effort available, we only inspected and labeled 100 comments with the help of \textbf{Hard}, interactively. Among these 100 inspected comments, 87 were found to be SATDs. Therefore, the precision@100 for \textbf{Hard} on Apache httpd-2.4.6 is 87\%. The full inspection results are available at the GitHub repository\footnote{\url{https://github.com/ai-se/Jitterbug/blob/master/httpd/httpd_rest_coded.csv}}.

\textbf{Summary: }on Apache httpd-2.4.6, the precision of \textbf{Easy} is 97\% and the precision@100 for \textbf{Hard} is 87\%. These results suggest that {\IT} can be successfully applied to unlabeled software projects.
As an existing work, Potdar and Shiha~\cite{potdar2014exploratory} applied their keyword-based method to identify SATDs on Apache httpd-2.4.6. They found 112 SATDs in total. The {\IT} result is better than Potdar and Shiha~\cite{potdar2014exploratory} since by manually inspecting 100 comments, {\IT} identified 144+87 = 231 SATDs, of which the 144 ``easy to find'' SATDs did not require any human effort. More details on how to apply {\IT} to extract SATDs from an unlabeled project are available on our GitHub repository\footnote{\url{https://github.com/ai-se/Jitterbug/blob/master/README.md}}.

\section{Threats to Validity}
\label{sec:Threats to Validity}

There are several validity threats~\cite{feldt2010validity} to the design of this study. Any conclusion made from this work must be considered with the following issues in mind:

\textbf{Conclusion validity} focuses on the significance of the treatment. To enhance   conclusion validity, we ran experiments on 10 different target projects and found that our proposed method always performed better than the state-of-the-art approaches.

\textbf{Internal validity }focuses on how sure we can be that the treatment caused the outcome. To enhance   internal validity, we heavily constrained our experiments to the same dataset, with the same settings, except for the treatments being compared.

\textbf{Construct validity }focuses on the relation between the theory behind the experiment and the observation. To enhance construct validity, we compared solutions with and without our strategies in Table~\ref{tab:overall} and showed that both components (\textbf{Easy} and \textbf{Hard}) improve the overall performance. However, we only showed that with our setting of featurization and default parameters of each learner, random forest learner is the best choice. What we have not shown is that whether the performance can get even better by tuning the parameters or using a different set of features. We plan to explore these in our future work.

\textbf{External validity} concerns how widely our conclusions can be applied. In order to test the generalizability of our approach, we always kept a project as the holdout test set and never used any information from it in training. In addition, we have applied {\IT} to identify SATDs from an unlabeled software project Apache httpd-2.4.6 with a real human inspecting the comments. The results shown in \tion{apply} demonstrated the success of {\IT} in this real-world application.

\section{Conclusion and Future work}
\label{sec:conclusion}

Identifying self-admitted technical debts (SATDs) from source code comments is important to maintain a healthy software project. Current solutions cannot automate this process due to the lack of precision of machine learning models in predicting the SATDs. To reduce the human effort required in identifying SATDs, this paper first showed that there are two types of SATDs--- (1) the ``easy to find'' SATDs that can be automatically identified with close to 100\% precision; and (2) the ``hard to find'' SATDs that only human experts can make the final decisions on. Then a half-automated two-step approach was proposed---  \textbf{Step 1}: apply a novel pattern recognition technique to learn and utilize strong patterns of the ``easy to find'' SATDs to identify them automatically; \textbf{Step 2}: (a) train/update a continuous learning model incrementally on both historically labeled data and new human decisions, and (b) guide the human experts to screen the comments that most likely contain the ``hard to find'' SATDs according to the model's prediction, iterate (a) and (b) until a target recall has been reached according to the model's estimation. Based on simulation results on ten software projects, we conclude that
\bi
\item
\textbf{Step 1} solution \textbf{Easy} can find 20-90\% of the SATDs with close to 100\% precision automatically.
\item
\textbf{Step 2} solution \textbf{Hard} outperforms the state-of-the-art methods in finding the remaining ``hard to find'' SATDs (\textbf{Hard} finds more SATDs with less human effort).
\item
\textbf{Step 2} solution \textbf{Hard} can also provide an accurate estimation of the number of ``hard to find'' SATDs undiscovered, thus offers a practical way to stop at the target recall.
\item
Overall, the proposed two-step solution {\IT} is most efficient in identifying SATDs. 
\ei
That said, {\IT} still suffers from the validity threats discussed in \tion{Threats to Validity}. To further reduce those threats and to move forward with this research, we propose the following future work:
\begin{enumerate}
\item
Apply hyper-parameter tuning on data preprocessing and model configuration to see if our current conclusions still hold and whether tuning can further improve the performance.
\item
Prototype {\IT} as a tool to make it more user-friendly.
\item
Explore more complex patterns (other than just single word patterns \textbf{Easy} has explored) in Step 1.
\item
Explore more advanced feature engineering in Step 2 for finding the ``hard to find'' SATDs. E.g. explore N-gram patterns~\cite{wattanakriengkrai2019automatic} and word embeddings with deep neural networks~\cite{flisar2019identification}.
\item
Explore whether replacing the random forest model in {\IT} with a deep learning model (CNN~\cite{ren2019neural}) will further improve its performance.
\item
Extend the work to other types of technical debts and compare it with other state-of-the-art methods which continue to appear.
\end{enumerate}
One important message this paper tries to convey is that--- do not waste effort on finding the ``easy to find'' SATDs, future research on identifying SATDs should mostly focus on the ``hard to find'' SATDs.

\counterwithin{figure}{section}
\appendices
\section{}
\begin{figure*}[h]
    \centering
    \subfloat[Apache Ant]
    {
        \includegraphics[width=0.45\linewidth]{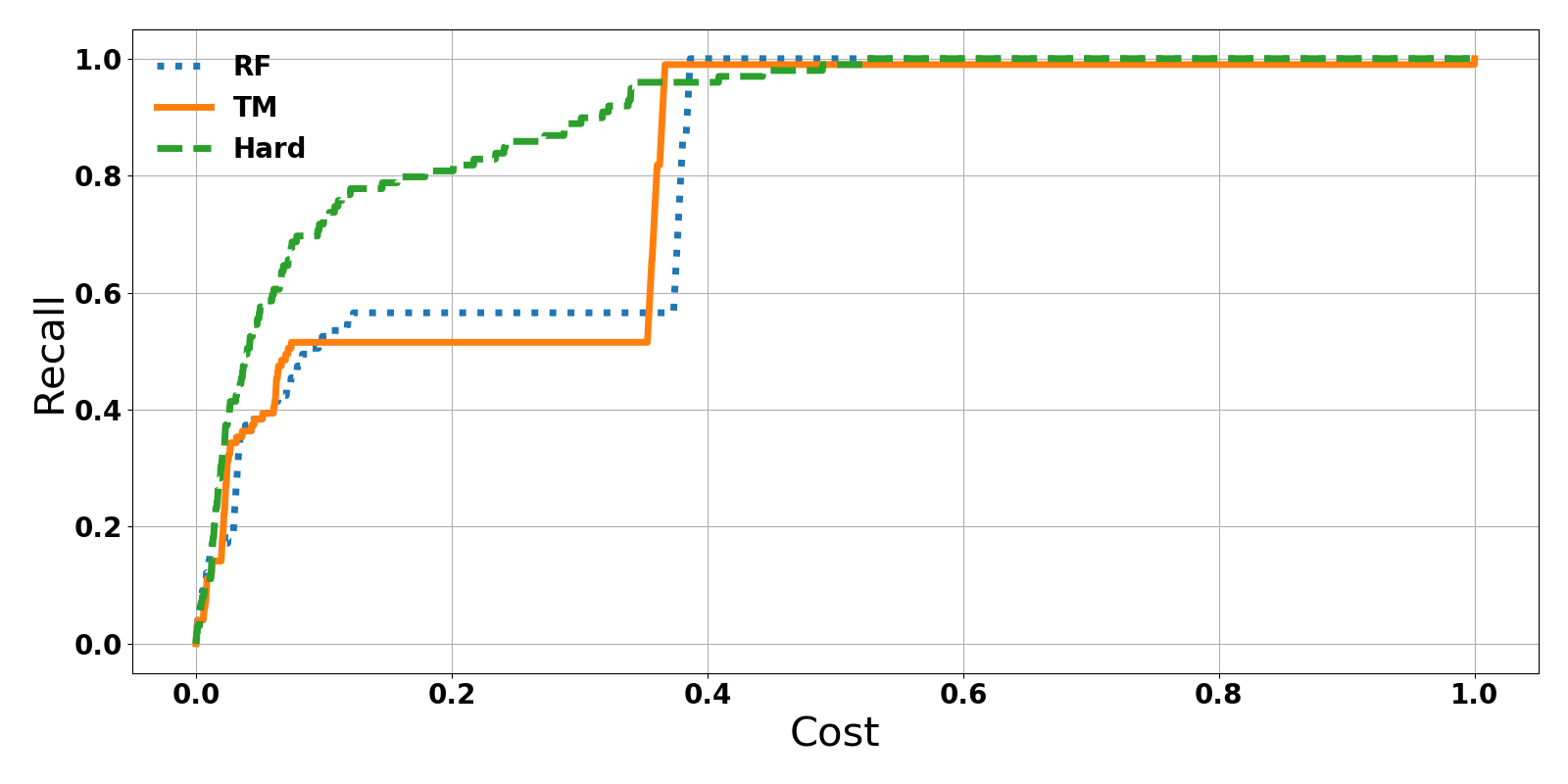}
    }\quad
    \subfloat[JMeter]
    {
        \includegraphics[width=0.45\linewidth]{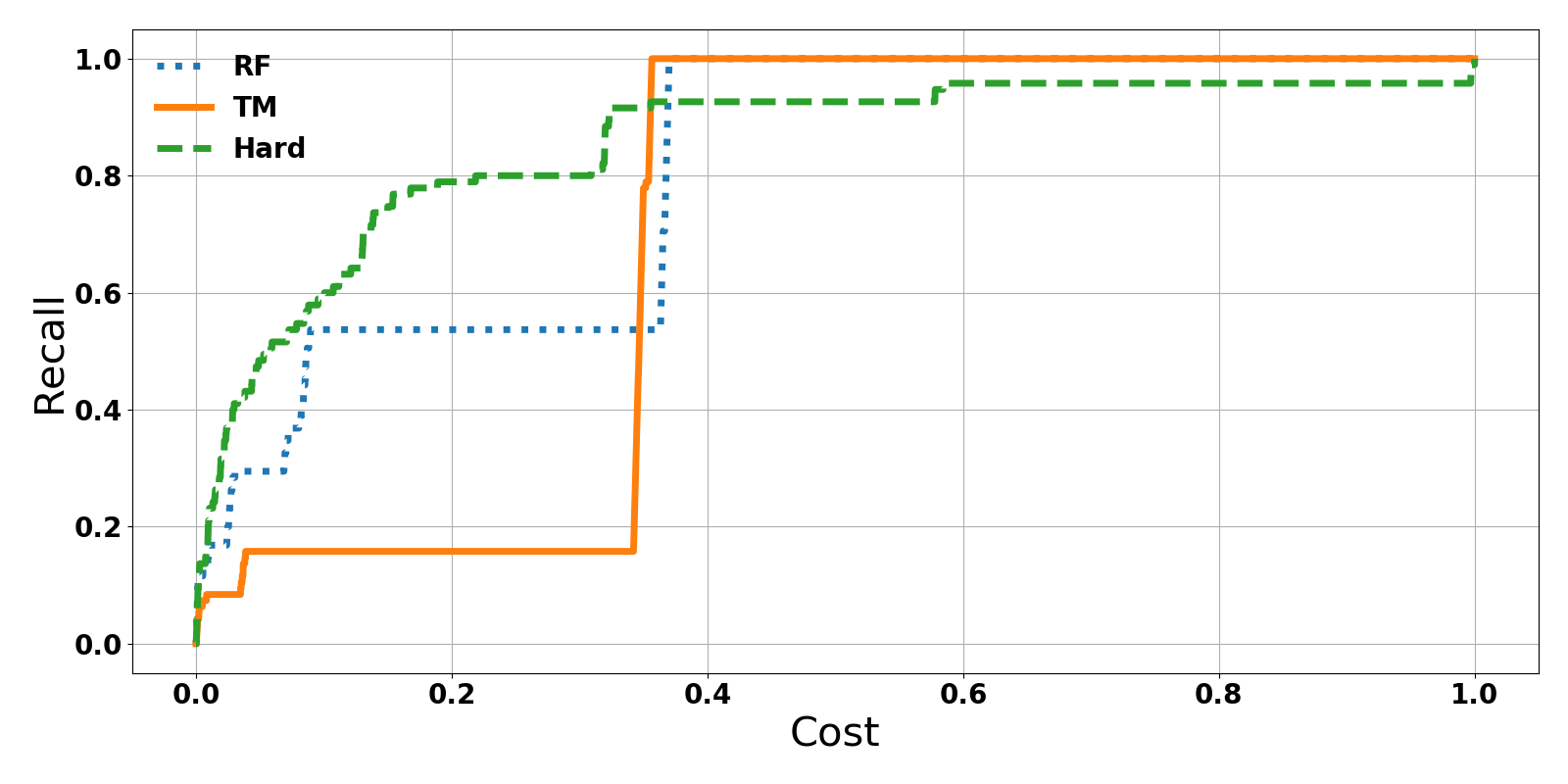}
    }\\
    \subfloat[ArgoUML]
    {
        \includegraphics[width=0.45\linewidth]{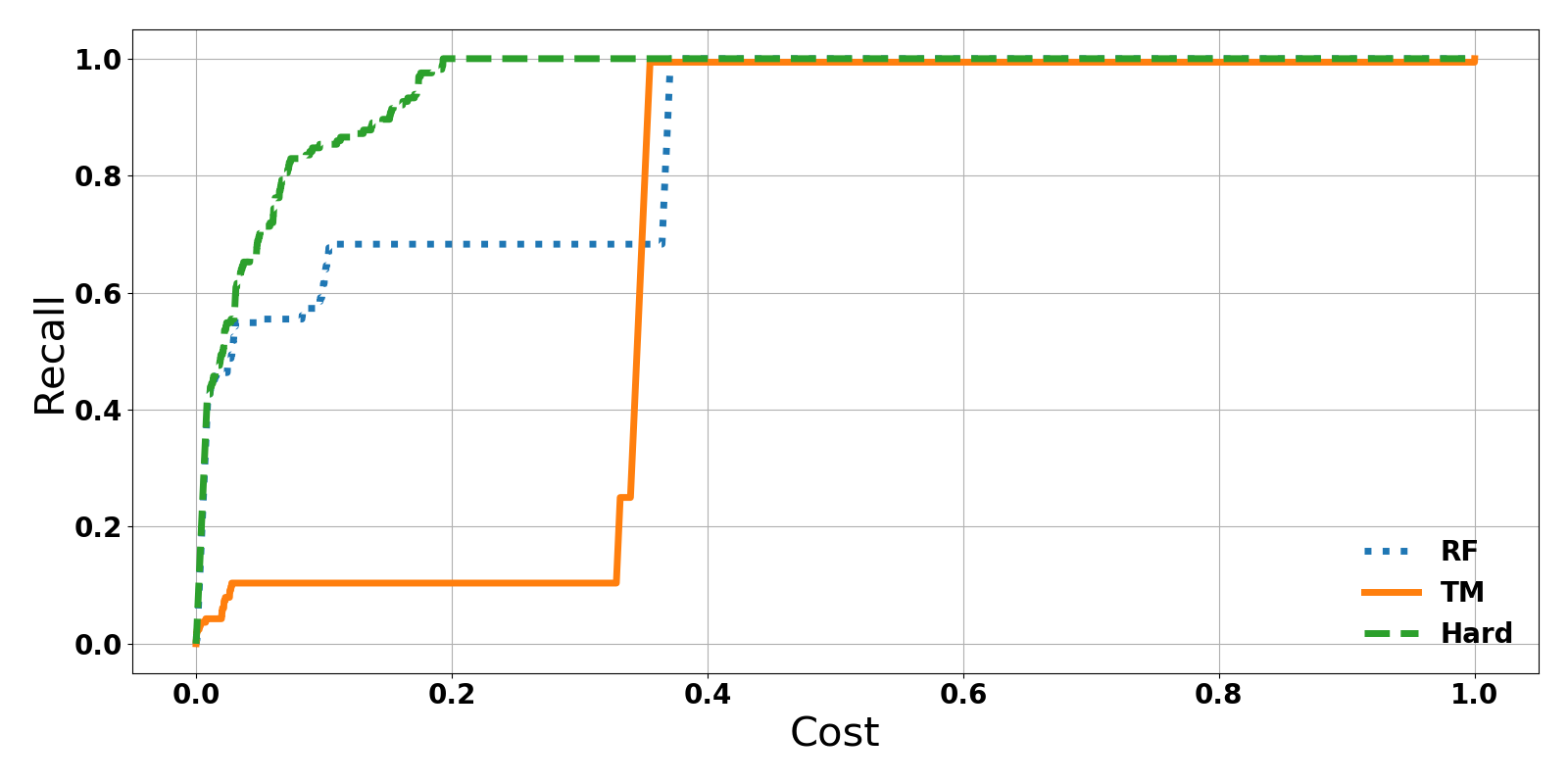}
    }\quad
    \subfloat[Columba]
    {
        \includegraphics[width=0.45\linewidth]{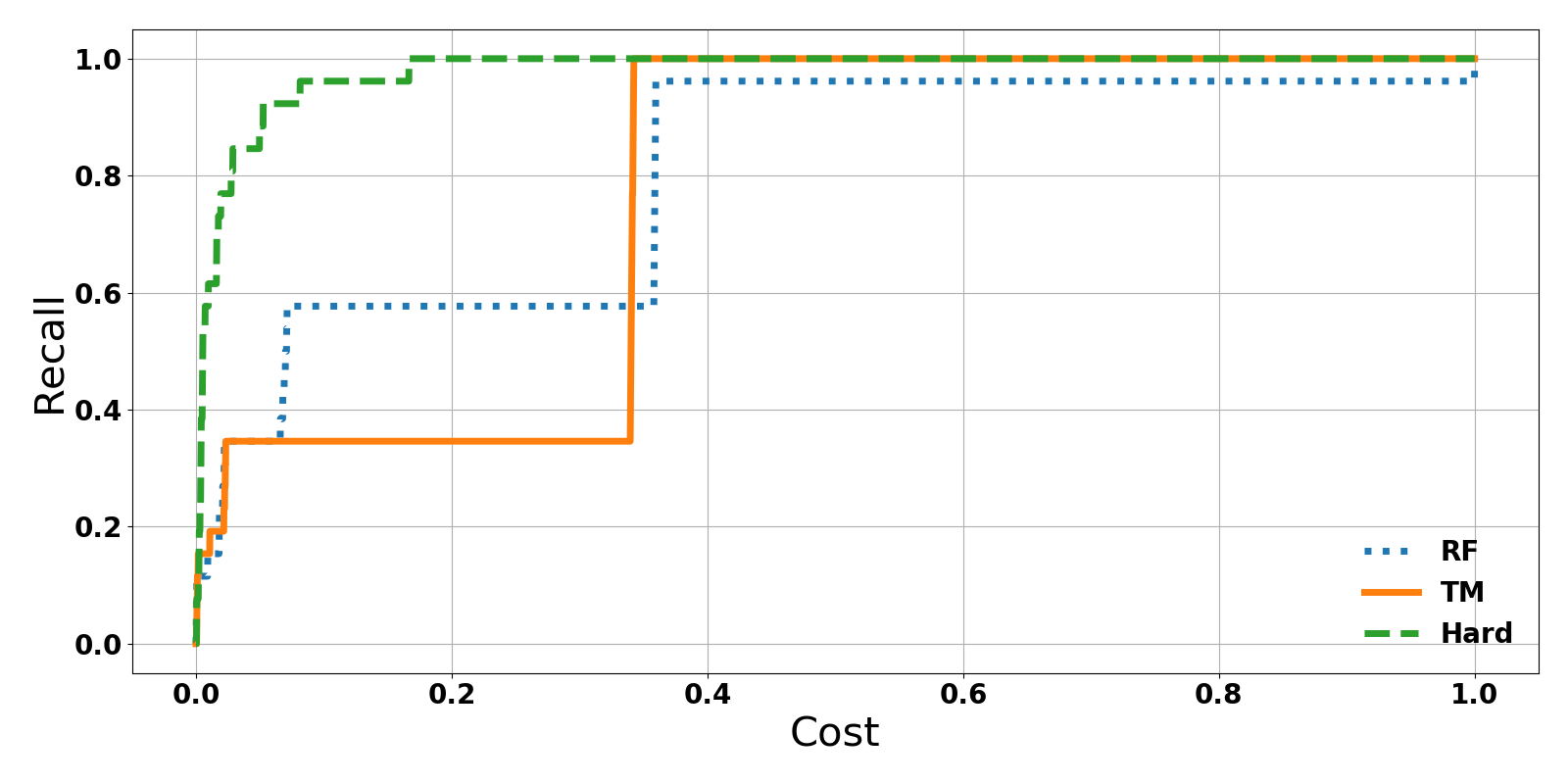}
    }\\
    \subfloat[EMF]
    {
        \includegraphics[width=0.45\linewidth]{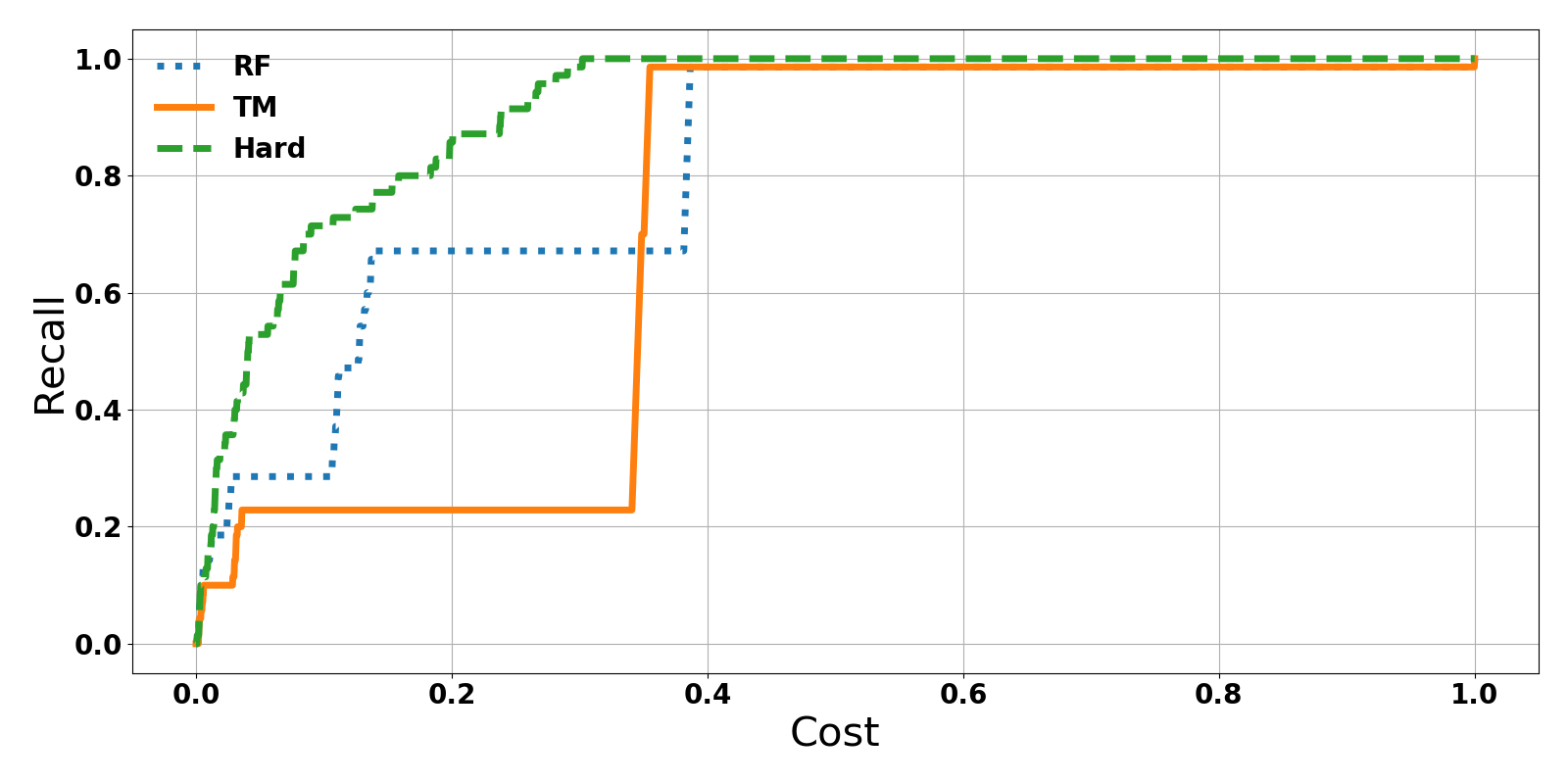}
    }\quad
    \subfloat[Hibernate]
    {
        \includegraphics[width=0.45\linewidth]{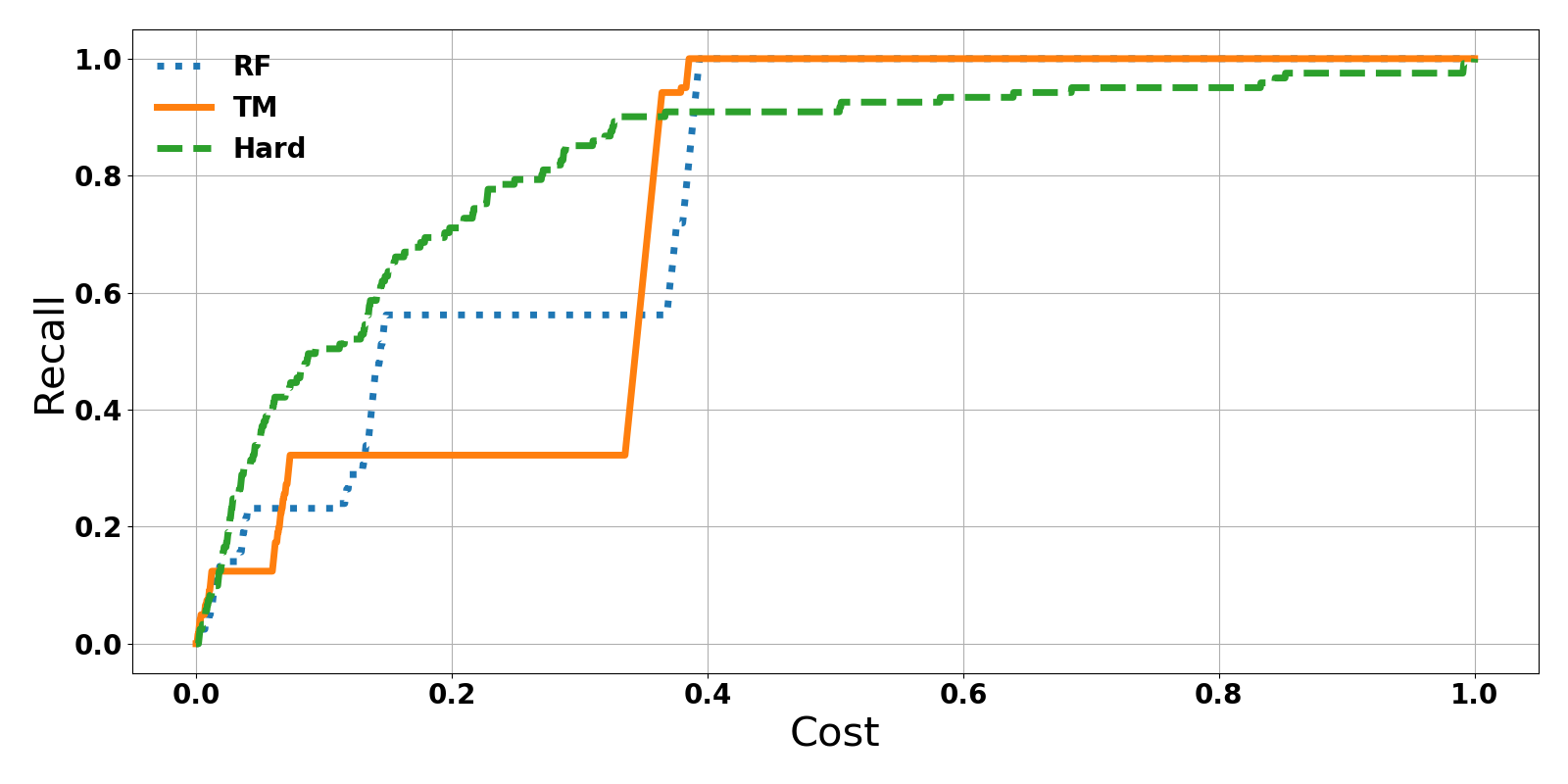}
    }\\
    \subfloat[JEdit]
    {
        \includegraphics[width=0.45\linewidth]{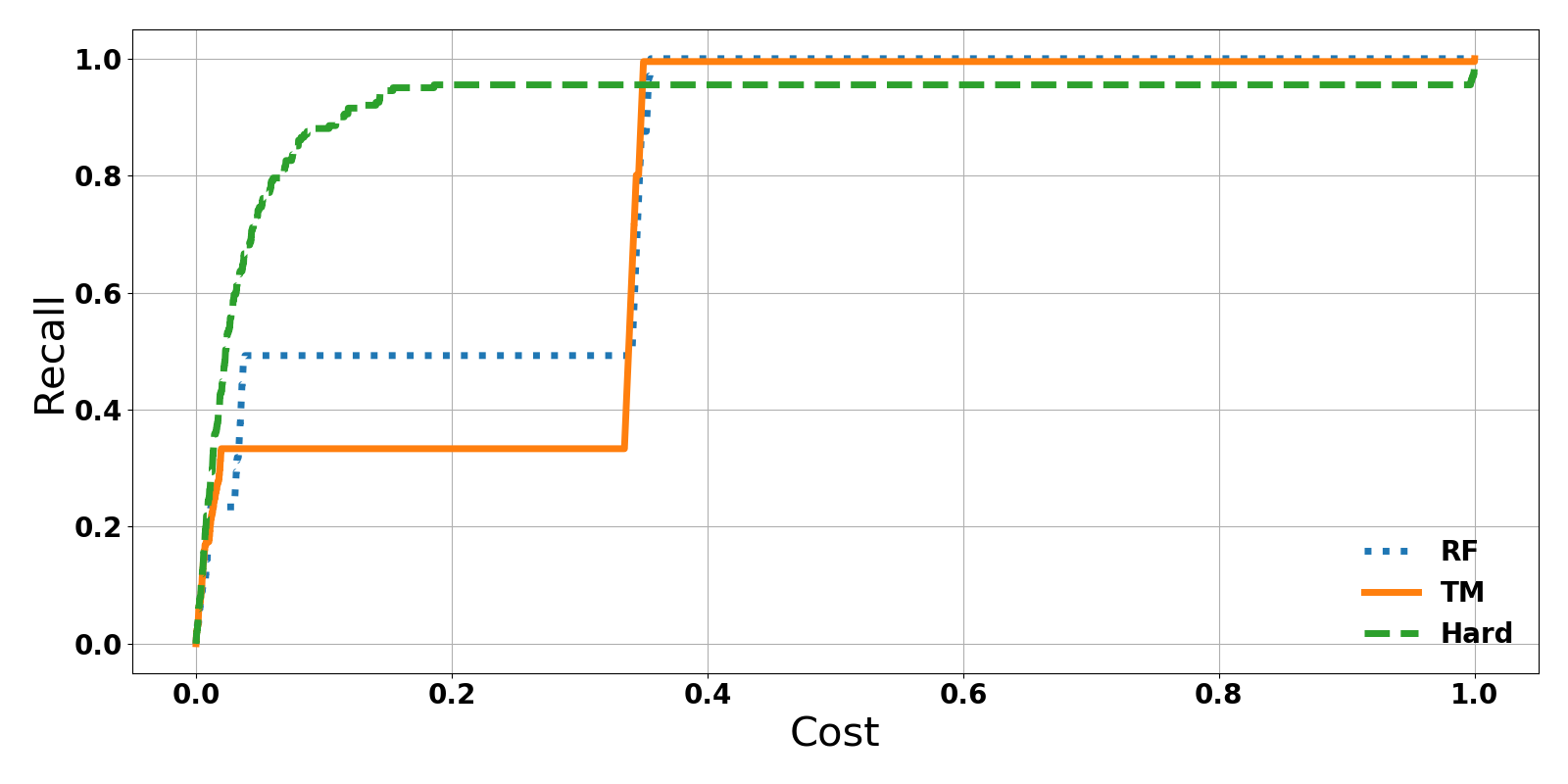}
    }\quad
    \subfloat[JFreeChart]
    {
        \includegraphics[width=0.45\linewidth]{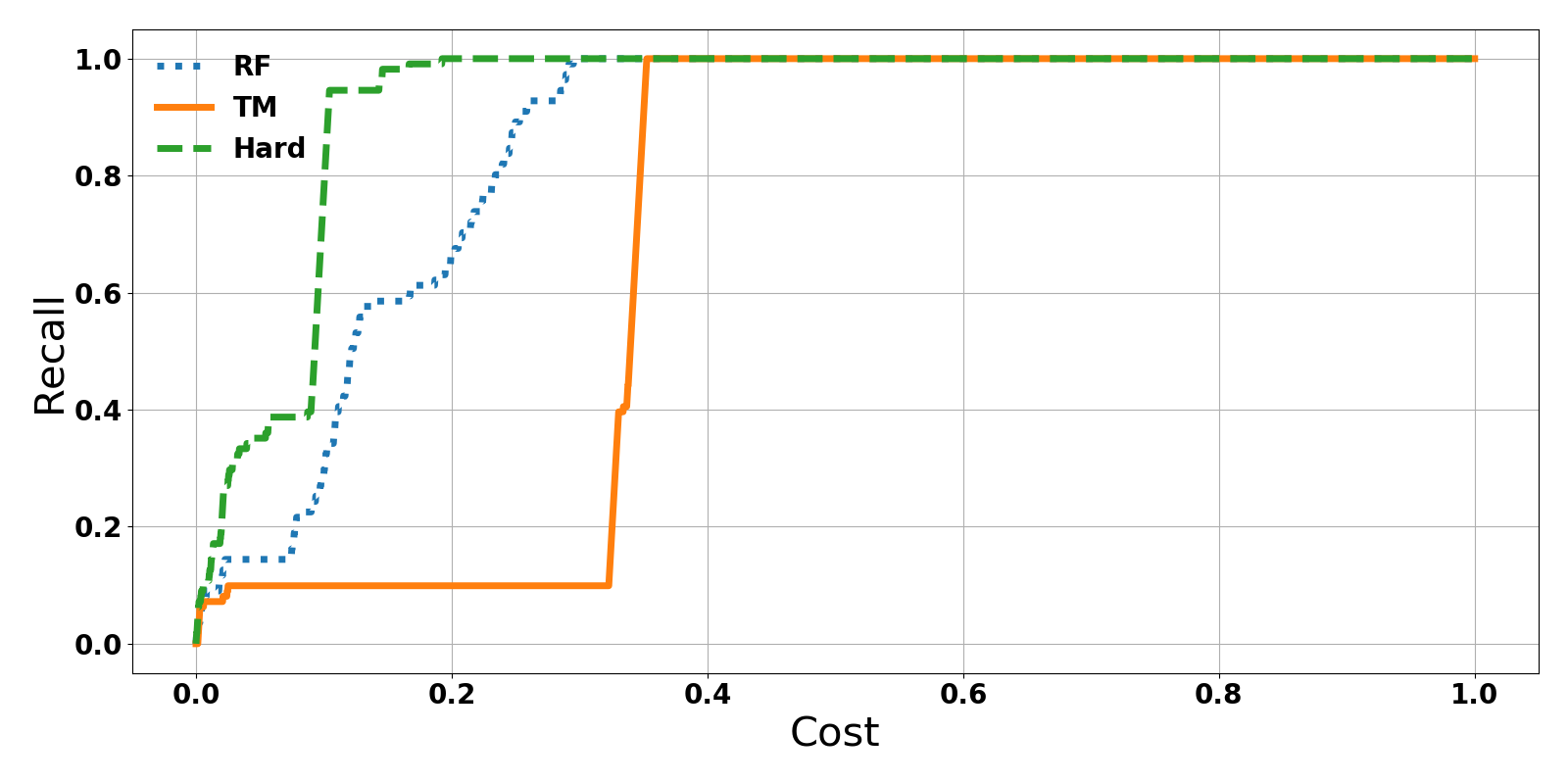}
    }\\
    \subfloat[JRuby]
    {
        \includegraphics[width=0.45\linewidth]{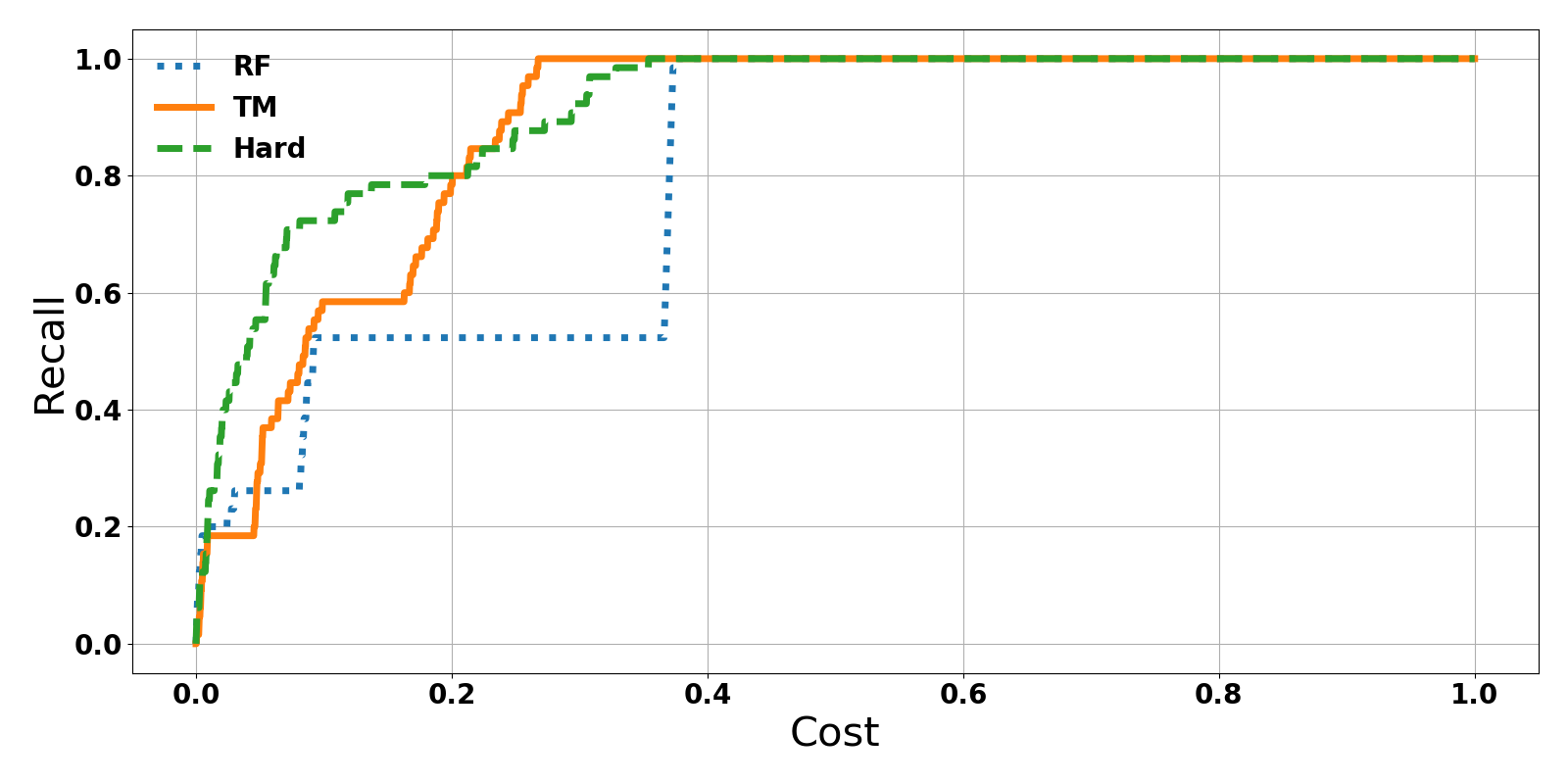}
    }\quad
    \subfloat[SQuirrel]
    {
        \includegraphics[width=0.45\linewidth]{figures/rest/sql12.png}
    }\\
    \caption{Recall-cost curves for finding ``hard to find'' SATDs on every target project.}
    \label{fig:rest_all}
\end{figure*}

\begin{figure*}[h]
    \centering
    \subfloat[Apache Ant]
    {
        \includegraphics[width=0.45\linewidth]{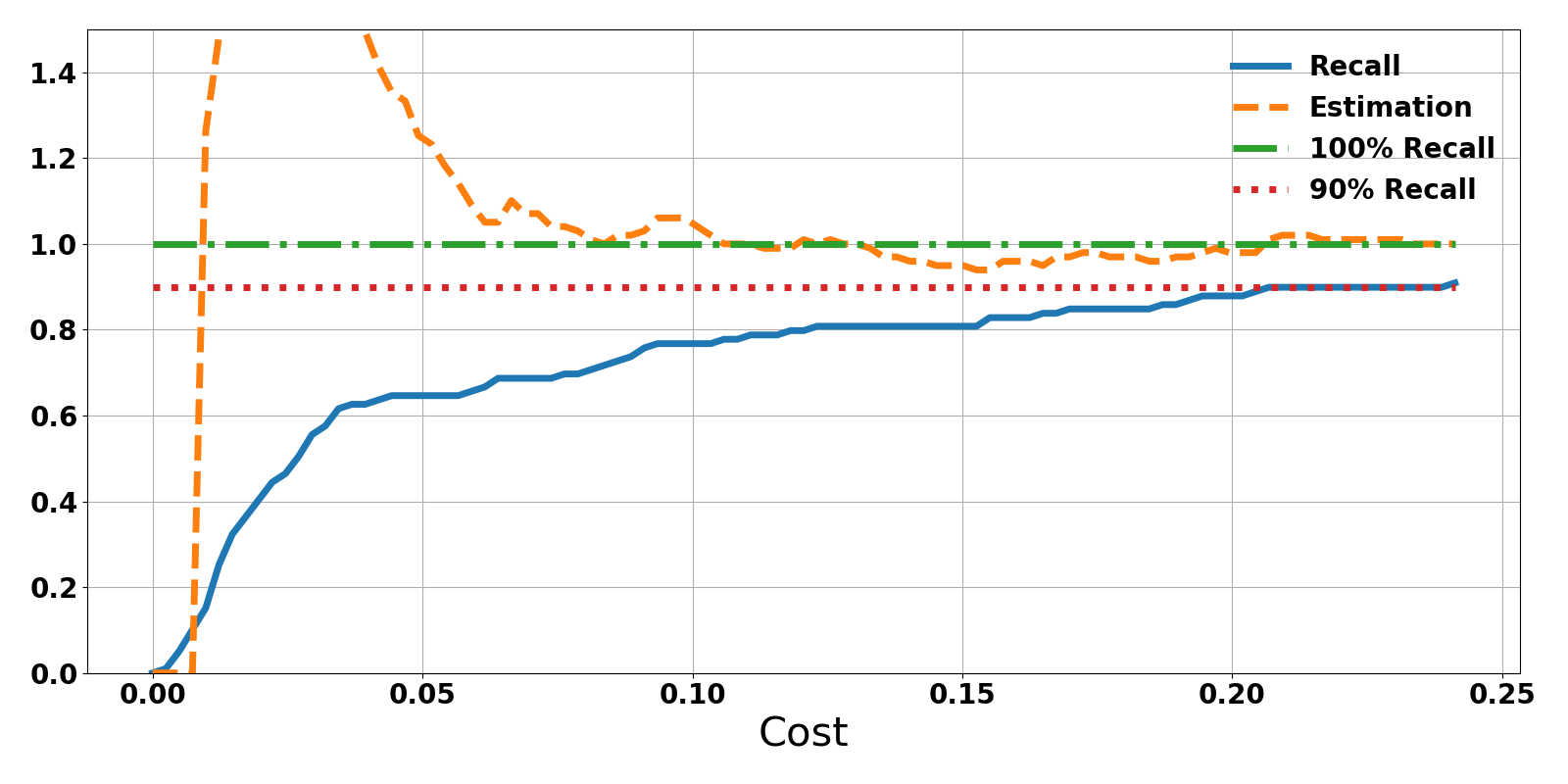}
    }\quad
    \subfloat[JMeter]
    {
        \includegraphics[width=0.45\linewidth]{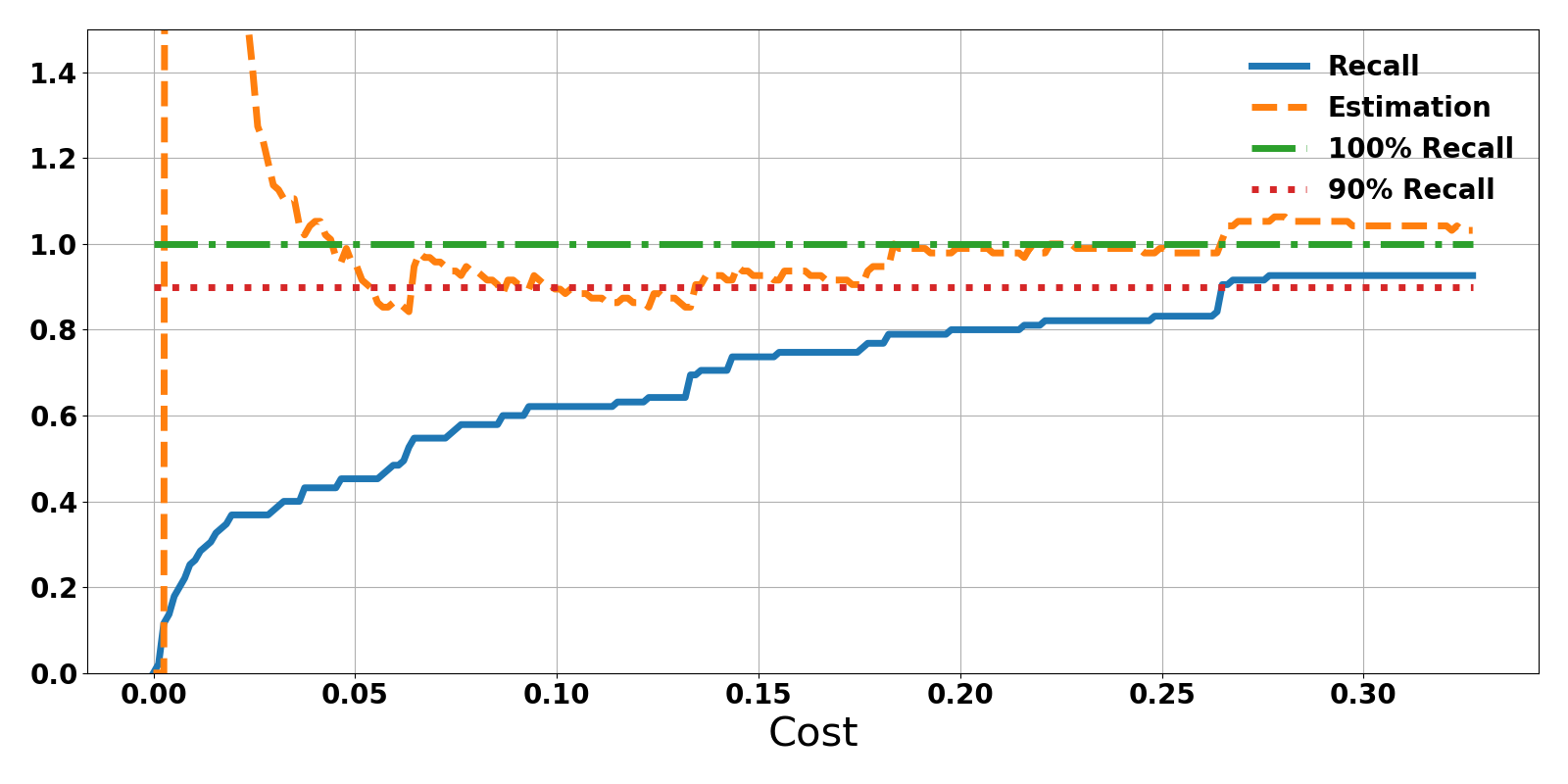}
    }\\
    \subfloat[ArgoUML]
    {
        \includegraphics[width=0.45\linewidth]{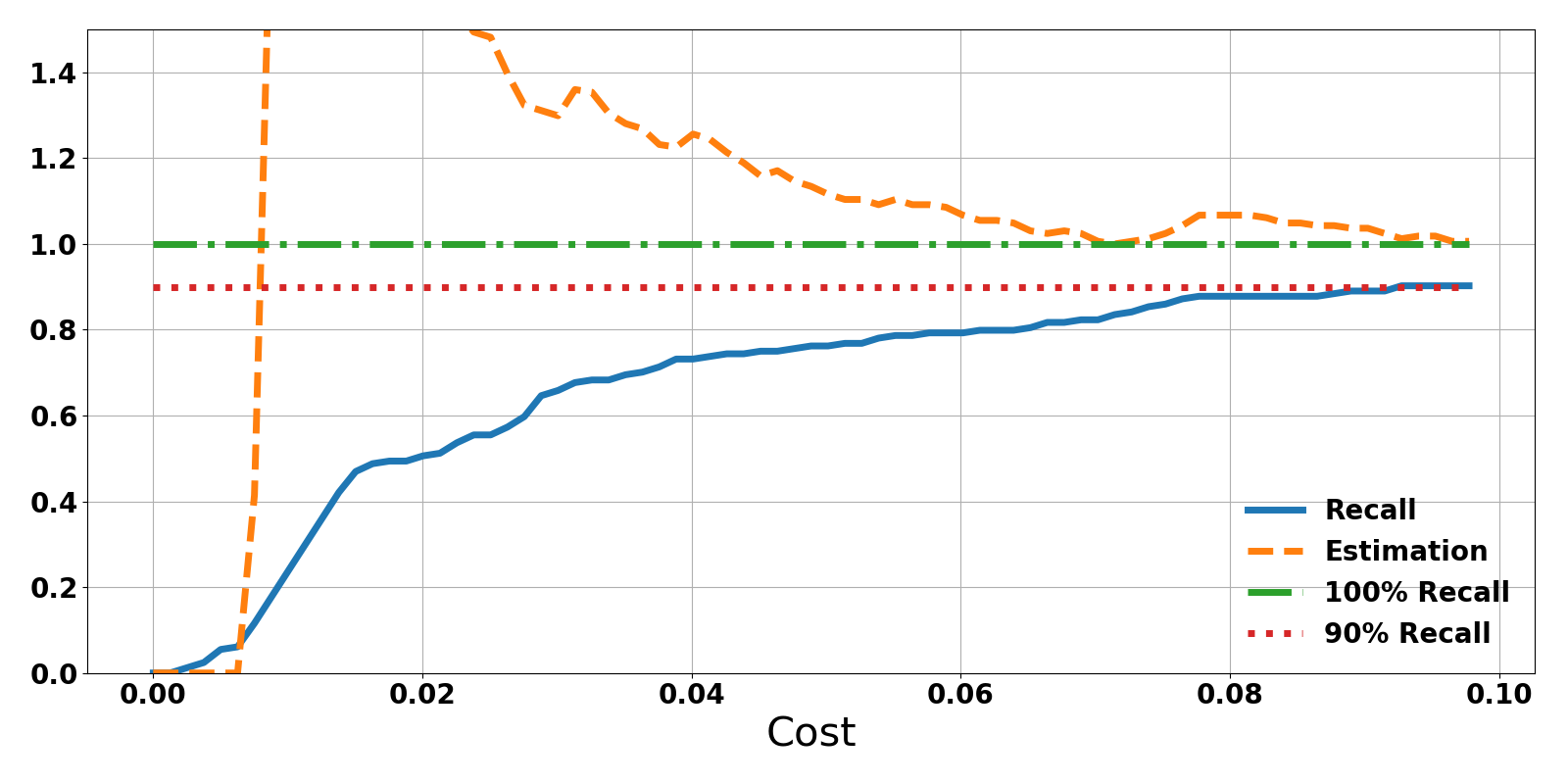}
    }\quad
    \subfloat[Columba]
    {
        \includegraphics[width=0.45\linewidth]{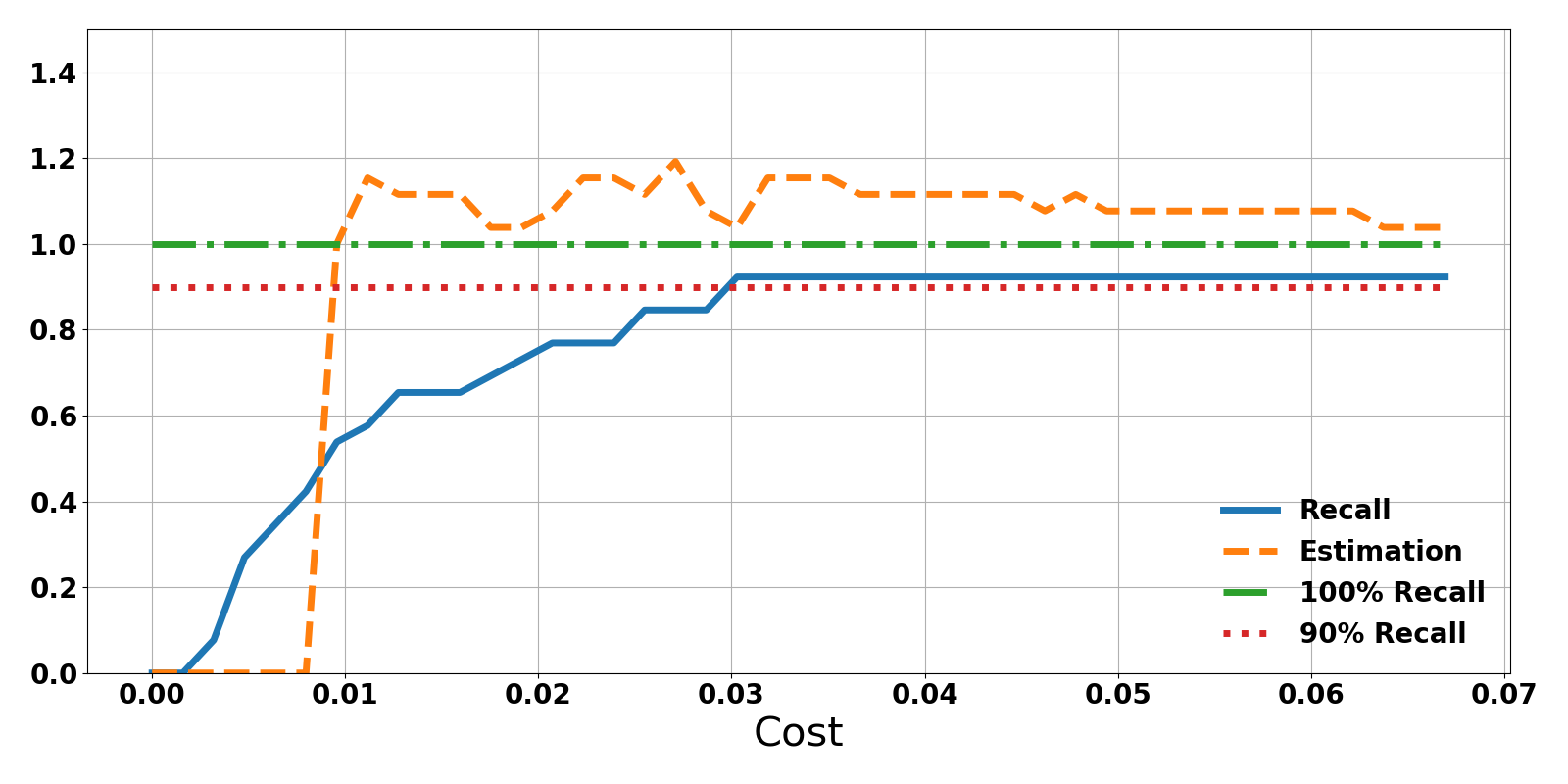}
    }\\
    \subfloat[EMF]
    {
        \includegraphics[width=0.45\linewidth]{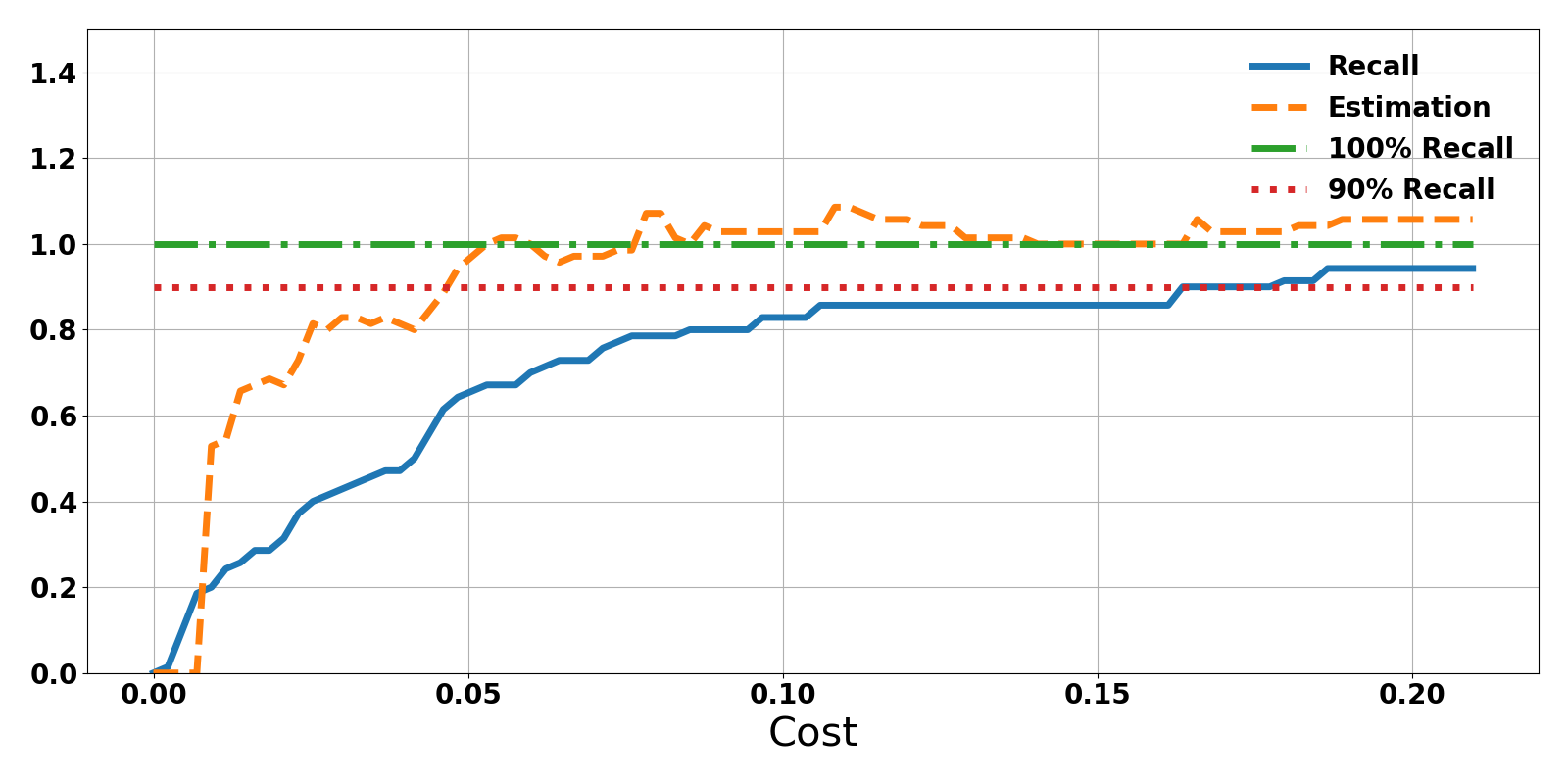}
    }\quad
    \subfloat[Hibernate]
    {
        \includegraphics[width=0.45\linewidth]{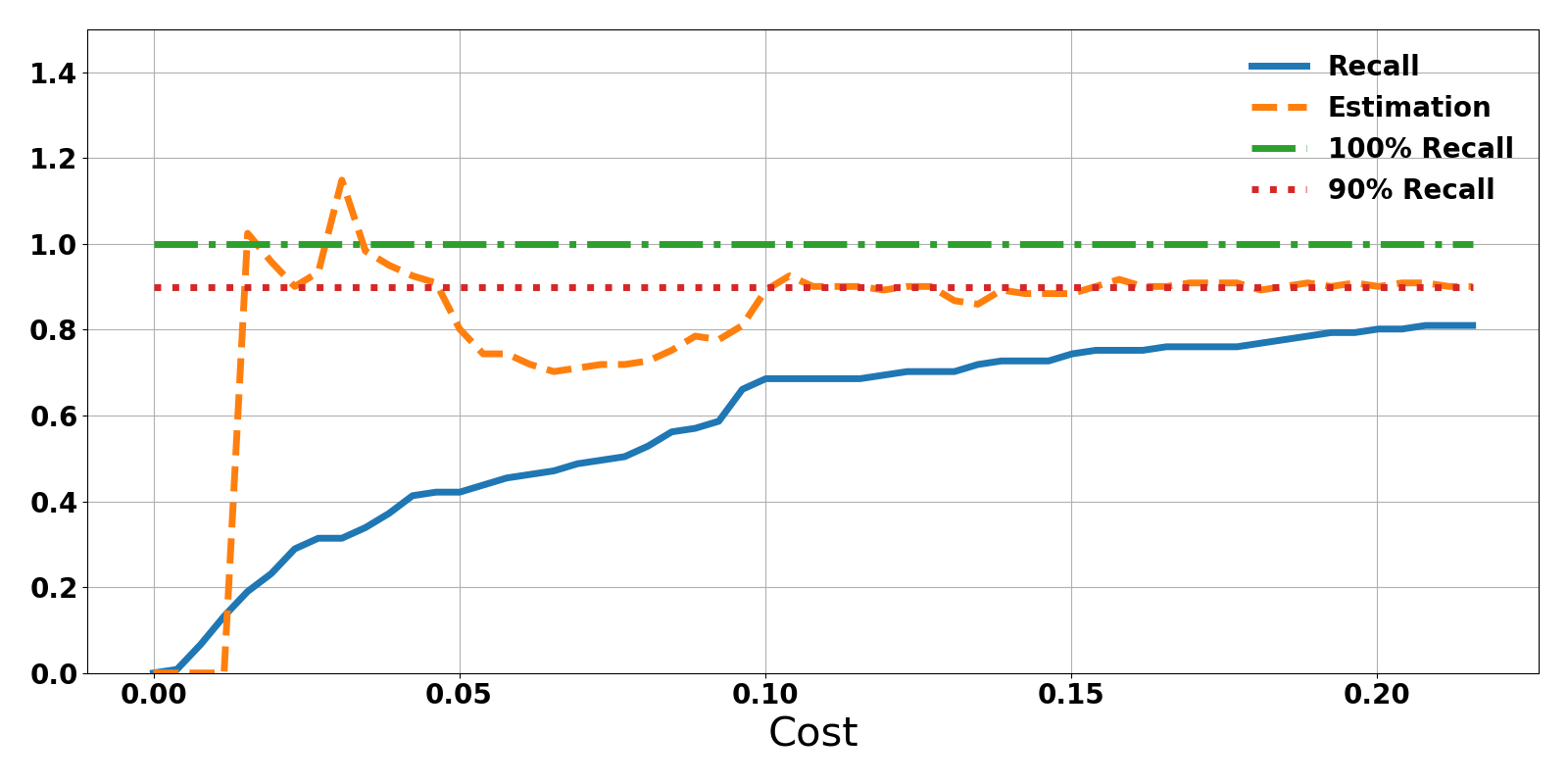}
    }\\
    \subfloat[JEdit]
    {
        \includegraphics[width=0.45\linewidth]{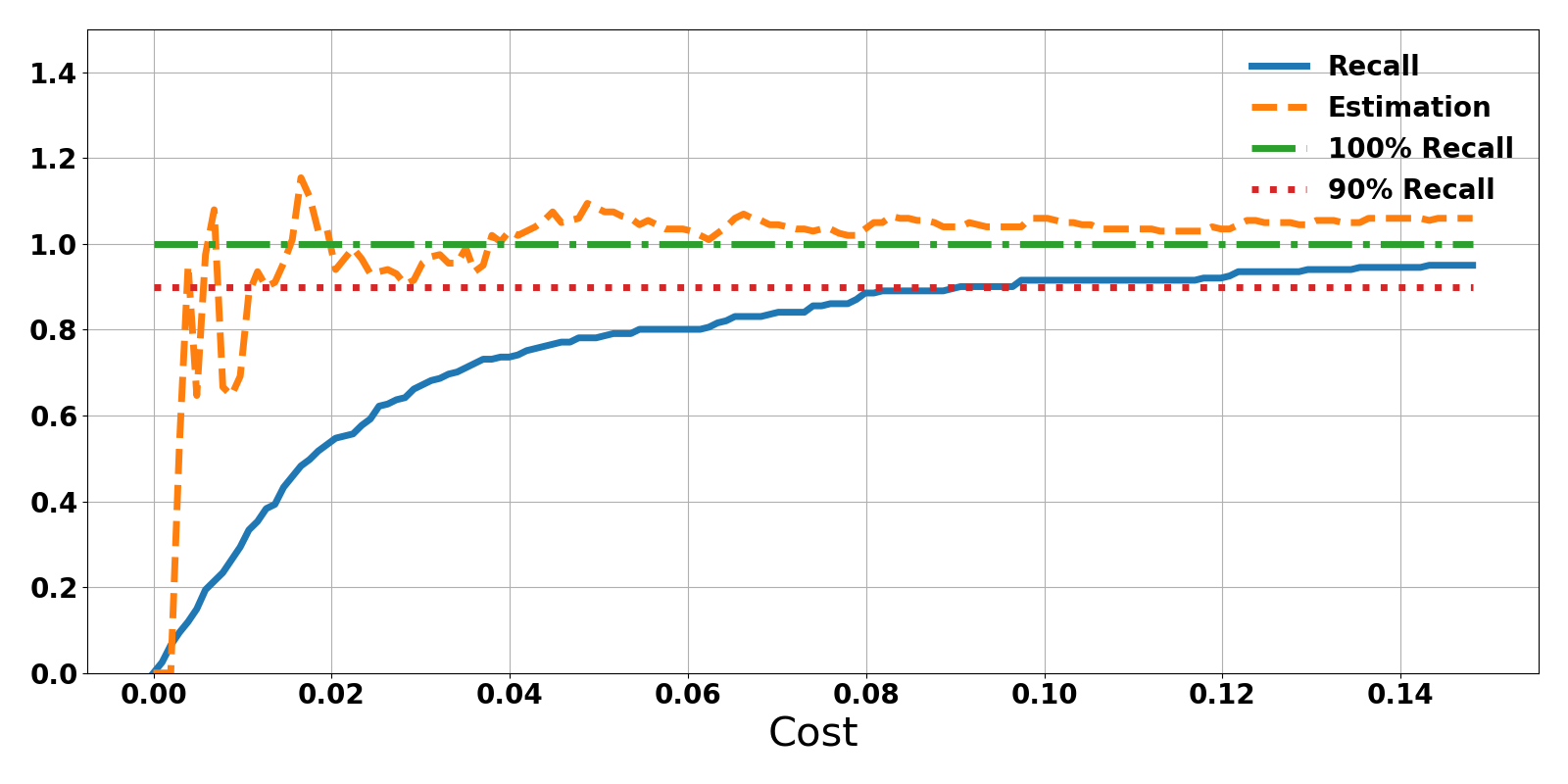}
    }\quad
    \subfloat[JFreeChart]
    {
        \includegraphics[width=0.45\linewidth]{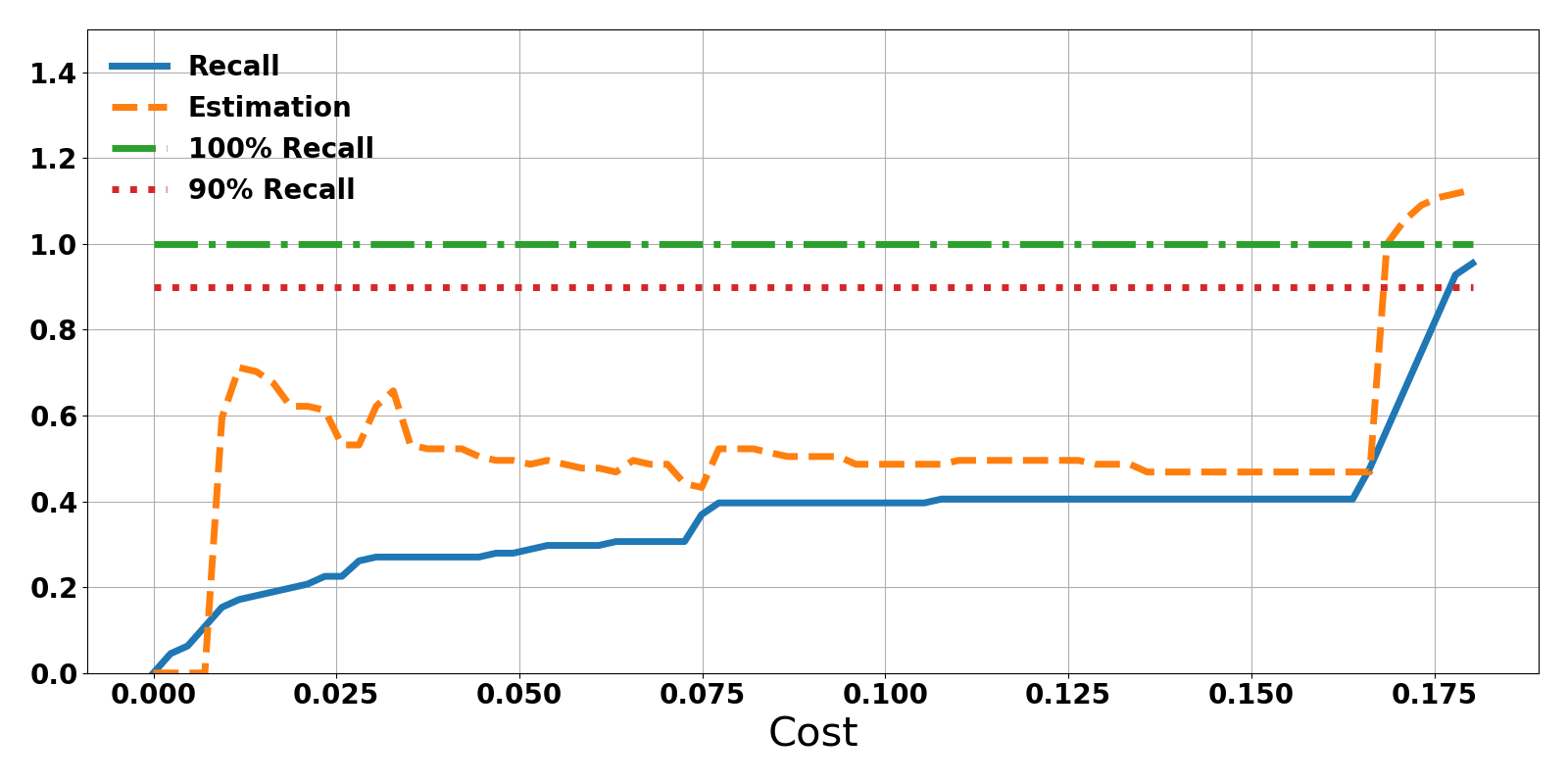}
    }\\
    \subfloat[JRuby]
    {
        \includegraphics[width=0.45\linewidth]{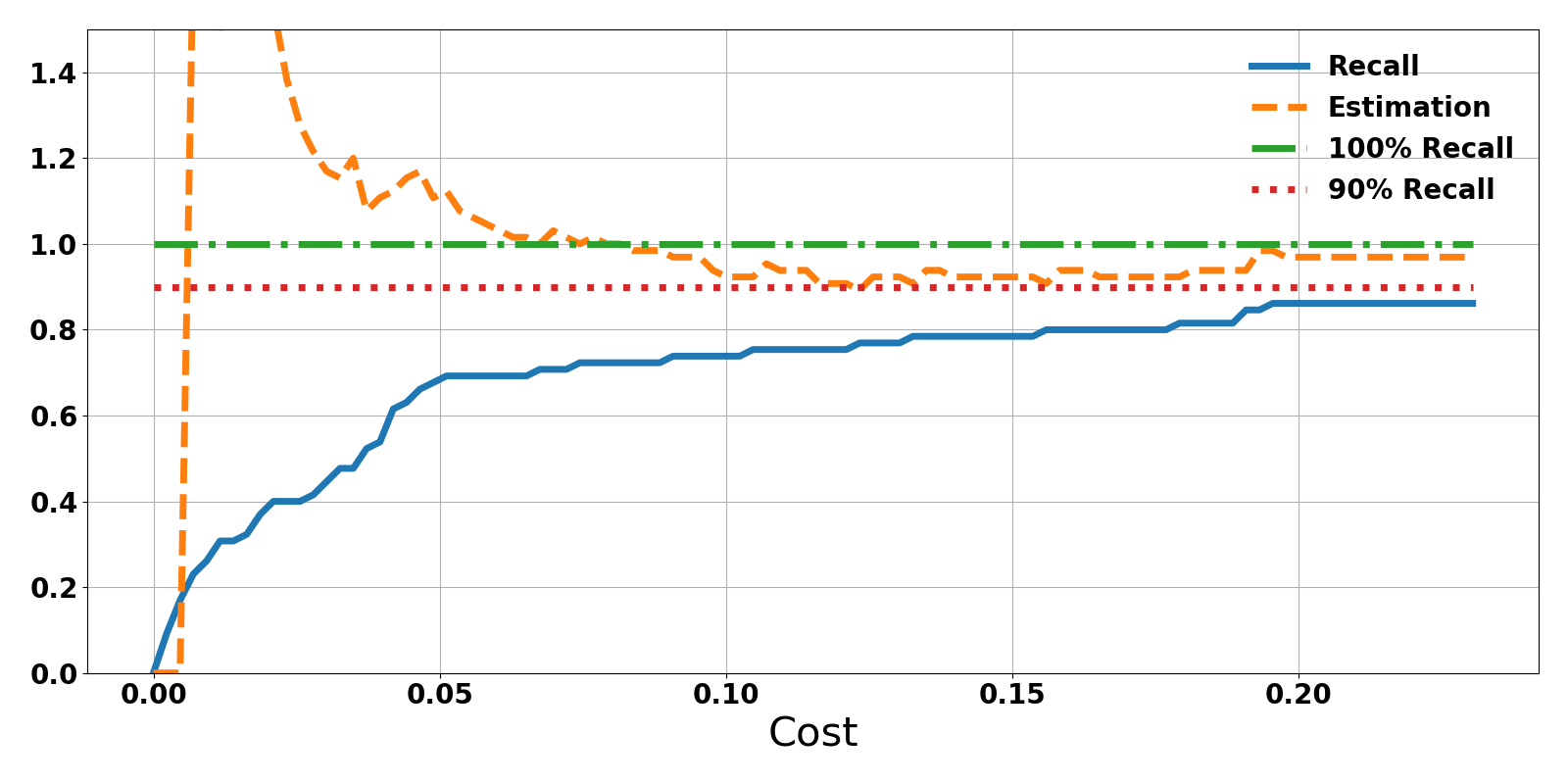}
    }\quad
    \subfloat[SQuirrel]
    {
        \includegraphics[width=0.45\linewidth]{figures/est/sql12.png}
    }\\
    \caption{Recall-cost and estimation-cost curves for finding 90\% of the ``hard to find'' SATDs with \textbf{Hard} on every target project.}
    \label{fig:est_all}
\end{figure*}

\begin{figure*}[h]
    \centering
    \subfloat[Apache Ant]
    {
        \includegraphics[width=0.45\linewidth]{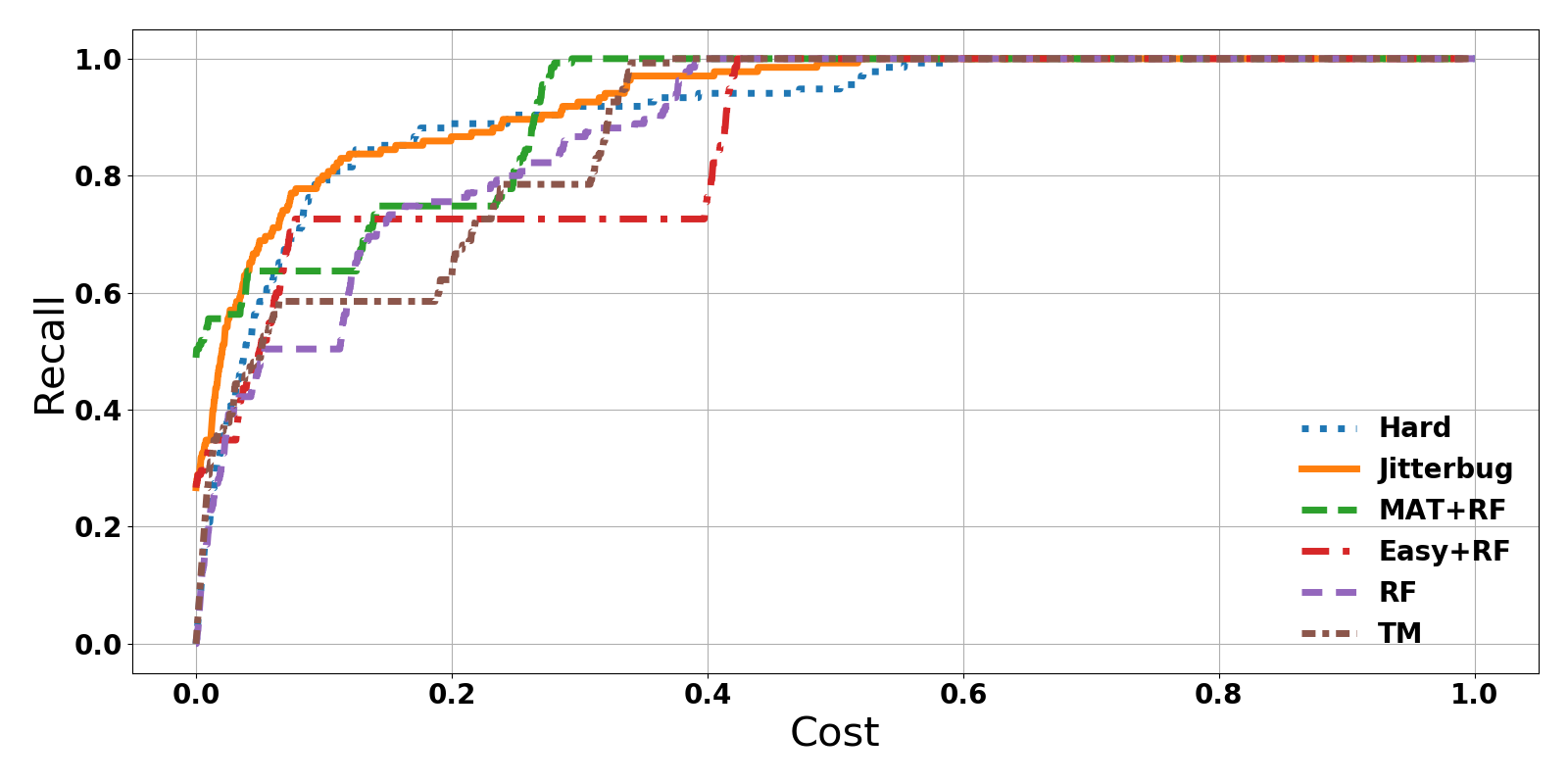}
    }\quad
    \subfloat[JMeter]
    {
        \includegraphics[width=0.45\linewidth]{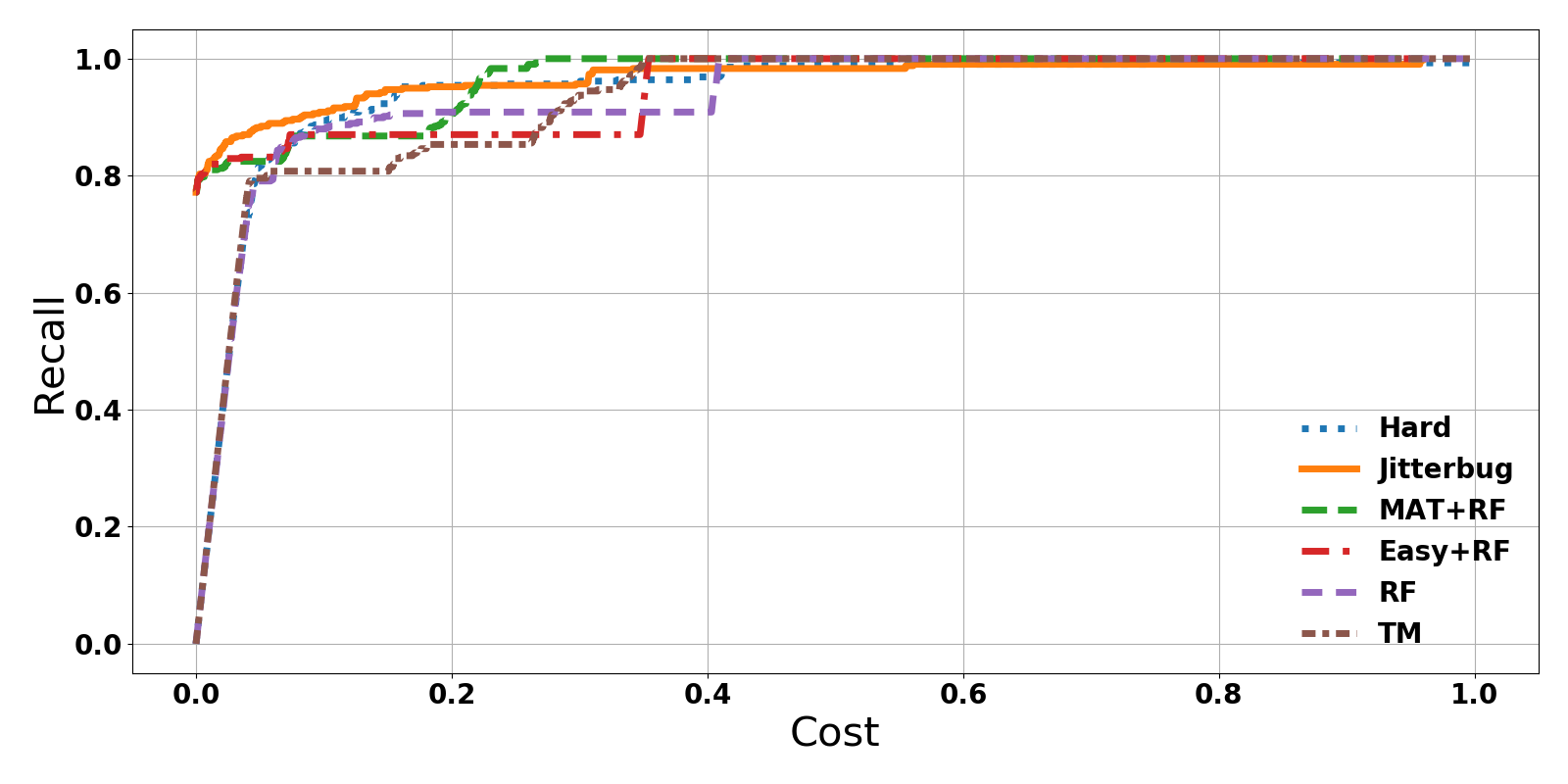}
    }\\
    \subfloat[ArgoUML]
    {
        \includegraphics[width=0.45\linewidth]{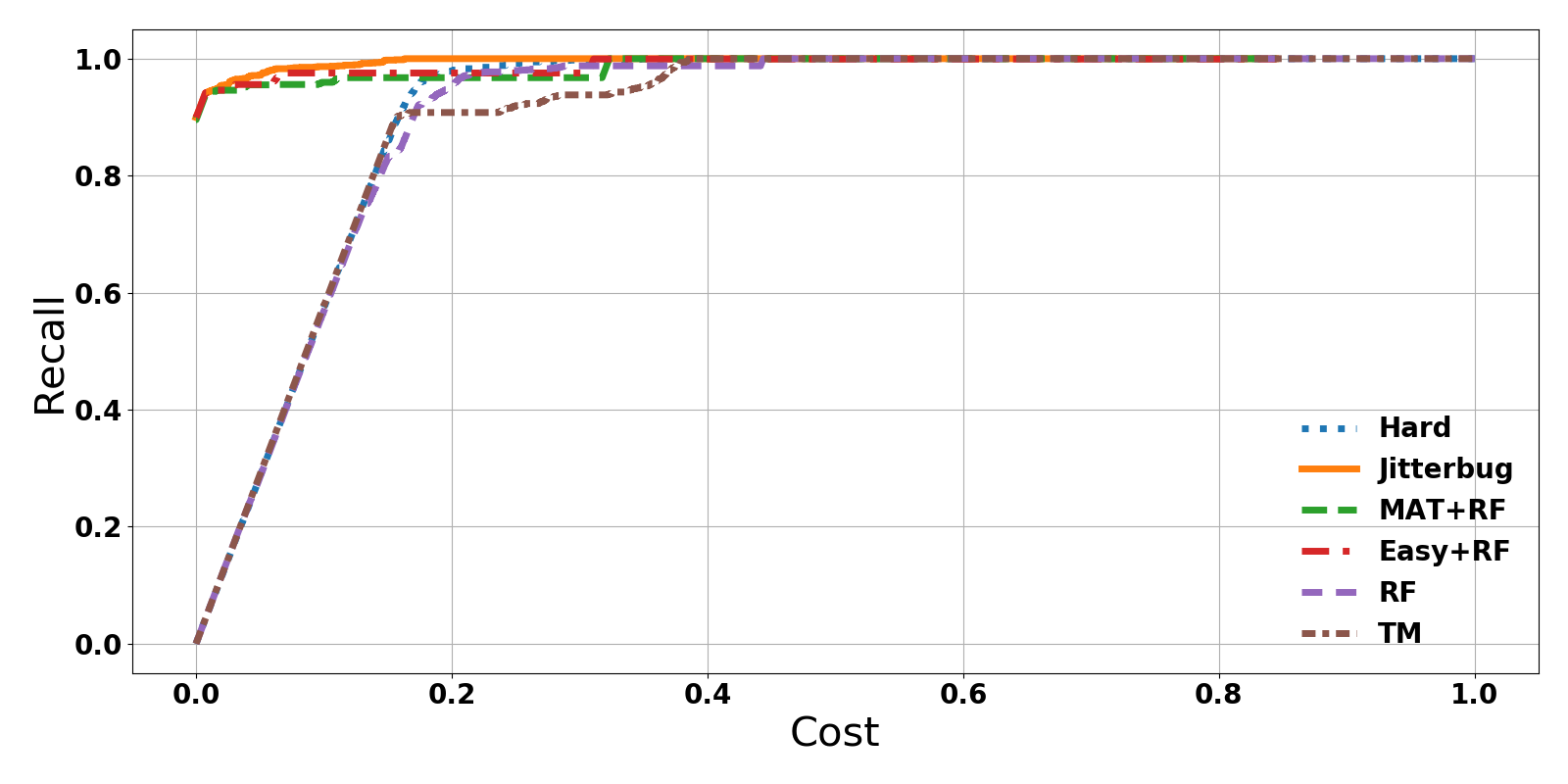}
    }\quad
    \subfloat[Columba]
    {
        \includegraphics[width=0.45\linewidth]{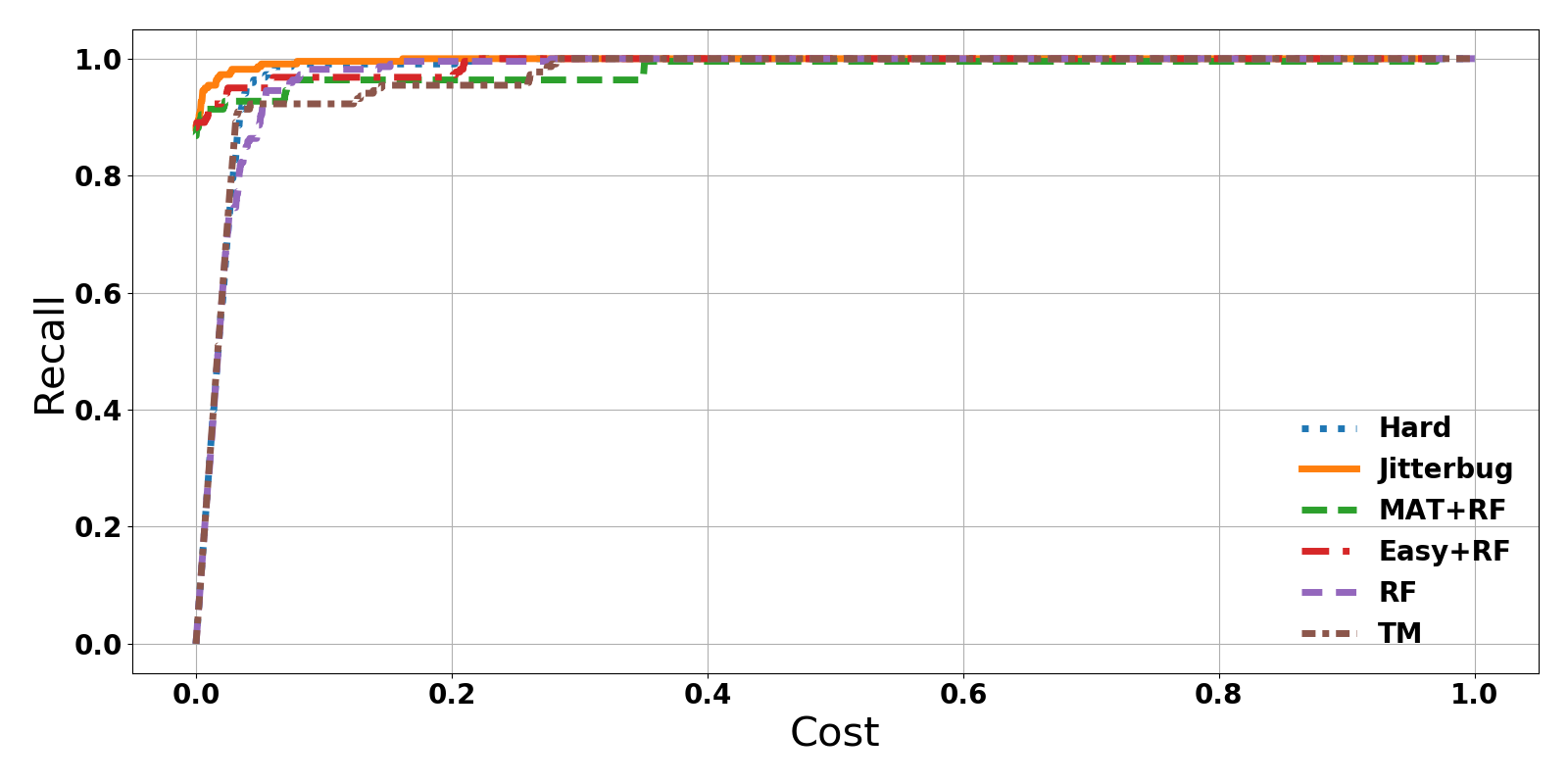}
    }\\
    \subfloat[EMF]
    {
        \includegraphics[width=0.45\linewidth]{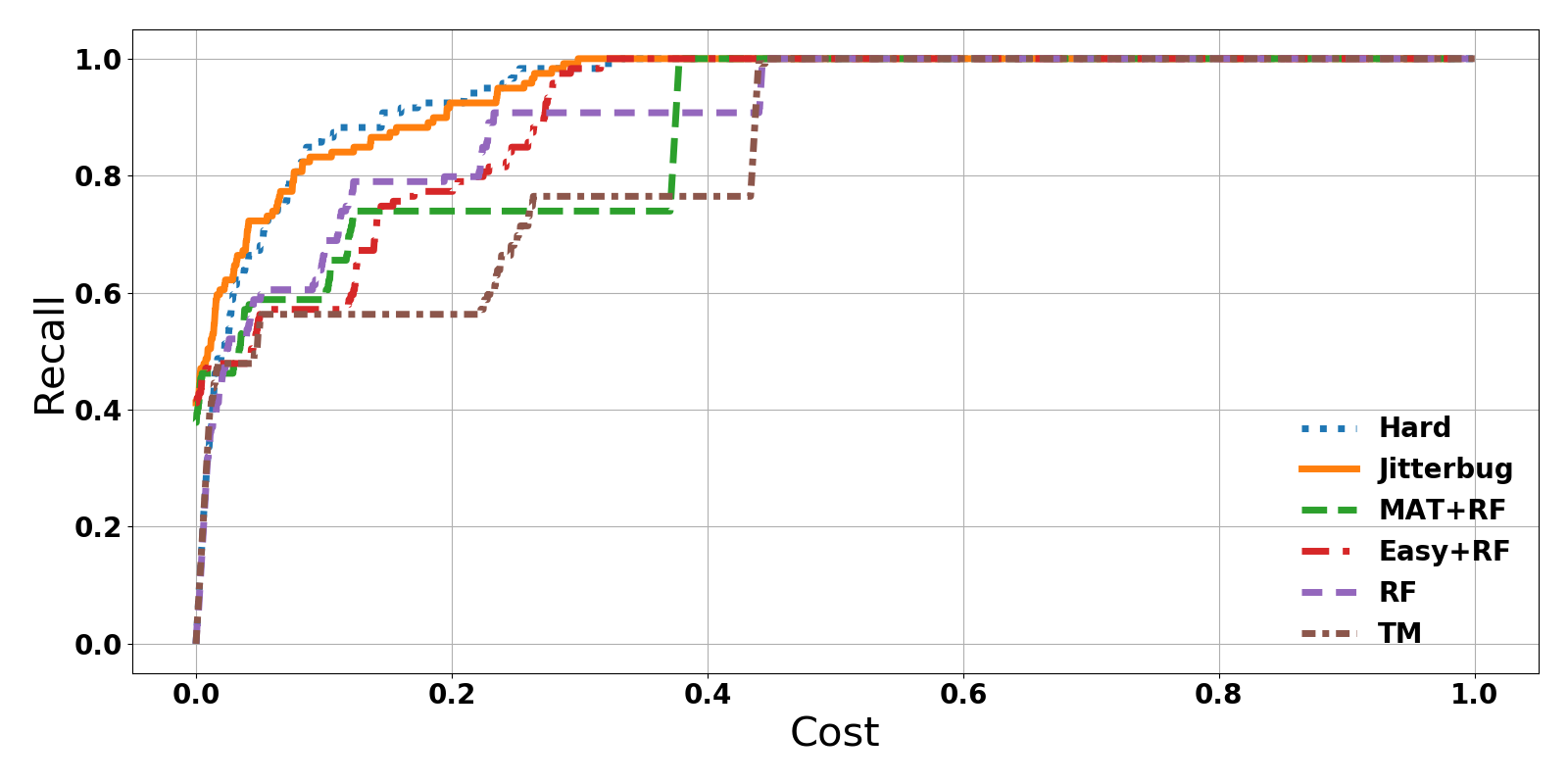}
    }\quad
    \subfloat[Hibernate]
    {
        \includegraphics[width=0.45\linewidth]{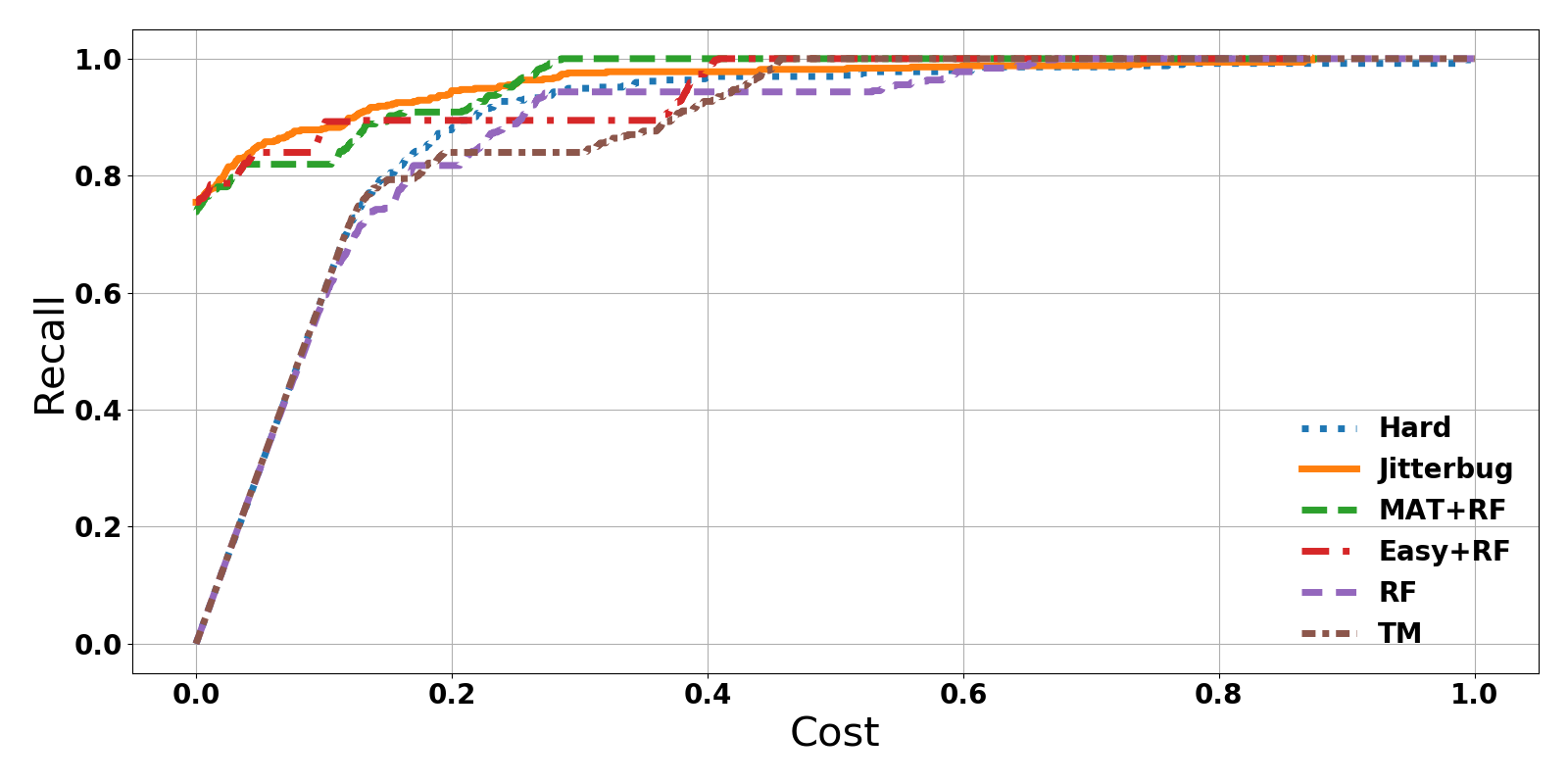}
    }\\
    \subfloat[JEdit]
    {
        \includegraphics[width=0.45\linewidth]{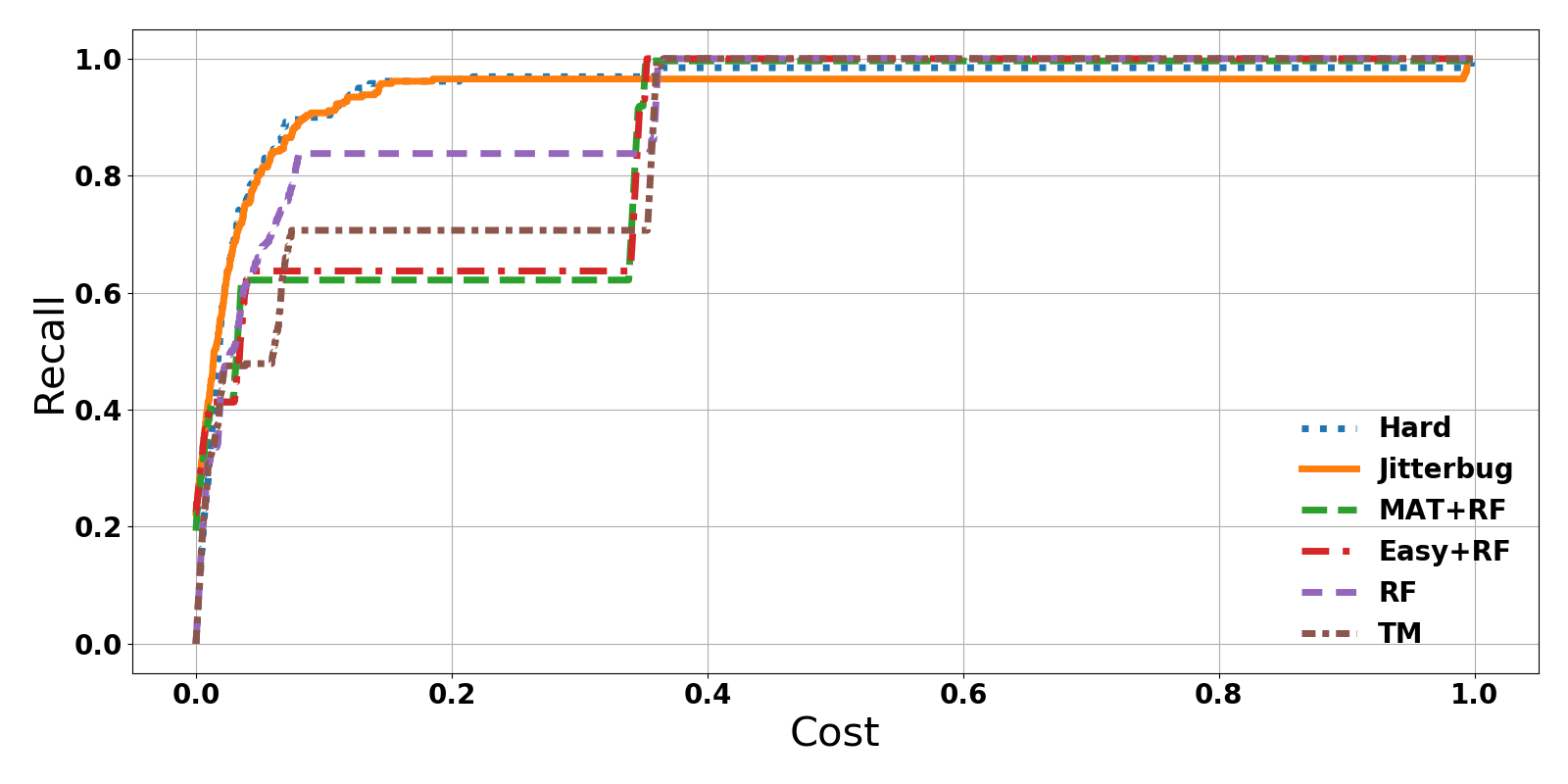}
    }\quad
    \subfloat[JFreeChart]
    {
        \includegraphics[width=0.45\linewidth]{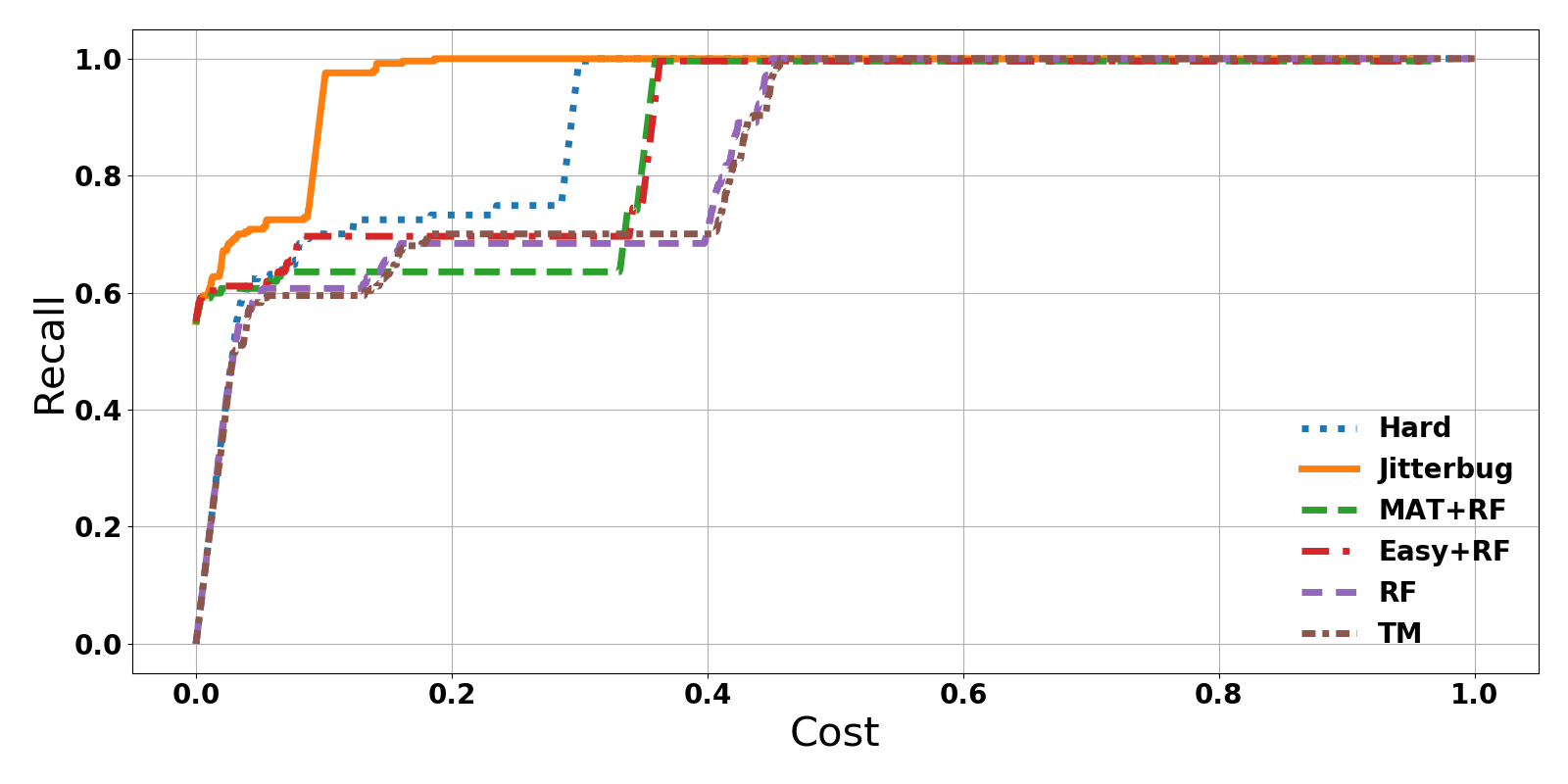}
    }\\
    \subfloat[JRuby]
    {
        \includegraphics[width=0.45\linewidth]{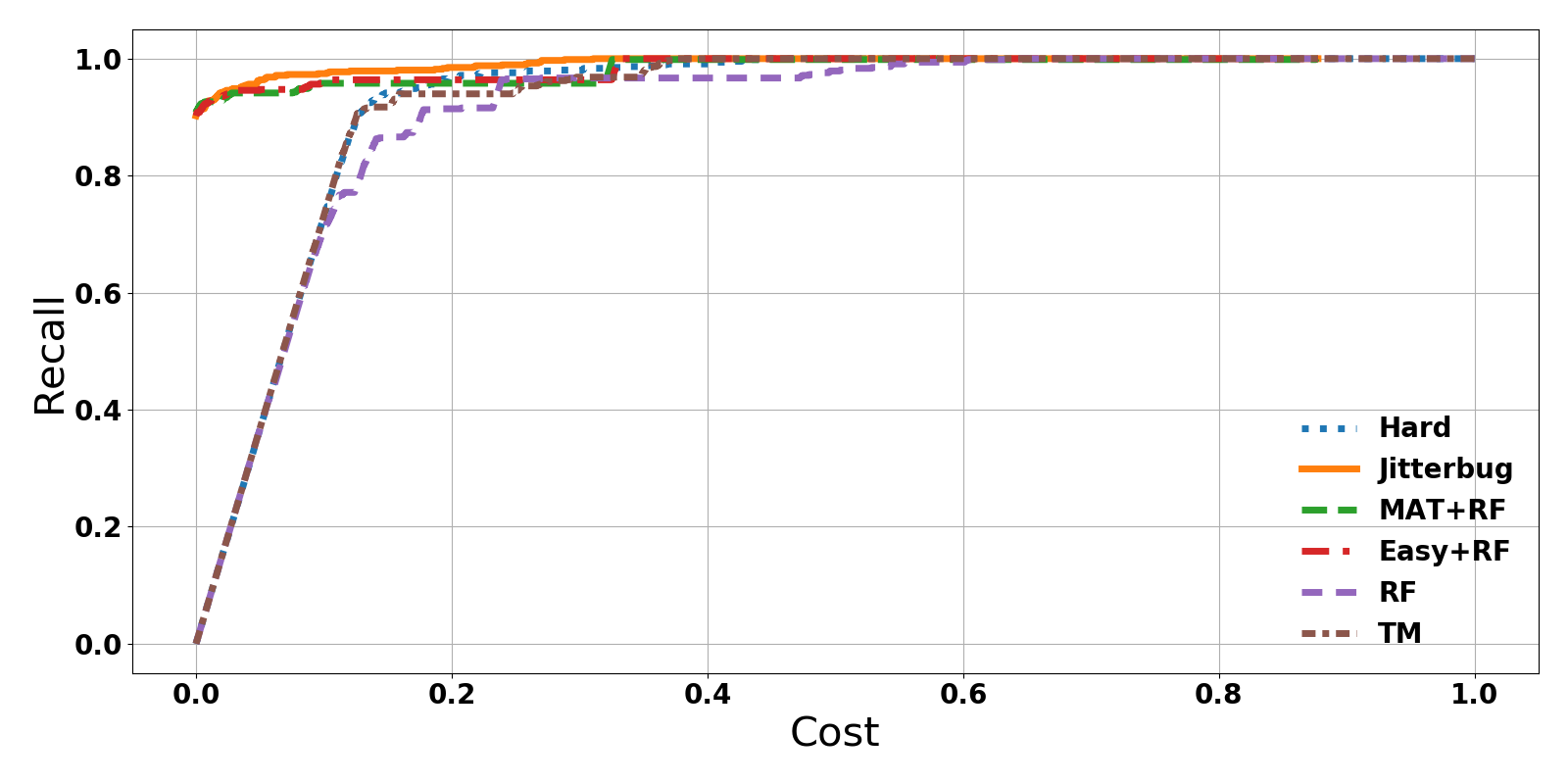}
    }\quad
    \subfloat[SQuirrel]
    {
        \includegraphics[width=0.45\linewidth]{figures/overall/sql12.png}
    }\\
    \caption{Recall-cost curves for finding all SATDs on every target project.}
    \label{fig:overall_all}
\end{figure*}

This appendix shows the recall-cost curves on every target project in \fig{rest_all}, \fig{est_all}, and \fig{overall_all}.

\section*{Acknowledgements}\label{sec:acks}
This research was partially funded by a National Science Foundation Grant \#1703487. The authors would like to thank Maldonado and Shihab for making their SATD data available to the public, which makes this research possible.

\bibliographystyle{IEEEtran}

\bibliography{mybib}

\vspace{-10 mm}
\begin{IEEEbiography}[{\includegraphics[width=1in,clip,keepaspectratio]{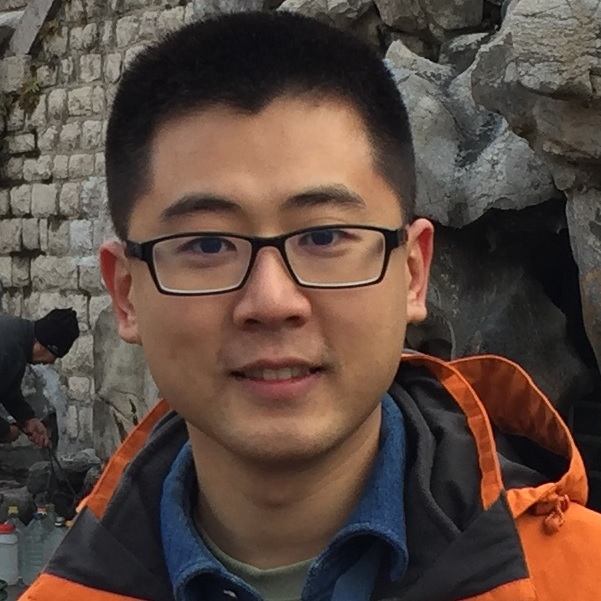}}]{Zhe Yu} (Ph.D. NC State University, 2020) is an
assistant professor in the Department of Software Engineering at Rochester Institute of Technology, where he teaches data mining
and software engineering. His research explores collaborations of human and machine learning algorithms that leads to
better performance and higher efficiency.
For more information, please visit \url{http://zhe-yu.github.io/}.
\end{IEEEbiography}
\vspace{-10 mm}
\begin{IEEEbiography}[{\includegraphics[width=1in,clip,keepaspectratio]{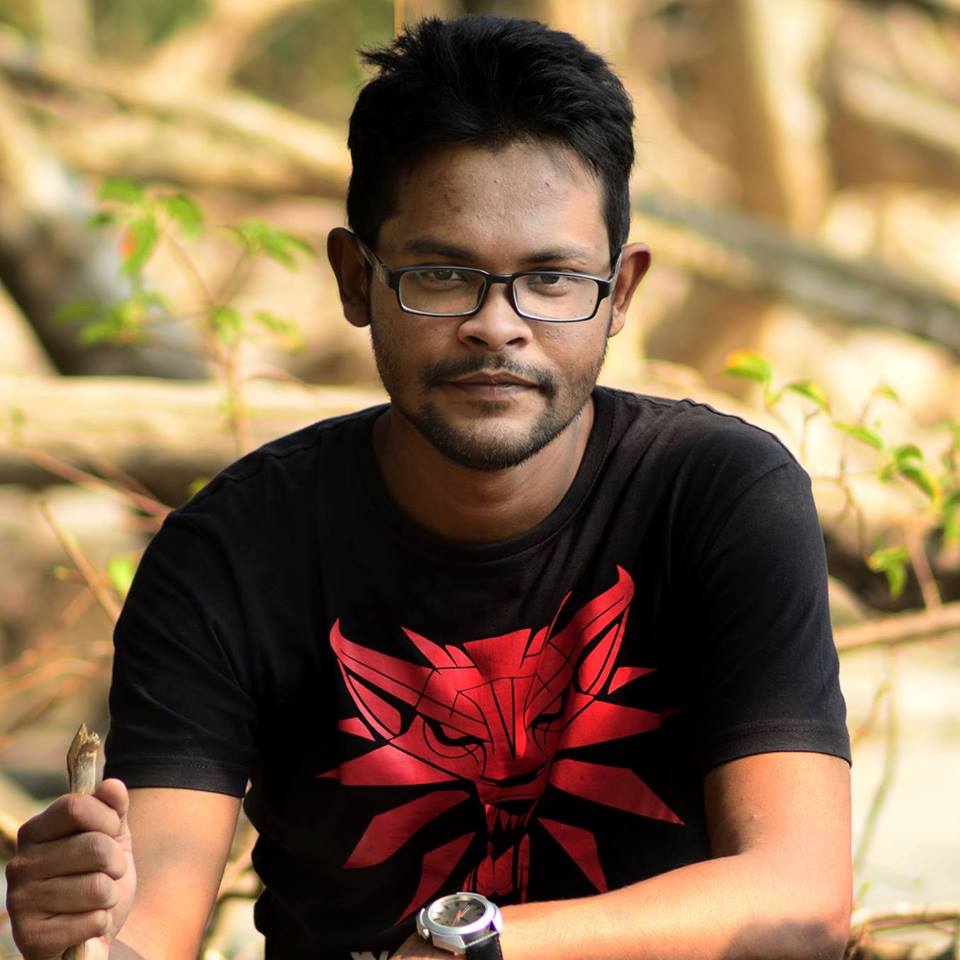}}]{Fahmid Morshed Fahid}
is a Ph.D. student in the IntelliMedia Lab at North Carolina State University, under the supervision of Dr. James Lester. His research interest is Intelligent Learning Environment. Before joining NC State, He worked as a Software Engineer in Research and Development department of Reve System Ltd. and received his Bachelors Degree in Computer Science and Engineering from Bangladesh University of Engineering and Technology (BUET) . \url{http://fahmidmorshed.github.io/}
\end{IEEEbiography}
\vspace{-10 mm}
\begin{IEEEbiography}[{\includegraphics[width=1in,clip,keepaspectratio]{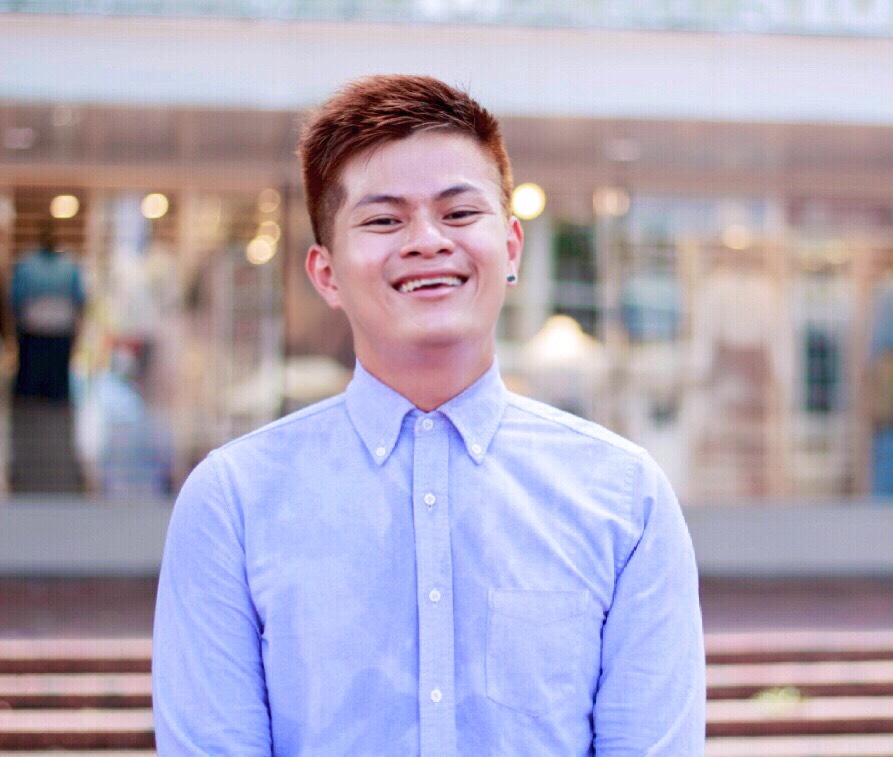}}]{Huy Tu}
is a Ph.D. student in the department of Computer Science at North Carolina State University. He explores machine learning models that support and leverage from the human experience to solve real-world problems in software engineering. For more information, please visit \url{http://kentu.us}.
\end{IEEEbiography}
\vspace{-10 mm}
\begin{IEEEbiography}[{\includegraphics[width=1in,clip,keepaspectratio]{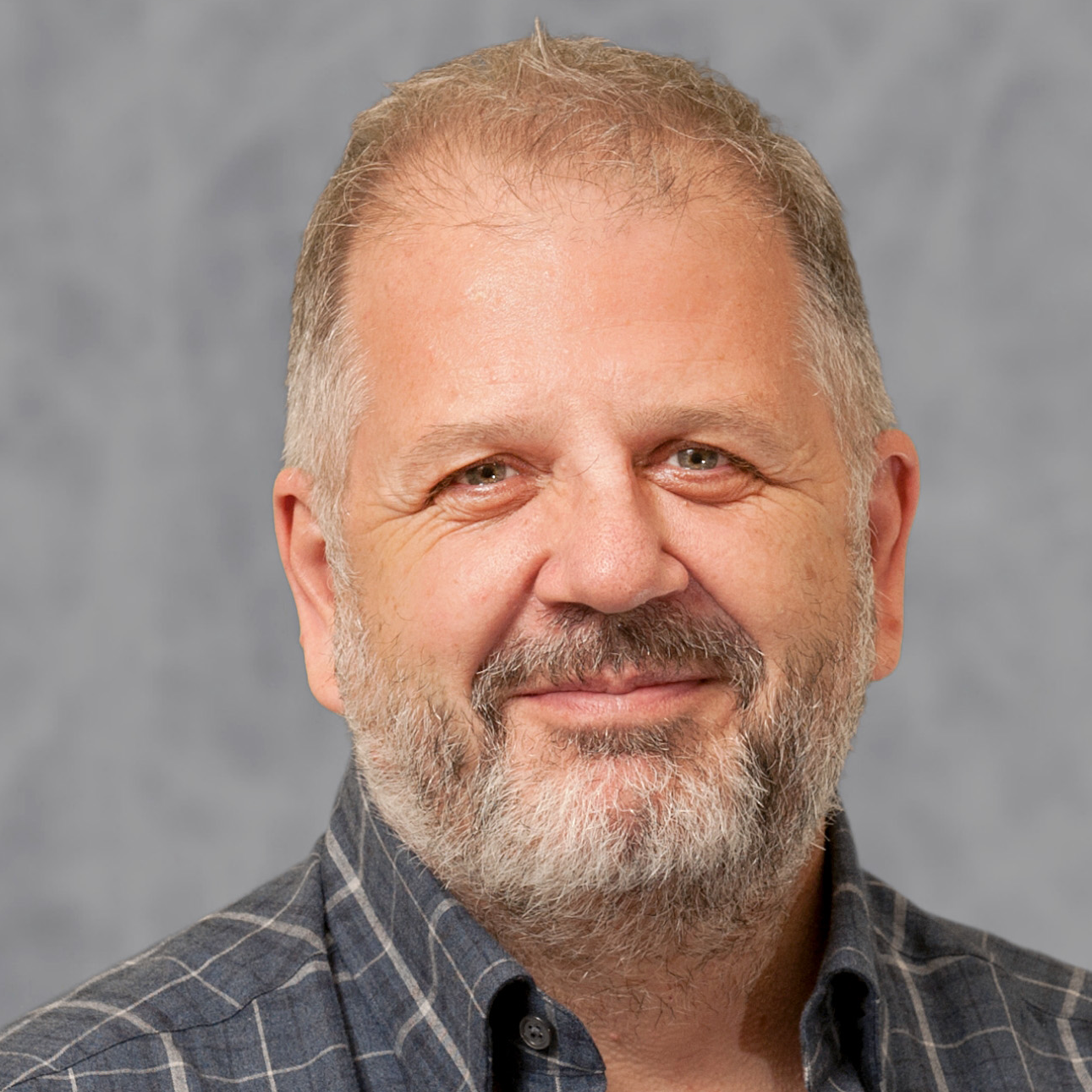}}]{Tim Menzies} (IEEE Fellow)
is a Professor in CS at NcState  His research interests include software engineering (SE), data mining, artificial intelligence, search-based SE, and open access science. \url{http://menzies.us}
\end{IEEEbiography}

\end{document}